\documentclass[a4paper,fleqn,usenatbib]{mnras}
\usepackage{txfonts}
\usepackage[T1]{fontenc}
\usepackage{ae,aecompl}
\usepackage{graphicx}
\usepackage{float}
\usepackage{caption}
\usepackage{appendix}
\usepackage{chngcntr}
\usepackage{lscape}
\usepackage{comment}

\title[Radio morphology of southern NLS1s with VLA observations]{Radio morphology of southern narrow-line Seyfert 1 galaxies with Very Large Array observations}

\author[S. Chen et al.]{
S. Chen, $^{1}$ \thanks{E-mail: sina.chen.120@gmail.com}
E. J{\"a}rvel{\"a}, $^{2}$
L. Crepaldi, $^{3}$
M. Zhou, $^{4,5}$
S. Ciroi, $^{3,6}$
M. Berton, $^{7,8}$
\newauthor
P. Kharb, $^{9}$
L. Foschini, $^{10}$
M. Gu, $^{4}$
G. La Mura, $^{11}$
and A. Vietri. $^{3}$
\\
$^{1}$ Physics Department, Technion, Haifa 32000, Israel \\
$^{2}$ European Space Agency, European Space Astronomy Centre, C/Bajo el Castillo s/n, 28692 Villanueva de la Ca{\~n}ada, Madrid, Spain \\
$^{3}$ Dipartimento di Fisica e Astronomia "G. Galilei", Universit{\`a} di Padova, Vicolo dell'Osservatorio 3, 35122 Padova, Italy \\
$^{4}$ Key Laboratory for Research in Galaxies and Cosmology, Shanghai Astronomical Observatory, Chinese Academy of Sciences, 80 \\ Nandan Road, Shanghai 200030, China \\
$^{5}$ University of Chinese Academy of Sciences, 19A Yuquan Road, Beijing 100049, China \\
$^{6}$ INAF - Osservatorio Astronomico di Padova, Vicolo dell'Osservatorio 5, 35122 Padova, Italy \\
$^{7}$ Finnish Centre for Astronomy with ESO (FINCA), University of Turku, Quantum, Vesilinnantie 5, 20014 Turku, Finland \\
$^{8}$ Aalto University Mets{\"a}hovi Radio Observatory, Mets{\"a}hovintie 114, FIN-02540 Kylm{\"a}l{\"a}, Finland \\
$^{9}$ National Centre for Radio Astrophysics - Tata Institute of Fundamental Research, Post Bag 3, Ganeshkhind, Pune 411007, India \\
$^{10}$ INAF - Osservatorio Astronomico di Brera, Via E. Bianchi 46, 23807, Merate (LC), Italy \\
$^{11}$ LIP - Laboratory of Instrumentation and Experimental Particle Physics, Av. Prof. Gama Pinto 2, 1649-003 Lisboa, Portugal \\
}

\date{Accepted XXX, Received YYY, in original form ZZZ.}

\pubyear{2020}

\begin{document}
\label{firstpage}
\pagerange{\pageref{firstpage}--\pageref{lastpage}}
\maketitle

\begin{abstract}

We present the results of new radio observations carried out with the Karl G. Jansky Very Large Array C-configuration at 5.5 GHz for a sample of southern narrow-line Seyfert 1 galaxies (NLS1s). This work increases the number of known radio-detected NLS1s in the southern hemisphere, and confirms that the radio emission of NLS1s is mainly concentrated in a central region at kpc-scale and only a few sources show diffuse emission. In radio-quiet NLS1s, the radio luminosity tends to be higher in steep-spectrum sources and be lower in flat-spectrum sources, which is opposite to radio-loud NLS1s. This may be because the radio emission of steep NLS1s is dominated by misaligned jets, AGN-driven outflows, or star formation superposing on a compact core. Instead the radio emission of flat NLS1s may be produced by a central core which has not yet developed radio jets and outflows. We discover new NLS1s harboring kpc-scale radio jets and confirm that a powerful jet does not require a large-mass black hole to be generated. We also find sources dominated by star formation. These NLS1s could be new candidates in investigating the radio emission of different mechanisms.

\end{abstract}

\begin{keywords}
galaxies: active; galaxies: nuclei; galaxies: Seyfert; galaxies: jets; radio continuum: galaxies
\end{keywords}

\section{Introduction}

Narrow-line Seyfert 1 galaxies (NLS1s), which are a subclass of active galactic nuclei (AGN), are identified by their unusual optical spectral properties. They exhibit Balmer lines from the broad-line region (BLR) with a full width at half maximum of FWHM(H$\beta$) $<$ 2000 km s$^{-1}$ which is only slightly broader than forbidden lines, and a flux ratio of [O III]$\lambda$5007 / H$\beta$ $<$ 3 \citep{Osterbrock1985,Goodrich1989}. Besides, NLS1s show several extreme optical and X-ray properties, such as strong Fe II multiplets emission \citep{Boroson1992}, frequently observed blueshifted line profiles \citep{Zamanov2002,Boroson2005,Leighly2004}, soft X-ray spectra \citep{Wang1996,Boller1996,Leighly1999a}, and rapid X-ray variability \citep{Pounds1995,Leighly1999b}. Many studies suggest that NLS1s are powered by a relatively low-mass central black hole with $M_{\rm{BH}} \sim 10^{6-8} M_{\odot}$, with respect to broad-line Seyfert 1 galaxies (BLS1s) with $M_{\rm{BH}} \sim 10^{7-9} M_{\odot}$ \citep{Jarvela2015,Cracco2016,Chen2018a}. Since the bolometric luminosity of NLS1s is comparable to that of BLS1s, a low black hole mass corresponds to a high Eddington ratio \citep{Boroson1992,Collin2004}. These properties suggest that NLS1s are likely to be a young and fast-growing phase of AGN \citep{Mathur2000,Grupe2000}. Another possibility is that the low black hole mass of NLS1s is just an inclination effect, due to the pole-on orientation of a disk-like BLR which prevents us from seeing any Doppler broadening \citep{Decarli2008}. However, several studies reveal that this hypothesis is likely not true in the majority of NLS1s and the differences between BLS1 and NLS1 populations should be intrinsic rather than just due to orientation effects \citep{Kollatschny2011,Kollatschny2013,Jarvela2017,Komossa2018a,Vietri2018,Berton2020a}.

The majority of NLS1 population is radio-quiet (RQ) with a radio loudness of $R <$ 10 \footnote{The radio loudness is defined as the ratio of 5 GHz radio flux density and optical B-band flux density $R = S_{5\rm{GHz}} / S_{4400\rm{\AA}}$ \citep{Kellermann1989}.}. The radio-loud (RL) fraction of NLS1s is very small, only $\sim$ 7$\%$ with $R >$ 10 and $\sim$ 2.5$\%$ with $R >$ 100 \citep{Komossa2006}, compared with that of BLS1s and quasars about $\sim$ 10$\%$-15$\%$ \citep{Ivezic2002}. However, the definition of radio loudness could be misleading, especially for NLS1s \citep{Foschini2011c,Berton2020b}. For instance, star formation can produce strong radio emission at low frequencies in late-type galaxies. AGN with such kind of host galaxies may be classified as RL even without launching a relativistic jet \citep{Caccianiga2015,Ganci2019}. On the contrary, if the relativistic jet is faint, or misaligned, or absorbed by ionized material at low frequencies, the radio emission is weak thus the AGN may be classified as RQ \citep{Foschini2011a,Foschini2012c}. Moreover, \citet{Sikora2007} found a similar dependence of the radio loudness increases with the decreasing Eddington ratio in both spiral-hosted and elliptical-hosted AGN, which implies that the radio loudness alone has little meaning if the brightness and accretion rate are not considered. Indeed, radio observations found some jet-like structures in RQ Seyfert galaxies \citep{Mundell2009,Giroletti2009,Doi2013}. Furthermore, \citet{Lahteenmaki2018} found strong radio emission at $\sim$ 1 Jy level at 37 GHz from sources formerly classified as RQ. Such strong radio emission at such a high frequency can not be explained by radiation from supernova remnants and can only be produced by a relativistic jet. Indeed, \citet{Padovani2017} suggested that AGN should be classified as jetted and non-jetted AGN, instead of using the $R$ parameter. Jetted sources are characterized by powerful relativistic jets. Instead, non-jetted sources do not launch a jet but may have outflows with considerably less power compared to jetted sources. This classification is based on physical differences rather than observational phenomena.

RL NLS1s are likely to harbor powerful relativistic jets, which are the predominant origin of the radio, optical, partially X-ray, and in some cases, $\gamma$-ray radiation. The discovery of $\gamma$-ray emission from RL NLS1s detected by the \textit{Fermi} Large Area Telescope (LAT) confirmed the presence of relativistic beamed jets in this type of AGN along with blazars and radio galaxies \citep{Abdo2009a,Abdo2009b,Abdo2009c,Foschini2011b,Foschini2015,Yao2015,Liao2015,D'Ammando2015,D'Ammando2016,Paliya2016,Berton2017,Paliya2018,Lahteenmaki2018,Yao2019}. In addition, radio observations of several RL NLS1s with the Very Long Baseline Array (VLBA) showed that they exhibit pc-scale radio jets, compact radio morphologies, flat or inverted radio spectra, high brightness temperatures, and a significant degree of polarization \citep{Abdo2009a,Abdo2009b,Gu2015,Lister2018}. Some radio monitoring campaigns also revealed the presence of fast and intense variability which is a common feature of relativistic jets \citep{Foschini2012a,Angelakis2015,Fuhrmann2016}. In some cases, an apparent superluminal motion was observed as well \citep{Lister2013,Lister2016,Lister2019}. These findings pose a challenging question about the generation and evolution of the jet system in NLS1s harboring a relatively undermassive black hole.

The origin of the radio emission in RQ NLS1s is not established yet. The possible mechanisms might include a low-power jet, an AGN-driven outflow, an accretion disk corona, star formation, and a combination of them \citep{Panessa2019}. Radio observations can partly distinguish these physical scenarios. A compact radio core at pc-scale with a flat or inverted spectral slope is probably associated with the corona \citep{Guedel1993,Laor2008}, the jet-base \citep{Blandford1979,Reynolds1982}, or the outflow-base \citep{Lena2015}. This central core is thought to be driven by the non-thermal processes with a high brightness temperature of $T_{\rm{b}} > 10^7$ K \citep{Blundell1998}. The jet/outflow base may physically coincide with the corona \citep{Merloni2002,King2017}, and they are indistinguishable in the radio images. The linear extended radio emission resolved at various scales from pc to kpc could be related to different components of the jet and/or outflow and be characterized by steep spectral indexes. The difference between a jet and an outflow depends on whether the ejected plasma is collimated or not \citep{Begelman1984}. At present, it is hard to separate them based on radio information only. Star formation activity can produce both free-free emission and synchrotron emission \citep{Condon1992}. The radio images exhibit host-like distributed diffuse emission surrounding the central core with a steep spectral slope and a low surface brightness of $T_{\rm{b}} < 10^5$ K \citep{Orienti2015}. However, these radio properties are limited by the resolution and sensitivity of the radio observations.

In order to study the radio properties of NLS1s and to identify sources with large radio jets, we proposed new observations with the Karl G. Jansky Very Large Array (VLA, Proposal ID: VLA/18B-126). This paper presents the main results of this survey and is organized as follows. In Section 2 we describe the data reduction, in Section 3 we present the data analysis, in Section 4 we discuss the main results, and in Section 5 we provide the summary. Throughout this work, we adopt a standard $\Lambda$CDM cosmology with a Hubble constant $H_{0}$ = 70 km s$^{-1}$ Mpc$^{-1}$, $\Omega_{\Lambda}$ = 0.73 and $\Omega_{\rm{M}}$ = 0.27 \citep{Komatsu2011}. We assume the flux density and spectral index convention at observed frequency is $S_{\nu} \propto \nu^{-\alpha}$.

\section{Data reduction}

The targets were selected from \citet{Chen2018a}, who presented a catalog of NLS1s in the southern hemisphere. It includes 168 \footnote{There are 167 NLS1s in \citet{Chen2018a}. We included one more source IRAS 13224$-$3809 as it was classified as NLS1 in \citet{Boller1993} and also detected with the 6dFGS.} NLS1s which were classified according to their optical spectra from the Six-degree Field Galaxy Survey (6dFGS) \footnote{\url{http://www-wfau.roe.ac.uk/6dFGS/}}. We observed 62 sources which are higher than the declination limit of the VLA ($-30^{\circ}$). The redshifts of these targets range from 0.01 to 0.44.

We were granted 55 hours observing time in C-configuration centered at 5.5 GHz with a bandwidth of 2 GHz and an angular resolution of 3.5 arcsec for this NLS1 sample. These observations were carried out between November 2018 and February 2019 with an exposure time of 30 minutes on each target, yielding an image sensitivity of $\sim$ 7 $\mu$Jy beam$^{-1}$. Such a resolution and sensitivity enable us to see diffuse emission that is missing in A-configuration and discover extended structure. We reduced the data using the VLA calibration pipeline version 5.4.0 and the Common Astronomy Software Applications (CASA) version 5.5.0. A standard flux density calibrator, 3C 147 or 3C 286, was used for every target. The measurement set is split into 16 spectral windows (centered at 4.55, 4.68, 4.81, 4.94, 5.06, 5.19, 5.32, 5.45, 5.55, 5.68, 5.81, 5.94, 6.06, 6.19, 6.32, 6.45 GHz).

To produce the radio maps, we used a cell size of 0.5 arcsec to properly sample the beam that has a FWHM of 3.5 arcsec. The maps were created in a region of 2048 $\times$ 2048 pixels centered on the source coordinates to check for the presence of nearby sources. We modeled the main target along with the nearby sources using the CLEAN algorithm in all spectral windows to avoid the contamination from sidelobes. If a bright source is outside the mapped region and its sidelobes do affect the target, we enlarged the image to include the bright source. In this way, we can model the extra source to reduce its effect. We used a natural weighting to create the first tentative image. For some faint sources, the noise level approaches the predicted image sensitivity of $\sim$ 7 $\mu$Jy beam$^{-1}$ after the first cleaning, and we did not proceed any further. Conversely, the noise level of some bright sources remained high after the first cleaning, and we applied iterative cycles of phase-only self-calibration with a Briggs weighting, intermediate between natural and uniform, on the visibilities to improve the dynamic range of the final maps.

We modeled the source with a Gaussian fit on the image plane and deconvolved it from the beam, to recover the radio position, the core size and its position angle, and the integrated and peak flux densities, $S_{\rm{int}}$ and $S_{\rm{p}}$, centered at 5.5 GHz. If the core size is too small to be determined compared to the beam size, we adopted a half beam size as an upper limit for the core size. We only modeled the central core if the source is resolved in more than one component. Thus we remark that our flux densities could be underestimated if the source displays some resolved structures. For a few sources, the integrated flux density is less than the peak flux density due to either the small core size with respect to the beam size or errors in the estimation of the integrated flux density. In this case, we treated $S_{\rm{p}}$ as the upper limit of $S_{\rm{int}}$. The root-mean-square (RMS) was estimated in a source-free region. The errors of each measurement were produced by CASA. We obtained the integrated flux densities ranging from 0.05 to 21.28 mJy and the average RMS of 10 $\mu$Jy beam$^{-1}$.

In addition, we split the data in 0-7 and 8-15 spectral windows with a bandwidth of 1 GHz and created another two maps centered at 5 and 6 GHz following the reduction procedures mentioned above. We did not use a narrower bandwidth in order to minimize the uncertainty of each measurement. We then smoothed the 6 GHz image with a Gaussian kernel to match the resolution of the 5 GHz image using the task IMSMOOTH, with the aim of evaluating the in-band spectral indexes.

In total, 49 sources have a detection in these observations. 13 sources are not detected or only detected at 3$\sigma$ level because (a) the target has either weak radio emission or high noise background, or (b) the image is affected by strong sidelobes of a nearby bright source. The detection rate at 5.5 GHz of this NLS1 sample is 79$\%$ (49/62), which suggests that most NLS1s are radio emitters when observed with an adequate sensitivity. The radio maps of southern NLS1s observed with the VLA are shown in Figs. \ref{J0354} and \ref{radiomap}.

\section{Data analysis}

\subsection{5.5 GHz}

We calculated the radio loudness according to the definition
\begin{equation}
R = S_{5\rm{GHz}} / S_{4400\rm{\AA}}
\label{eqr}
\end{equation}
where $S_{4400\rm{\AA}}$ is the flux density in optical B-band derived from the optical spectra \citep{Chen2018a}. Despite the boundary of $R$ = 10 being arbitrary, it can give us an idea of the fraction of radio flux with respect to optical flux and make our results comparable to other studies using this parameter. There are 45 RQ NLS1s in the sample. Only four sources (J0122$-$2646, J0452$-$2953, J0846$-$1214, and J2021$-$2235) have $R >$ 10, but not very high with $R \sim$ 14-38. The RL fraction is 6.5$\%$ (4/62) in the whole sample. This is consistent with \citet{Komossa2006} who found that the fraction of RL NLS1s is about 7$\%$. However, the radio loudness strongly depends on whether the optical and radio observations include the whole galaxy (both AGN and host) or only the nucleus \citep{Ho2001b,Kharb2014}. Besides, the RL fraction tends to be lower in the nearby universe and be higher with an increasing large-scale environment density \citep{Cracco2016,Jarvela2017}.

To study the concentration of radio emission, we estimated the fraction of peak to integrated flux density following \citet{Berton2018}
\begin{equation}
f = S_{\rm{p}} / S_{\rm{int}}.
\end{equation}
We considered that the radio emission is concentrated (C) in a compact core at the center if $f \geq$ 0.5 and is diffuse (D) in a region slightly larger than the angular resolution if $f <$ 0.5. In the case of the upper limit of $S_{\rm{int}}$ equal to $S_{\rm{p}}$, we get $f$ = 1 which means that all the flux densities are concentrated in a central beam. These fractions range from 0.3 to 1. For the majority of NLS1s (45 objects), the radio emission is mainly concentrated in a central core. Only in four RQ sources (J0447$-$0403, J0622$-$2317, J0845$-$0732, and J1937$-$0613), the radio emission spreads over a region somewhat larger than the core. This could be because Seyfert galaxies and low luminosity AGN are typically unresolved at arcsec-scale \citep{Ho2001a} and confirms that RL sources do not show a large fraction of diffuse emission.

In addition, we measured the in-band spectral index between 5 and 6 GHz $\alpha_{\rm{in-band}}$ by modeling the spectrum with a power-law and giving the definition
\begin{equation}
\alpha = - \frac{\log(S_2/S_1)}{\log(\nu_2/\nu_1)}
\label{eqa}
\end{equation}
where $S_1$ and $S_2$ are the integrated flux densities at the observing frequencies $\nu_1$ = 5 GHz and $\nu_2$ = 6 GHz respectively. There are 18 sources having a flat (F) radio spectrum with $\alpha <$ 0.5 and 31 sources having a steep (S) radio spectrum with $\alpha \geq$ 0.5. The in-band spectral indexes range from $-$1.91 to 3.11 with a mean value of $\alpha_{\rm{in-band}}$ = 0.53. Nevertheless, we remark that the in-band spectral indexes have large errors due to the narrow bandwidth between 5 and 6 GHz.

The integrated and peak luminosities, $L_{\rm{int}}$ and $L_{\rm{p}}$, at the observing frequency $\nu$ = 5.5 GHz were derived by
\begin{equation}
L_{\nu} = 4 \pi D_{\rm{L}}^2 \nu S_{\nu} (1+z)^{(\alpha-1)}
\label{eql}
\end{equation}
where $D_{\rm{L}}$ is the luminosity distance at a cosmological redshift and $\alpha$ is the in-band spectral index. Throughout this work, the luminosity $L$ is measured in unit erg s$^{-1}$. The integrated luminosities at 5.5 GHz have a range of $\log L_{\rm{int}} \sim$ 37.5-40.7 with an average value of $\log L_{\rm{int}}$ = 39.0.

Besides, we also collected the central black hole masses from \citet{Chen2018a} who estimated the virial black hole mass via the BLR size and the H$\beta$ line dispersion. The BLR size was calculated by the relation between the BLR size and the 5100 $\rm{\AA}$ continuum luminosity \citep{Bentz2013}, and this relation was derived from reverberation mapping method \citep{Peterson2004a,Peterson2014}. The black hole masses have a range of $\log (M_{\rm{BH}} / M_{\odot}) \sim$ 6.0-7.4 with a mean value of $\log (M_{\rm{BH}} / M_{\odot})$ = 6.8, which are typical for NLS1s. During this estimation, the errors of the black hole mass are 0.1-0.2 dex. Without taking into account the radiation pressure, which is likely to be important in highly accreting AGN such as NLS1s, there will be an additional average scatter of 0.2 dex \citep{Marconi2008}. Thus the uncertainty of the black hole mass is about 0.5 dex.

\begin{figure}
\centering
\includegraphics[width=0.5\textwidth]{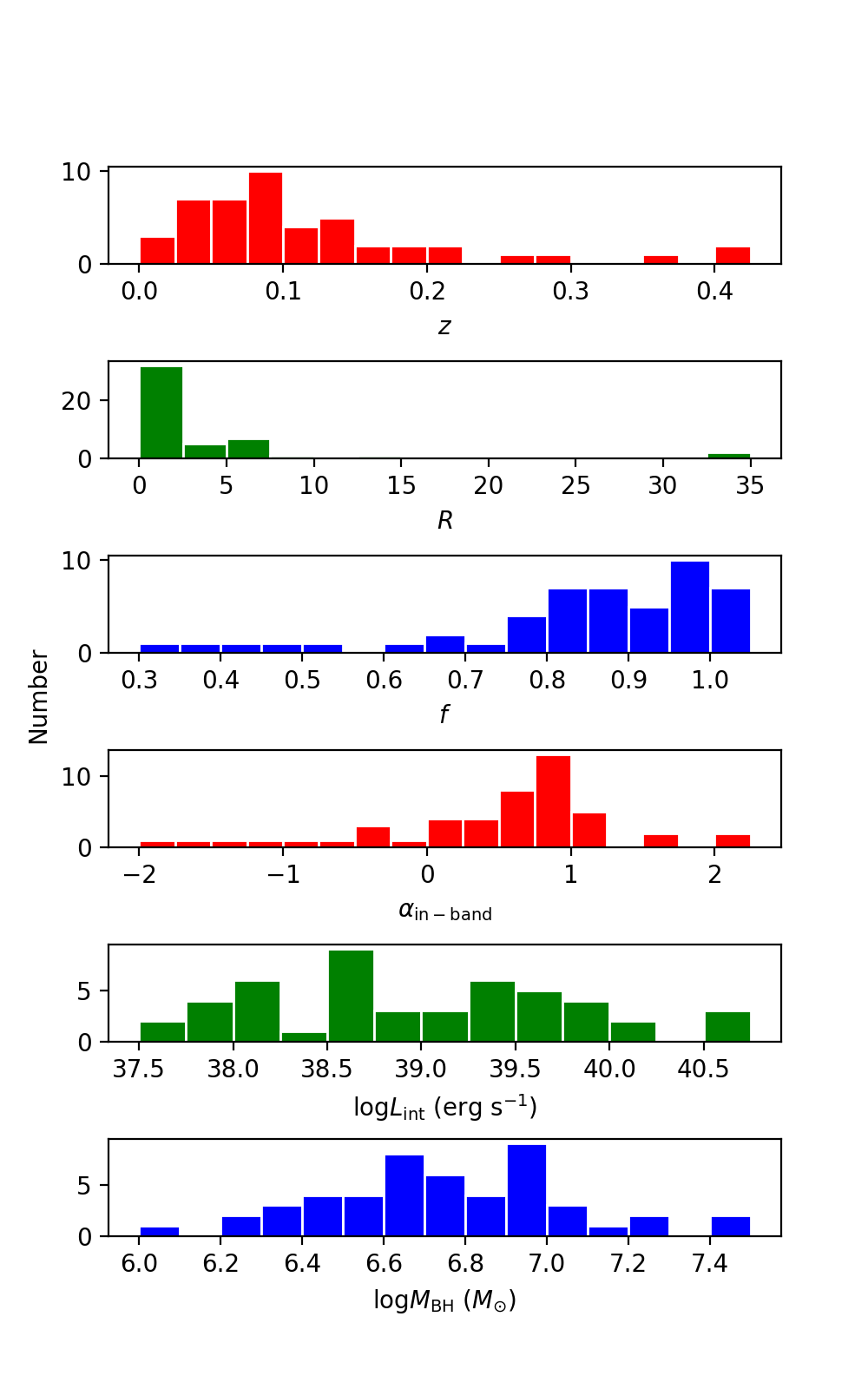}
\caption{The distributions of redshift, radio loudness, flux concentration, in-band spectral index, integrated luminosity at 5.5 GHz, and black hole mass (from top to bottom panels) using 0.025, 2.5, 0.05, 0.25, 0.25, and 0.1 bin width respectively.}
\label{hist}
\end{figure}

The distributions of redshift, radio loudness, flux concentration, in-band spectral index, integrated luminosity at 5.5 GHz, and black hole mass are plotted in Fig. \ref{hist}. The coordinates, redshifts, scales, core sizes, position angles, and integrated and peak flux densities and luminosities at 5.5 GHz of the sample are listed in Tab. \ref{radio_measurement}. The radio loudness, flux concentration, in-band spectral index between 5 and 6 GHz, different classifications, flux density in optical B-band, and black hole mass are reported in Tab. \ref{classify+optical}.

\subsection{1.4 GHz}

For comparison of the classifications derived at different frequencies, we collected the radio data from other sky surveys. There are 21 NLS1s detected at 1.4 GHz, including five sources from the Faint Images of the Radio Sky at Twenty-Centimeters (FIRST) with a resolution of 5 arcsec \citep{Helfand2015} and 16 sources from the NRAO VLA Sky Survey (NVSS) with a large restoring beam of 45 arcsec \citep{Condon1998}. If a source was detected with both surveys, we applied the flux density from the FIRST as it has a better angular resolution. Their flux densities range from 1.2 to 59.8 mJy.

We computed the radio loudness for these 21 sources using Eq. \ref{eqr}. The radio flux density at 5 GHz is derived from that at 1.4 GHz under the hypothesis of a power-law spectrum. As the spectral slope at 5.5 GHz can not precisely represent that at 1.4 GHz, we applied a conservative spectral index of $\alpha$ = 0.5 which is consistent with the average in-band spectral index, to avoid the large scatters for individual sources. The RL and RQ classifications at the two frequencies appear to be robust, as they are consistent in 20 objects and differ in just one case, J1511$-$2119, a source lying close to the $R \sim$ 10 threshold, which is classified as RQ using the 5.5 GHz flux but becomes RL using the 1.4 GHz flux.

\begin{figure}
\centering
\includegraphics[width=0.5\textwidth]{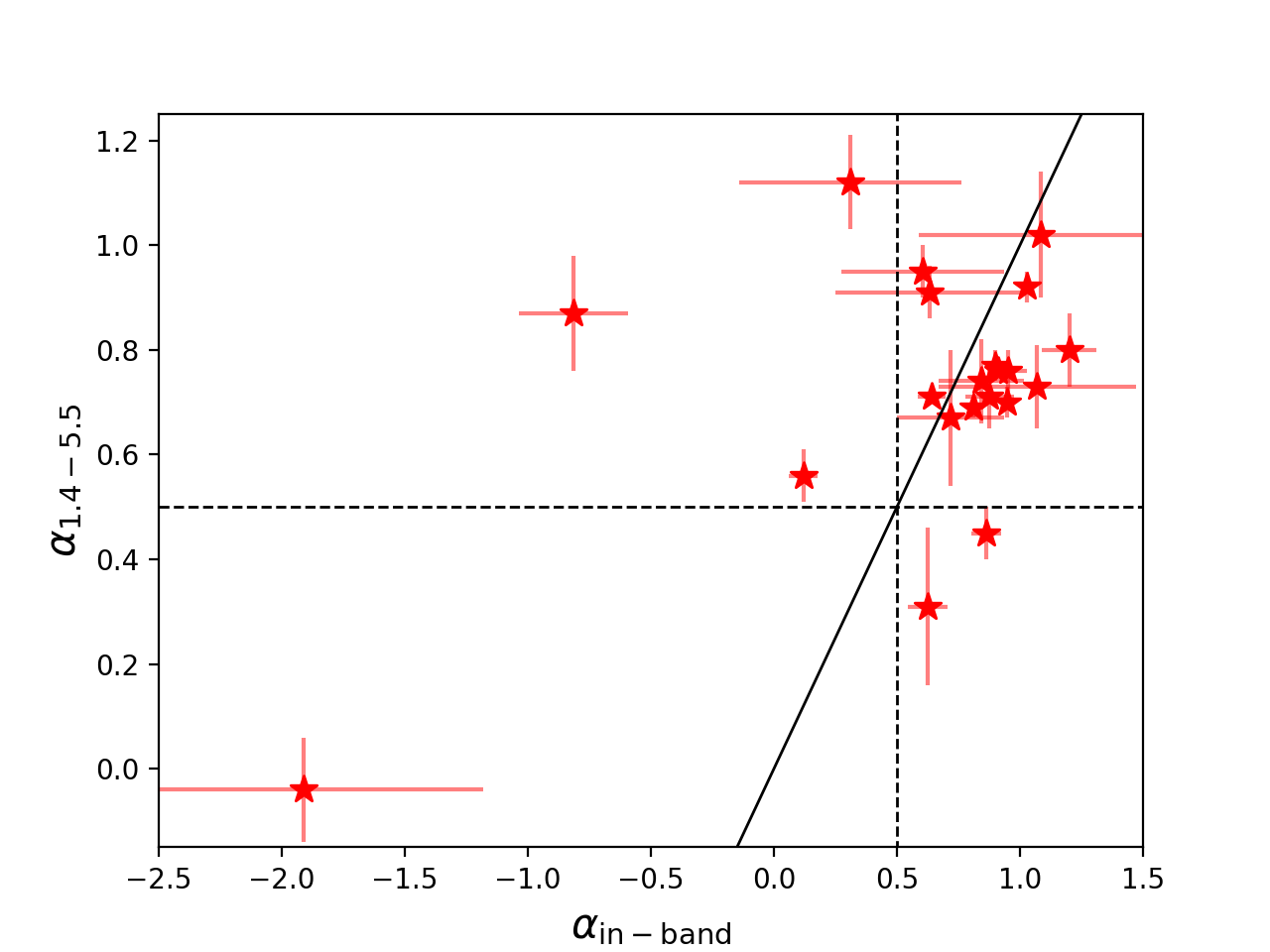}
\caption{The spectral index between 1.4 and 5.5 GHz vs. the in-band spectral index. The solid black line indicates the 1:1 ratio. The dashed lines are $\alpha_{\rm{1.4-5.5}}$ = 0.5 and $\alpha_{\rm{in-band}}$ = 0.5.}
\label{alpha}
\end{figure}

We also measured the spectral index between 1.4 and 5.5 GHz $\alpha_{\rm{1.4-5.5}}$ using Eq. \ref{eqa}. They range from $-$0.04 to 1.12 and have an average of 0.72. As a comparison, we plotted $\alpha_{\rm{1.4-5.5}}$ versus $\alpha_{\rm{in-band}}$ for 21 NLS1s shown in Fig. \ref{alpha}. We note that 16 sources keep the same spectral class, while five sources (J0354$-$1340, J0400$-$2500, J0549$-$2425, J0622$-$2317, and J1638$-$2055) show a difference in the spectral class. Besides, 14 and seven objects stay below and above the 1:1 ratio line indicating that the spectra become steeper and flatter toward higher frequencies respectively. Nevertheless, we remark that the spectral index between 1.4 and 5.5 GHz could be biased due to flux density variability and different beam sizes during these non-simultaneous observations.

The luminosity at 1.4 GHz of 21 sources was computed using Eq. \ref{eql} as well, where the spectral index between 1.4 and 5.5 GHz was used. They come within a range of $\log L_{1.4} \sim$ 37.0-40.5 and have a mean value of $\log L_{1.4}$ = 39.2. The flux density, luminosity, radio loudness, spectral index between 1.4 and 5.5 GHz, and different types based on the 1.4 GHz detections are listed in Tab. \ref{first+nvss}.

\section{Discussion}

\subsection{Radio emission concentration}

\begin{figure}
\centering
\includegraphics[width=0.5\textwidth]{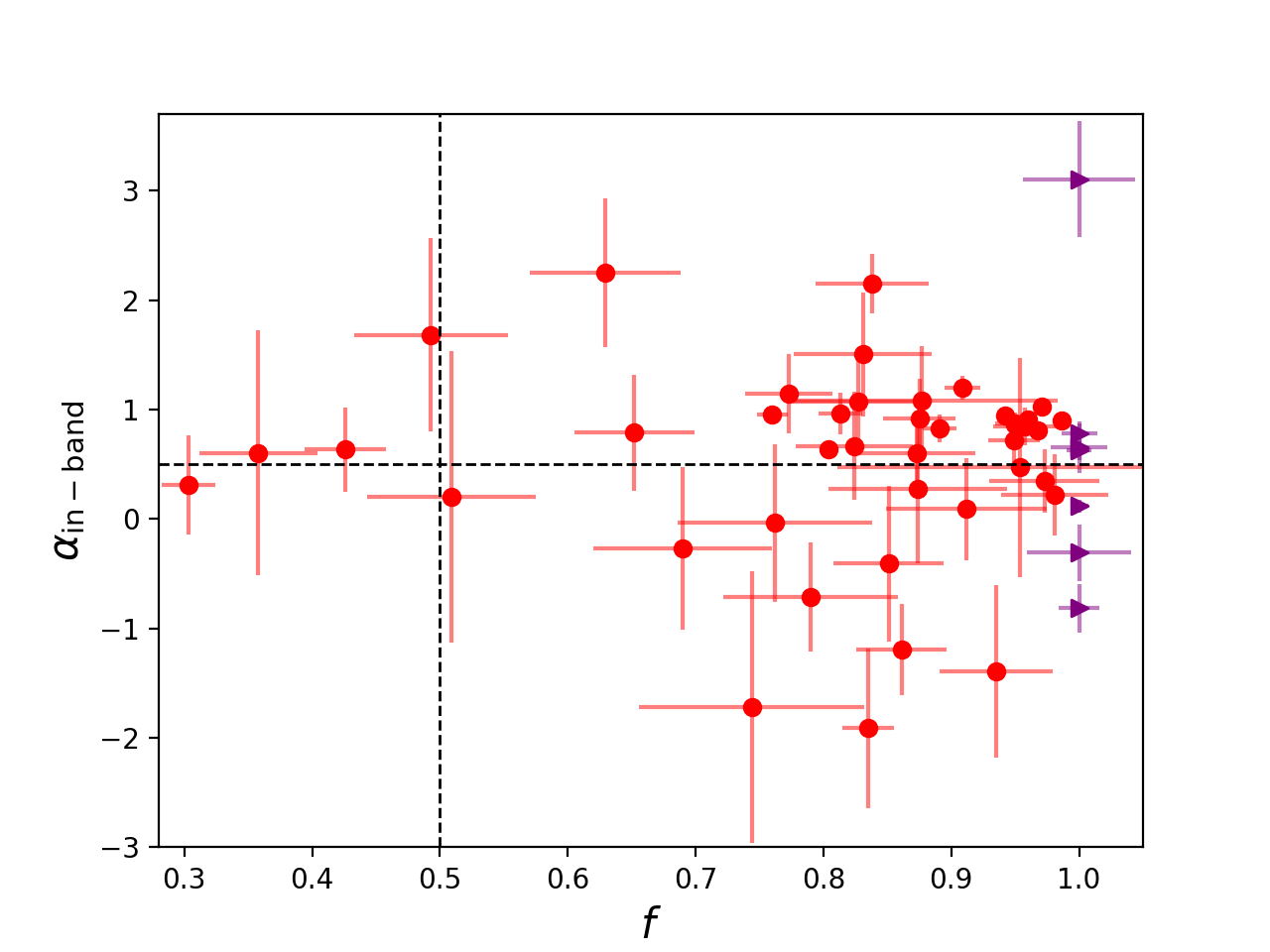}
\caption{The distribution of the in-band spectral index and the concentration of radio emission. Sources with $f$ = 1 are marked with right triangles in purple, the others are marked with circles in red. The dashed lines are $\alpha_{\rm{in-band}}$ = 0.5 and $f$ = 0.5.}
\label{alpha+fc}
\end{figure}

We plotted the distribution of the in-band spectral index $\alpha_{\rm{in-band}}$ and the concentration of radio emission $f$ shown in Fig. \ref{alpha+fc}. As expected, those sources whose flux densities are not centrally concentrated, basically have steep spectral indexes indicating that the diffuse emission is mainly from an optically thin region. For the majority of NLS1s, their flux densities are concentrated in a central core and their spectral indexes have a wide range from flat to steep. This suggests that the radio emission of NLS1s originated from either the core or the extended structure, is mainly concentrated in a region at arcsec-scale. In addition, \citet{Berton2018} found that RQ NLS1s usually show diffuse emission surrounding a central core at the resolution of the VLA A-configuration. At mas-scale, \citet{Orienti2010} found that the radio emission of Seyfert galaxies is concentrated in the central core if a flat spectrum is present, and is extended or diffuse in a large region if a steep spectrum is present. Hence it may reveal that the core emission is mainly concentrated at pc-scale, and the extended or diffuse emission can spread over a region at kpc-scale.

\subsection{Radio spectral slope}

\begin{figure*}
\centering
\includegraphics[width=0.45\textwidth]{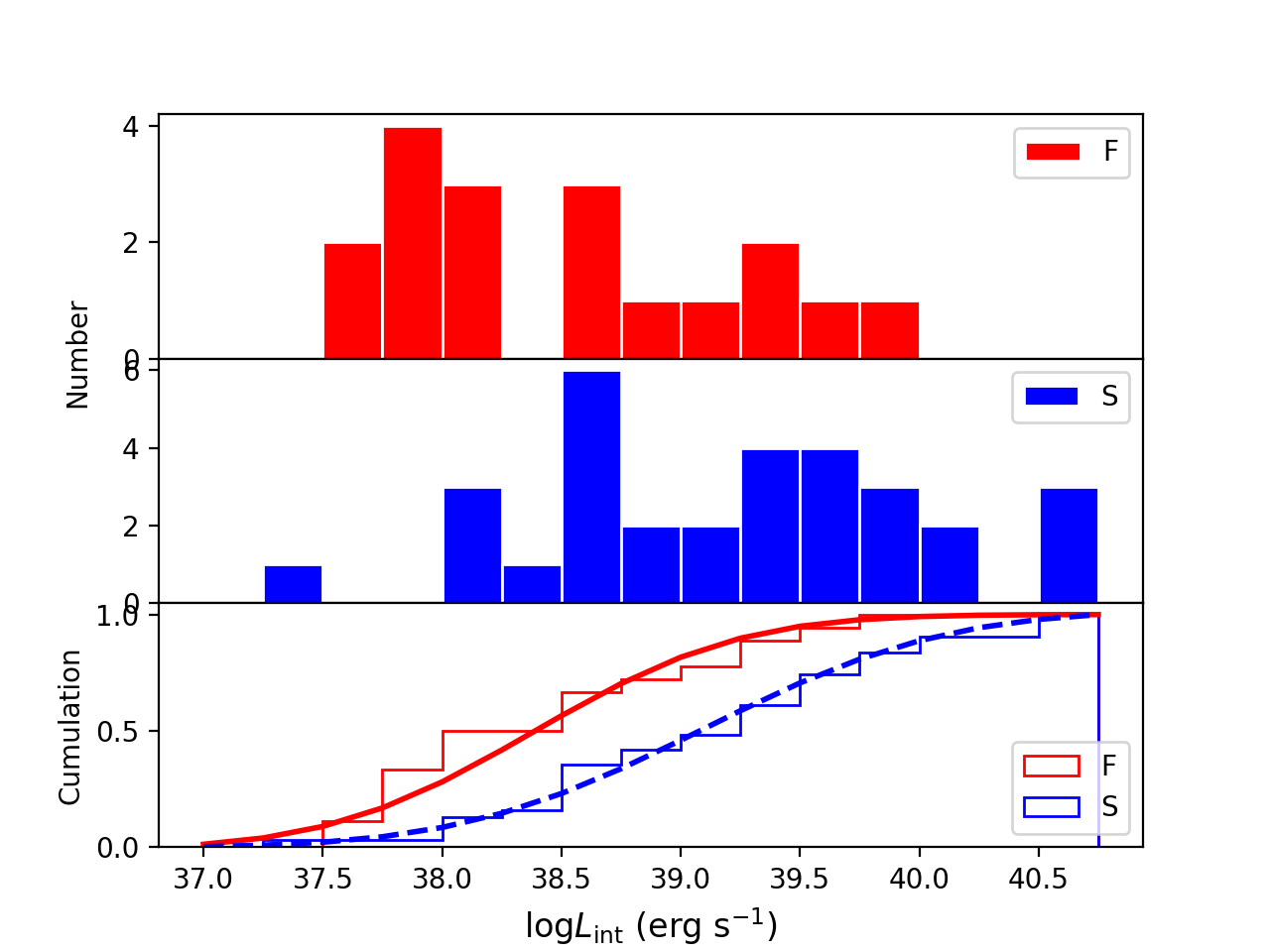}
\includegraphics[width=0.45\textwidth]{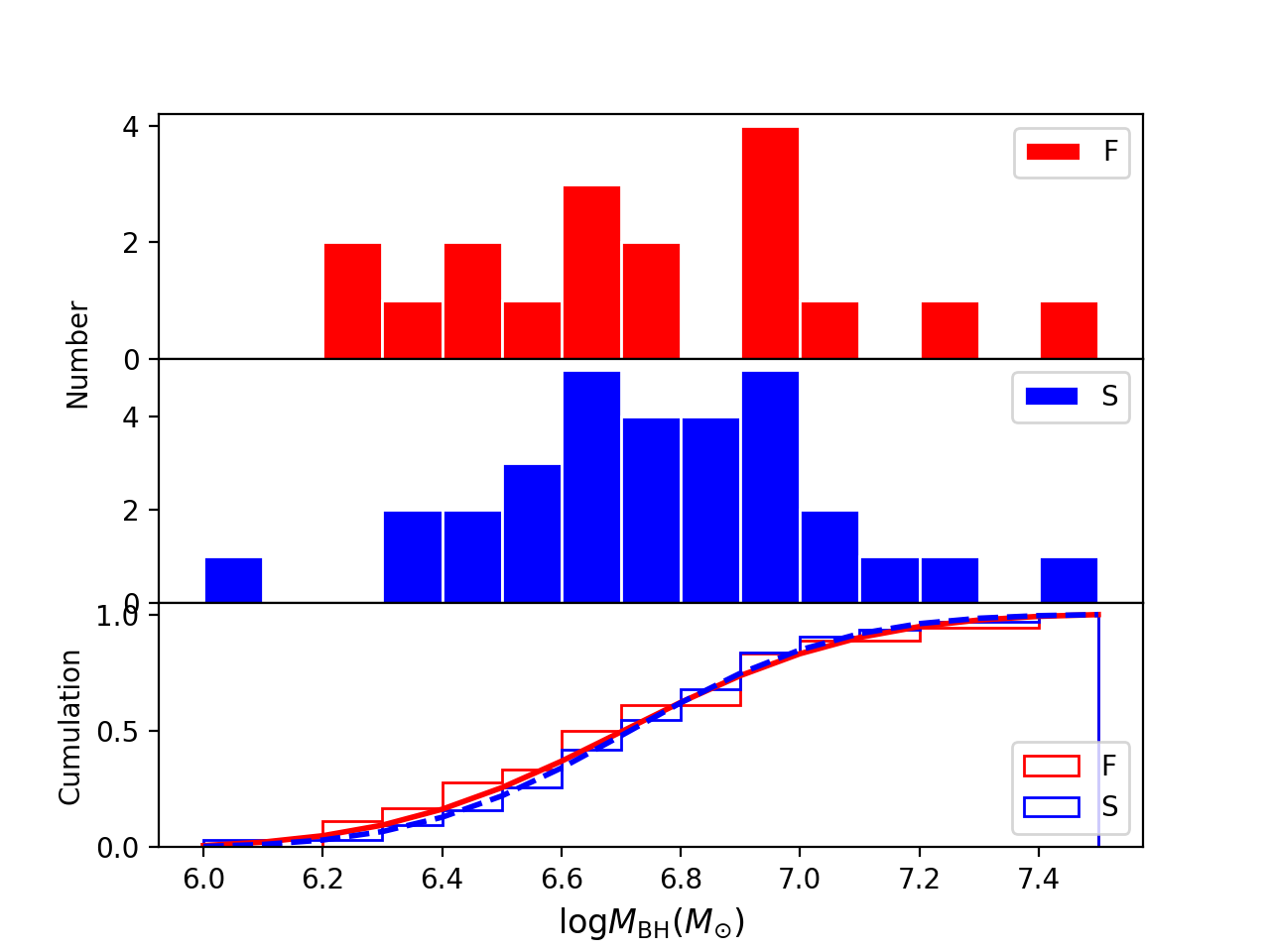}
\caption{The numerical distributions of radio luminosity (\textit{left panel}) and black hole mass (\textit{right panel}) for the flat (\textit{top panels}) and steep (\textit{middle panels}) subsamples. \textit{Bottom panels:} the red solid and blue dashed lines represent the cumulative distributions of the flat and steep subsamples respectively.}
\label{hist_lumin+mass}
\end{figure*}

We divided this sample into two subsamples: flat and steep spectral slopes based on the in-band spectral index. The distributions of radio luminosity and black hole mass for these two subgroups are shown in Fig. \ref{hist_lumin+mass}. We applied the Kolmogorov-Smirnov (K-S) test on these parameters to verify if the flat and steep sources are actually statistically different. The null hypothesis is that two distributions originate from the same parent population. We applied the rejection of the null hypothesis at a 99$\%$ confidence level corresponding to a value of $p \leq$ 0.01. The K-S test of radio luminosity ($p \sim$ 7.6 $\times$ 10$^{-3}$) suggests that the flat and steep objects have different origins. Conversely, the K-S test of black hole mass ($p \sim$ 0.86) suggests that these two subsamples originate from the same parent population. We further applied the K-S test on the redshift and found that the flat and steep sources have similar distributions ($p \sim$ 0.77), which suggests that the different origins of these two subgroups on the radio luminosity are not a selection effect and could be intrinsic.

On one hand, these distributions imply that the radio luminosity may be not related to the black hole mass. On the other hand, unlike RL NLS1s, flat-spectrum objects have a higher radio luminosity compared to steep-spectrum objects \citep{Berton2018}. The majority of NLS1s in our sample are RQ, the flat sources tend to have a lower radio luminosity than the steep sources. This may be related to the different radio mechanisms. The steep-spectrum radio emission may be originated from an optically thin synchrotron region, where a misaligned jet, an AGN-driven outflow, or star-forming activity may be present and superpose on an optically thick core. The flat-spectrum radio emission may be produced in a central compact core, which has not yet developed a radio jet or outflow. Therefore, the steep sources tend to have a higher radio luminosity than the flat sources in the RQ NLS1 population. However, these results should be taken with caution, since the separation of the flat and steep sources is somewhat arbitrary, due to a large fraction of sources cluster around the threshold of $\alpha$ = 0.5 with relatively large uncertainties.

\subsection{Evidences for radio jets}

We find one source J0354$-$1340 that is a jetted NLS1 (details seen in \ref{app_j0354}). Its morphology favors an Fanaroff-Riley type II (FRII) radio galaxy \citep{Fanaroff1974}. The central core is compact and bright at radio frequencies. Two components located to the south and north of the central core, are probably the radio lobes showing diffuse emission. Additionally, a hotspot where the relativistic jet terminates in, may be present in the end of the southern lobe. The projected linear sizes of the radio lobes are around a hundred kpc, which are consistent with those of FRII radio galaxies \citep{Hardcastle1998,Kharb2008}. The luminosity of the southern lobe is brighter than that of the northern one, indicating that the southern component is the approaching jet and the northern component is the receding jet. The NVSS map shows an elongated structure which is consistent with the VLA C-configuration map exhibiting a core-jet/lobe morphology. In the 5.5 GHz image, the radio lobes are resolved, and the central compact core is probably the jet-base having a flat in-band spectral index. In the 1.4 GHz image, the radio lobes are unresolved, and the radio emission from both the core and the lobes is included giving a slightly steep slope between 1.4 and 5.5 GHz.

The jet speed can be calculated based on the assumptions that the morphology and the physical parameters of the FRII radio galaxy are intrinsically symmetric. The only difference is caused by different Doppler enhancements, leading to an amplification of the approaching jet with an observed flux of $f_+$ and a dimming of the receding jet with an observed flux of $f_-$. The flux ratio can be derived from the Doppler factors \citep{Beckmann2012}
\begin{equation}
\frac{f_+}{f_-} = (\frac{1 + \beta \cos \theta}{1 - \beta \cos \theta}) ^{(2 + \alpha)}
\end{equation}
where $\beta = v / c$, $\theta$ is the angle of the jet with respect to the line of sight, and $\alpha$ is the spectral index of the continuum emission. If we assume that the jet propagates close to the speed of light ($\beta \sim$ 1), this gives the radial velocity of $v \cos \theta \sim 0.25c$ and the projected velocity of $v \sin \theta \sim 0.97c$. The age of the approaching jet can also be evaluated using the projected linear size over the projected velocity, resulting in a lower limit for the age of $\sim$ 3 $\times$ 10$^5$ years. This lies at the low tail of the age range of 10$^5$ - 10$^7$ years found for other RL NLS1s \citep{Doi2012,Richards2015,Rakshit2018}.

We further estimated the jet power following two methods. The first one is adopting the relationships based on the radio core luminosity at 15 GHz in \citet{Foschini2014}
\begin{equation}
\log P_{\rm{rad}} = (12 \pm 2) + (0.75 \pm 0.04) \log L_{15}
\end{equation}
and
\begin{equation}
\log P_{\rm{kin}} = (6 \pm 2) + (0.90 \pm 0.04) \log L_{15}
\end{equation}
where $L_{15}$ was derived from the radio core flux at 5.5 GHz and the in-band spectral index assuming a power-law radio spectrum. We obtained the same radiative and kinetic powers of $\log P_{\rm{rad}}$ = $\log P_{\rm{kin}}$ = 42.0 $\pm$ 2.6 (in unit erg s$^{-1}$), revealing that the particles and radiation have equivalent contributions to the jet power. The total jet power was calculated via the sum of the kinetic and radiative powers, resulting in $\log P_{\rm{jet}}$ = 42.27 $\pm$ 1.81 (in unit erg s$^{-1}$). The second one is using the relationship based on the radio power at 1.4 GHz in \citet{Cavagnolo2010}
\begin{equation}
\log P_{\rm{cav}} = (0.75 \pm 0.14) \log P_{1.4} + (1.91 \pm 0.18)
\end{equation}
where $P_{1.4}$ is in unit 10$^{40}$ erg s$^{-1}$ and $P_{\rm{cav}}$ is the cavity power in unit 10$^{42}$ erg s$^{-1}$. Assuming $P_{\rm{cav}}$ = $P_{\rm{jet}}$, this can give an estimate of the jet power with $\log P_{\rm{jet}}$ = 43.51 $\pm$ 0.20 (in unit erg s$^{-1}$). The jet powers based on two measurements are consistent within errors and generally less powerful than flat-spectrum RL NLS1s \citep{Foschini2015}.

In addition, J1615$-$0936 may also be a jetted NLS1 (details seen in \ref{app_j1615}). It displays a core-jet/lobe structure though the radio emission is not so powerful. The western component and the eastern diffuse emission may be the radio jets or lobes spreading to a projected linear size of $\sim$ 8 kpc. At such a scale, the radio emission is probably produced by a misaligned jet. A steep spectral slope indicates that the source is dominated by extended emission.

Two sources, J0952$-$0136 and J2021$-$2235, are well studied in literature and suggested to be jetted NLS1s (details seen in \ref{app_j0952} and \ref{app_j2021}). J0952$-$0136 exhibits a resolved and linear extended structure at pc-scale and a slightly elongated structure at kpc-scale. The radio spectrum is steep showing that the radiation is originated from an optically thin source. These indicate that the radio emission is dominated by a misaligned jet at pc-scale \citep{Doi2013,Doi2015}. J2021$-$2235 displays a compact core at kpc-scale and a steep spectral slope. It has a remarkably high star formation rate which only account for up to $\sim$ 30$\%$ of the total radio luminosity \citep{Caccianiga2015}. Thus the presence of a jet is required to explain the non-thermal synchrotron emission. These suggest that a pc-scale jet and star formation are co-existing in this source \citep{Berton2019}.

Furthermore, other sources resembling J0952$-$0136 and J2021$-$2235 (e.g. J0452$-$2953, J0846$-$1214, and J1511$-$2119), may be candidates of harboring a radio jet at pc-scale. They are characterized by a compact radio core at kpc-scale with relatively high luminosities and steep spectral slopes, which reveal that the strong radio emission is generated from an optically thin source and centrally concentrated at kpc-scale. However, high-resolution observations to resolve the central core are necessary to verify this interpretation.

These jetted NLS1s, especially the FRII-like NLS1 which has a smaller black hole mass than a regular FRII radio galaxy, confirm our current knowledge that the generation of relativistic jets does not require a massive black hole. Indeed, radio emission from kpc-scale jets was discovered in about a dozen NLS1s \citep{Whalen2006,Anton2008,Gliozzi2010,Doi2012,Doi2015,Richards2015,Berton2018,Rakshit2018}. Other powerful radio jets were also found in disk galaxies which are believed to be late-type galaxies \citep{Keel2006,Morganti2011,Kotilainen2016,Jarvela2018,Olguiniglesias2020}. How is a powerful jet generated in an AGN with a relatively low-mass black hole? Other factors, such as the spin of black hole, may also play an important role in the evolution of relativistic jets \citep{Laor2000,Foschini2012c}. Further studies will be helpful to unveil this picture.

\subsection{Peculiar sources}

J0000$-$0541 is a radio bright source (details seen in \ref{app_j0000}). It is currently unclear if the northwestern source is connected to the central target since no optical counterpart is associated with it on the image from the Panoramic Survey Telescope and Rapid Response System (Pan-STARRS) \footnote{\url{https://panstarrs.stsci.edu/}}, or if they are two independent sources since no diffuse emission is connecting them. A dedicated analysis is needed to clarify if this object is jetted or non-jetted NLS1.

In addition, we find two sources, J0436$-$1022 and J1937$-$0613, probably dominated by star formation (details seen in \ref{app_j0436} and \ref{app_j1937}). Their radio emission is extended or diffuse and has similar morphologies to their host galaxies in the optical images. Other sources whose radio emission is not centrally concentrated at kpc-scale, may also have a star-forming origin. They are characterized by extended or diffuse emission with low luminosities and steep spectral slopes. Further studies at infrared frequencies may help to clarify this hypothesis.

Another interesting source may be J0850$-$0318 (details seen in \ref{app_j0850}). The radio luminosity is comparable to jetted NLS1s. But it shows a slightly extended morphology and a very steep spectral index with the C-configuration. This may suggest that the radio emission is originated from a relic emission due to a previous AGN activity episode \citep{Congiu2017}. However, such extended emission could be out-resolved with the A-configuration, resulting in a compact core and a spectral slope that is broadly consistent with optically thin synchrotron emission. Besides, it is unclear whether the western source and the central target are interacting or not. More research is needed to find an answer.

\section{Summary}

In this work, we present a new VLA survey using the C-configuration at 5.5 GHz of a southern NLS1 sample. There are 49 sources detected in these observations. We reduced their radio maps, measured their flux densities, and calculated their radio loudness, radio flux concentration, spectral index, and luminosities. Other properties were also collected from literature. The main results of this work are summarized as follow.

1. This work increases the number of radio-detected NLS1s especially in the southern sky which remains largely unexplored. Among this sample, 47 sources are observed at 5 GHz for the first time, and 27 sources are new radio-emitting NLS1s without a previous observation at any radio frequencies. It provides good candidates for future observations with new facilities located in the southern hemisphere, such as the Atacama Large Millimeter/sub-millimeter Array (ALMA) and the Square Kilometre Array (SKA).

2. This study confirms that the majority of NLS1s are unresolved at kpc-scale. The radio emission is generally concentrated in a central region, independent of whether the origin is from a compact core or from extended structures. Only a few sources show diffuse emission with steep spectral slopes.

3. We find that the radio luminosity of RQ NLS1s tends to be higher in steep-spectrum sources and be lower in flat-spectrum sources, which is in contrast to RL NLS1s. This may be because the radio emission of steep NLS1s is dominated by misaligned jets, AGN-driven outflows, or star formation superposing on a compact core, while the radio emission of flat NLS1s is produced by a central core which has not yet developed large radio jets or outflows. However, we should take with caution since the distinction between the flat and steep spectral sources is blurred due to a large fraction of sources cluster around the threshold and the statistical uncertainty is relatively high.

4. We discovered new NLS1s harboring kpc-scale radio jets, such as J0354$-$1340 and J1615$-$0936. Such jetted NLS1s confirm that a powerful jet does not require a large-mass black hole to be generated. We also find sources dominated by star formation, such as J0436$-$1022 and J1937$-$0613. These NLS1s could be new candidates in investigating the radio emission of different mechanisms.

Future high-resolution observations will be necessary to assess whether the radio properties of this sample have been correctly interpreted in terms of intrinsic structure. Further broadband observations will also play an important role in unveiling the nature of NLS1s.

\section*{Acknowledgements}

We thank the anonymous referee for suggestions leading to the improvement of this work.
The National Radio Astronomy Observatory is a facility of the National Science Foundation operated under cooperative agreement by Associated Universities, Inc. This research has made use of the NASA/IPAC Extragalactic Database (NED), which is operated by the Jet Propulsion Laboratory, California Institute of Technology, under contract with the National Aeronautics and Space Administration.

\section*{Data availability}

The data underlying this article are available in the NRAO Science Data Archive at \url{https://archive.nrao.edu/archive/advquery.jsp}, and can be accessed with the project codes of VLA/18B-126 and VLA/19B-285.

\bibliographystyle{mnras}
%\bibliography{/Users/sinachen/Dropbox/Publication/bibliography.bib}
%\bibliography{/home/sina/Nutstore/Publication/bibliography.bib}
\bibliography{bibliography}

\appendix

\section{Individual sources}
\label{individual}
\counterwithin{figure}{section}

We present an investigation on individual sources exhibiting a resolved or slightly resolved morphology in the VLA maps or having well studied radio properties in literature.

\subsection{J0000$-$0541}
\label{app_j0000}

This target exhibits a compact core at kpc-scale (Fig. \ref{a} left panel) with a relatively large radio power but small black hole mass compared to the averages of this sample. The flux density, luminosity, and black hole mass are $S_{\rm{int}} \sim$ 1.16 mJy, $\log L_{\rm{int}} \sim$ 39.16, and $\log (M_{\rm{BH}} / M_{\odot}) \sim$ 6.36 respectively. It has a steep spectrum with $\alpha_{\rm{in-band}} \sim$ 1.20. This source was also detected at 1.4 GHz associating with a counterpart of FIRST J000040.2$-$054101. It has a flux density of $S_{1.4} \sim$ 3.5 mJy, a luminosity of $\log L_{1.4} \sim$ 39.03, and a steep slope of $\alpha_{\rm{1.4-5.5}} \sim$ 0.80.

A northwestern source (centered at R.A. 00:00:38.53 and Dec. $-$05:40:37.63) has a separation of 34.8 arcsec from the central target. We overlaid the radio contours on the optical image from the Pan-STARRS (Fig. \ref{aa} left panel). No optical counterpart is associated with it and no redshift is available. This nearby source shows a compact core with a flux density of $S_{\rm{int}} \sim$ 2.05 mJy. It has an associated counterpart of FIRST J000038.5$-$054037 at 1.4 GHz with a flux density of $S_{1.4} \sim$ 7.8 mJy. The spectral slope is steep with $\alpha_{\rm{in-band}} \sim$ 1.30 and $\alpha_{\rm{1.4-5.5}} \sim$ 0.97. We added a Gaussian taper with a radius of 10k$\lambda$ (the longest baseline is 60k$\lambda$) to increase the sensitivity of weak extended emission possibly connecting these two sources. However, no diffuse emission is seen between them in the tapered map.

\subsection{J0350$-$1025}

This source is the faintest one in the sample with a flux density of $S_{\rm{int}} \sim$ 0.06 mJy at redshift $z$ = 0.128. It is unresolved at kpc-scale (Fig. \ref{f} right panel). The luminosity of $\log L_{\rm{int}} \sim$ 38.27 is below the average, while the black hole mass of $\log (M_{\rm{BH}} / M_{\odot}) \sim$ 6.91 is above the average. It has a very steep spectrum with $\alpha_{\rm{in-band}} \sim$ 3.11. This object was not detected at 1.4 GHz.

A nearby source to the west of the target, is associated with an optical counterpart of 2MASS J03505598$-$1025556 at redshift $z$ = 0.071. We overlaid the radio contours on the optical image from the Pan-STARRS (Fig. \ref{aa} middle panel). It is weak at radio with a flux density of $S_{\rm{int}} \sim$ 0.07 mJy and a luminosity of $\log L_{\rm{int}} \sim$ 37.66. The spectrum is steep with $\alpha_{\rm{in-band}} \sim$ 1.4. No detection at 1.4 GHz is available.

\subsection{J0354$-$1340}
\label{app_j0354}

\begin{figure}
\centering
\includegraphics[width=0.5\textwidth, trim={4cm 13.5cm 7cm 2cm}, clip]{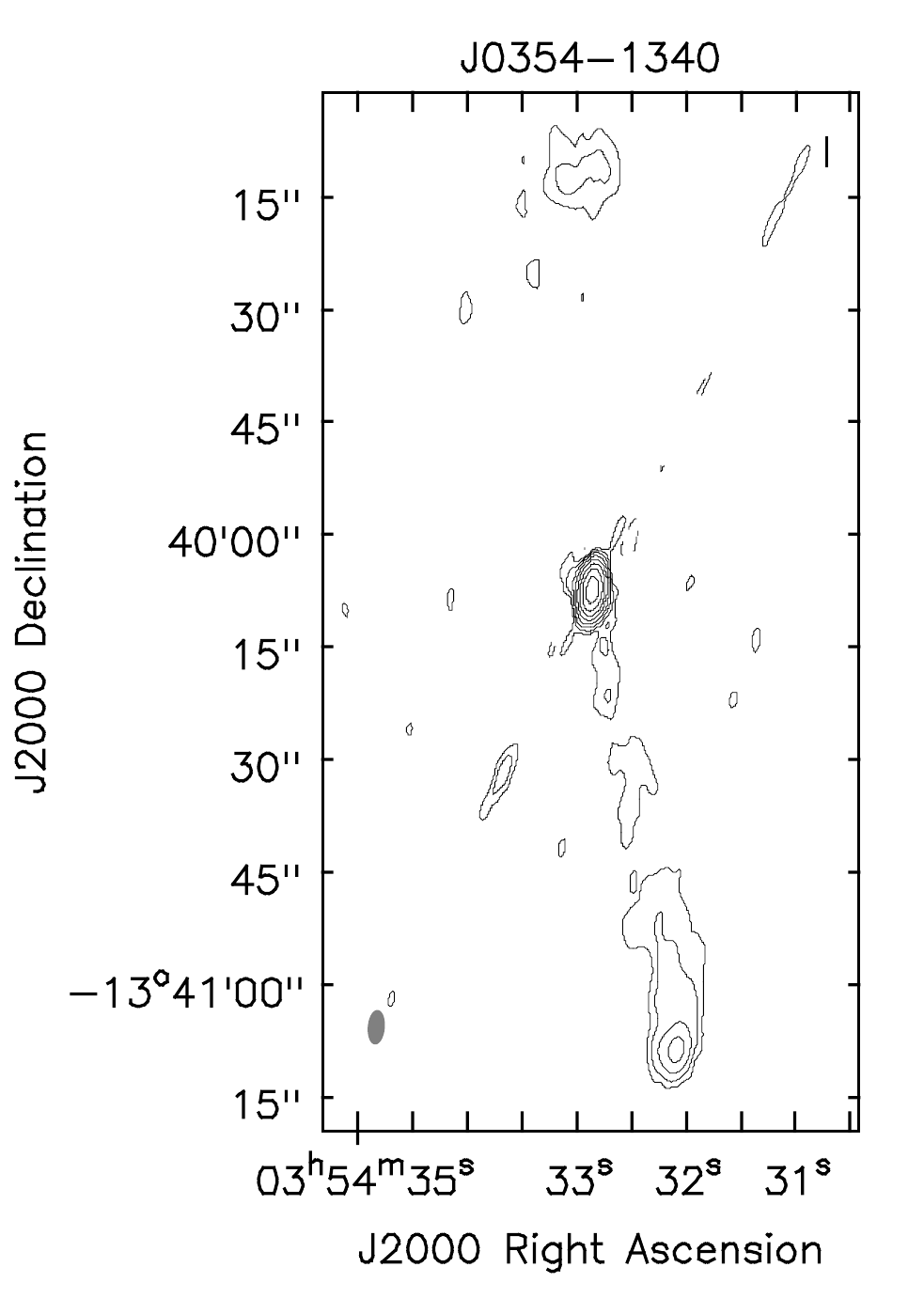}
\caption{J0354$-$1340 with the VLA observation, rms = 8 $\mu$Jy beam$^{-1}$, contour levels at $-$3, 3 $\times$ 2$^n$, $n \in$ [0,7], beam size 7.38 $\times$ 3.47 kpc.}
\label{J0354}
\end{figure}

This target has a very compact central core at kpc-scale with a flux density of $S_{\rm{int}} \sim$ 5.09 mJy. The VLA map is shown in Fig. \ref{J0354}. The luminosity and black hole mass are $\log L_{\rm{int}} \sim$ 39.58 and $\log (M_{\rm{BH}} / M_{\odot}) \sim$ 6.99 respectively, which are higher than the mean values of this sample. The in-band spectral slope is flat with $\alpha_{\rm{in-band}} \sim$ 0.12.

A southern component (centered at R.A. 03:54:32.09 and Dec. $-$13:41:08.46) has a separation of 62.2 arcsec from the central target. At redshift $z$ = 0.076, the projected linear size is 93.5 kpc. The flux density and luminosity are $S_{\rm{int}} \sim$ 1.35 mJy and $\log L_{\rm{int}} \sim$ 39.05 respectively, which are weaker than the central core. It shows diffuse emission. We measured a spectral index of $\alpha_{\rm{in-band}} \sim$ 1.7 which is steeper than the core.

A northern component (centered at R.A. 03:54:32.95 and Dec. $-$13:39:11.57) has a separation of 55.7 arcsec from the central target, corresponding to a projected linear size of 83.9 kpc. The flux density and luminosity are $S_{\rm{int}} \sim$ 0.45 mJy and $\log L_{\rm{int}} \sim$ 38.43 respectively, which are fainter and more diffuse than the southern one. We did not measure its spectral index because it is not detected in the 5 and 6 GHz images which have a higher RMS than the original one.

\begin{figure}
\centering
\includegraphics[width=0.5\textwidth, trim={2cm 13cm 4cm 1cm}, clip]{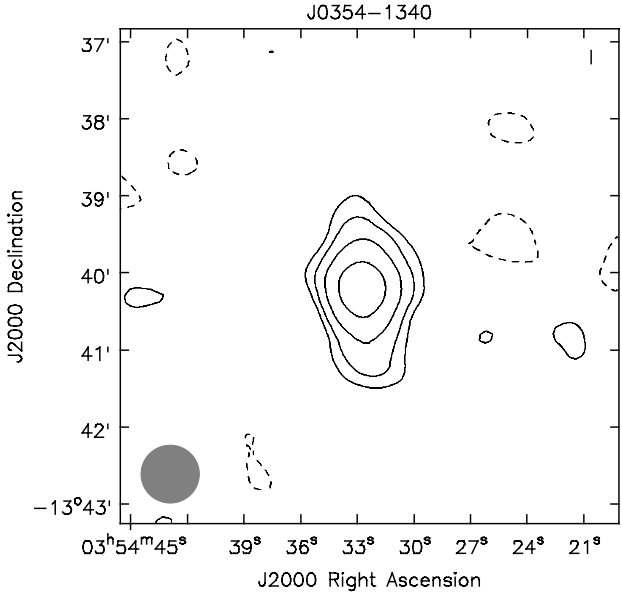}
\caption{J0354$-$1340 with the NVSS detection, rms = 300 $\mu$Jy beam$^{-1}$, contour levels at $-$3, 3 $\times$ 2$^n$, $n \in$ [0,3], beam size 60.41 $\times$ 60.41 kpc.}
\label{j0354_nvss}
\end{figure}

This source was also detected at 1.4 GHz associating with a counterpart of NVSS J035432$-$134012. It has a flux density of $S_{1.4} \sim$ 14.9 mJy and a luminosity of $\log L_{1.4} \sim$ 39.46. The NVSS map exhibits an elongated structure shown in Fig. \ref{j0354_nvss}. The southern and northern components are unresolved, thus the flux measurement includes the radio emission from both the core and the two components. The spectral index between 1.4 and 5.5 GHz was estimated using the NVSS flux and the VLA flux of a core plus two components giving $\alpha_{\rm{1.4-5.5}} \sim$ 0.56. We overlaid the radio contours on the optical image from the Pan-STARRS (Fig. \ref{aa} right panel) and found that no optical counterpart is associated with the southern and northern components.

\subsection{J0436$-$1022 (Mrk 618)}
\label{app_j0436}

Mrk 618 looks like a spiral galaxy on the optical image from the Pan-STARRS (Fig. \ref{bb} left panel). The VLA map shows an extended morphology at kpc-scale (Fig. \ref{i} right panel) with a flux density of $S_{\rm{int}} \sim$ 4.62 mJy. The southern extended structure is twice as bright as the northern one. We overlaid the radio contours on the optical image and found that the morphology of the extended structure in the radio map is coincident with that of the host galaxy in the optical image. The spectral slope is steep with $\alpha_{\rm{in-band}} \sim$ 0.60. It has a luminosity of $\log L_{\rm{int}} \sim$ 38.86 and a black hole mass of $\log (M_{\rm{BH}} / M_{\odot}) \sim$ 6.90, which are at an intermediate level.

\begin{figure}
\centering
\includegraphics[width=0.5\textwidth]{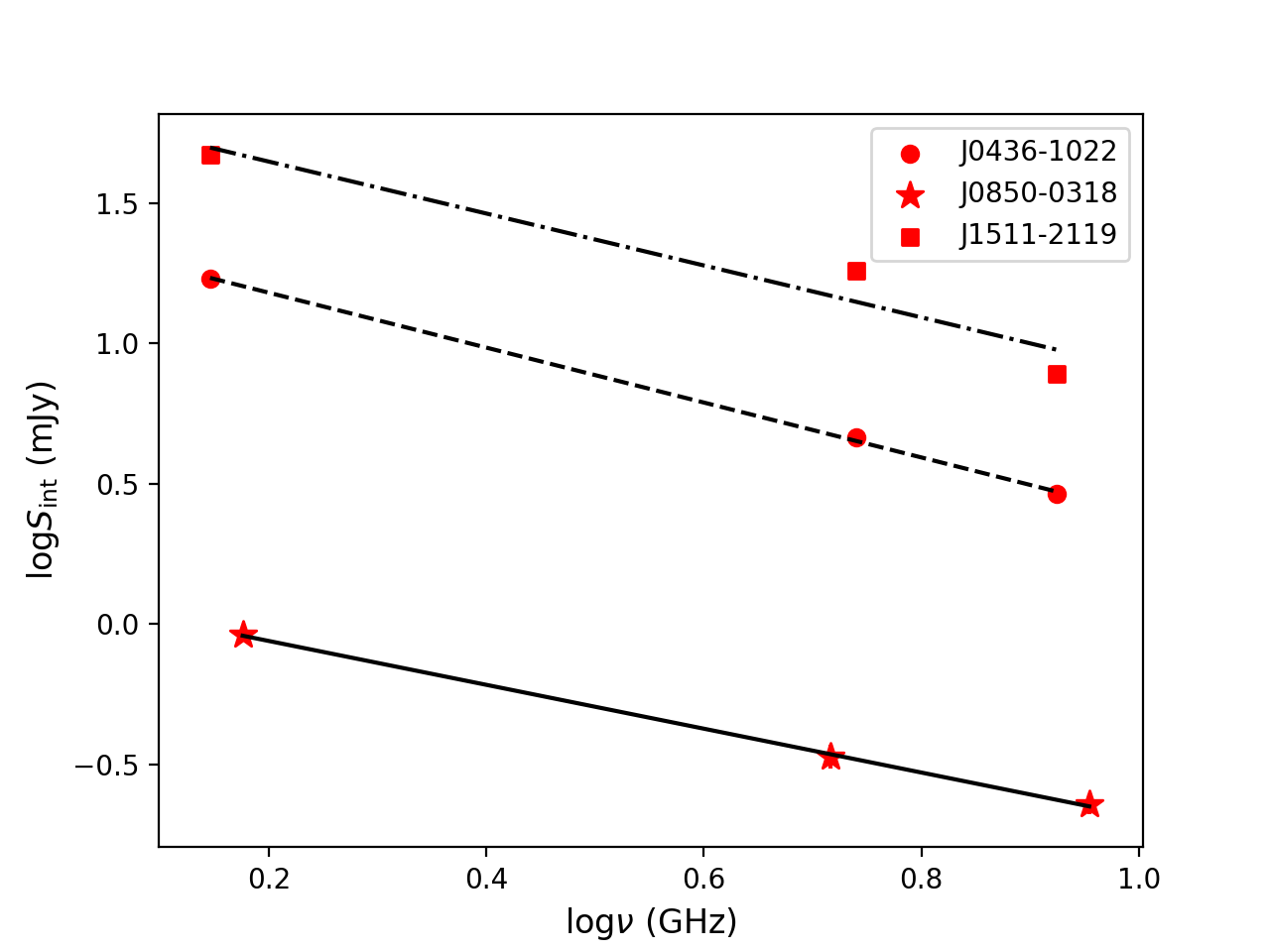}
\caption{The radio spectra with a power-law fit of J0436$-$1022 (circles and dashed line), J0850$-$0318 (stars and solid line), and J1511$-$2119 (squares and dashdot line).}
\label{spectra}
\end{figure}

This object was detected with the NVSS at 1.4 GHz giving a flux density of $S_{1.4} \sim$ 17.0 mJy, a luminosity of $\log L_{1.4} \sim$ 38.83, and a steep spectral index of $\alpha_{\rm{1.4-5.5}} \sim$ 0.95. It was also observed with the VLA A-configuration at 8.4 GHz having a flux density of $S_{8.4} \sim$ 2.9 mJy \citep{Thean2000}. The source displays a compact core in the 8.4 GHz map, where the northern and southern extended structure in the 5.5 GHz map is probably out-resolved at a high-resolution observation. We modeled the radio spectrum between 1.4 and 8.4 GHz with a power-law yielding a steep spectral index of $\alpha_{\rm{1.4-8.4}} \sim$ 0.98, as shown in Fig. \ref{spectra}.

\subsection{J0447$-$0508}

This source has a compact core at kpc-scale (Fig. \ref{j} right panel) with a flux density of $S_{\rm{int}} \sim$ 4.04 mJy. The luminosity is slightly higher than the average and the black hole mass is comparable to the average, having $\log L_{\rm{int}} \sim$ 39.56 and $\log (M_{\rm{BH}} / M_{\odot}) \sim$ 6.85 respectively. The spectral slope is steep with $\alpha_{\rm{in-band}} \sim$ 0.83. It was not detected at 1.4 GHz.

We note a very bright radio source to the northeast of the central target. It is a typical FRII radio galaxy according to the morphology. The central core is probably hidden. The radio emission is dominated by the radio lobes on both sides. Two hotspots are likely to be present at the end of the radio lobes. It has a flux density of $S_{\rm{int}} \sim$ 22.70 mJy and a steep slope of $\alpha_{\rm{in-band}} \sim$ 0.66. This radio galaxy has an associated counterpart of NVSS J044722$-$050750 at 1.4 GHz, with a flux density of $S_{1.4} \sim$ 68.4 mJy and a steep spectral index of $\alpha_{\rm{1.4-5.5}} \sim$ 0.81. We overlaid the radio contours on the optical image from the Pan-STARRS (Fig. \ref{bb} middle panel). It could be associated with an optical counterpart of 2MASS J04472253$-$0507494, but no redshift is available.

\subsection{J0850$-$0318}
\label{app_j0850}

This target shows a slightly extended structure at kpc-scale (Fig. \ref{o} left panel) with a flux density of $S_{\rm{int}} \sim$ 0.43 mJy. Its radio power is at a mean level with $\log L_{\rm{int}} \sim$ 39.30. But the black hole mass is large in this sample with $\log (M_{\rm{BH}} / M_{\odot}) \sim$ 7.17. It has a steep spectral index of $\alpha_{\rm{in-band}} \sim$ 2.15.

We overlaid the radio contours on the optical image from the Pan-STARRS (Fig. \ref{bb} right panel), and note that a nearby source to the west of the central target is associated with an optical counterpart of 2MASS J08502683$-$0318107. But no redshift is available. It has a flux density of $S_{\rm{int}} \sim$ 0.37 mJy and a very steep slope of $\alpha_{\rm{in-band}} \sim$ 2.38. To highlight the sensitivity of weak diffuse emission possibly connecting these two sources, we added a Gaussian taper with a radius of 30k$\lambda$ (the longest baseline is 170k$\lambda$). Even in the tapered map, the two sources do not appear to be connected. Besides, they were not detected with the 1.4 GHz surveys.

This target was also observed with the VLA A-configuration in L, C, and X bands (Proposal ID: VLA/19B-285) for the purpose of studying the steep in-band spectral index. The exposure time is 10 minutes on source in each band, yielding an average RMS of 30 $\mu$Jy beam$^{-1}$. It exhibits a compact core in the A-array images at all frequencies. The flux densities are $S_{1.5} \sim$ 0.92 mJy, $S_{5.2} \sim$ 0.34 mJy, and $S_{9.0} \sim$ 0.23 mJy. We modeled the radio spectrum from 1.5 to 9.0 GHz with a power-law giving a simultaneous steep slope of $\alpha_{\rm{1.5-9.0}} \sim$ 0.78, as shown in Fig. \ref{spectra}. The nearby source was not detected with the A-array due to either the emission being out-resolved or the high RMS compared with the C-array observation.

\subsection{J0952$-$0136 (Mrk 1239)}
\label{app_j0952}

Mrk 1239 is a well studied object in literature. It exhibits a very compact core with the C-configuration at 5.5 GHz (Fig. \ref{o} right panel) and has a flux density of $S_{\rm{int}} \sim$ 21.28 mJy. The radio luminosity is at an intermediate level with $\log L_{\rm{int}} \sim$ 39.01. The black hole mass is relatively small with $\log (M_{\rm{BH}} / M_{\odot}) \sim$ 6.33, which is consistent with other measurements in the range of $\log (M_{\rm{BH}} / M_{\odot}) \sim$ 5.89-7.02 \citep{Ryan2007,Berton2015}. The radio spectrum is steep with $\alpha_{\rm{in-band}} \sim$ 0.91. This object was also observed with the A-configuration at 5 GHz giving similar measurements and showing extended emission \citep{Berton2018}.

Other frequency observations are available as well for this target. At 1.4 GHz, it has a flux density of $S_{1.4} \sim$ 59.8 mJy with the FIRST detection, yielding a luminosity of $\log L_{1.4} \sim$ 38.86 and a steep slope of $\alpha_{\rm{1.4-5.5}} \sim$ 0.76. In addition, a VLBA observation at 1.7 GHz gave a flux density of $S_{1.7} \sim$ 41.8 mJy and provided an image exhibiting a resolved and linear extended structure \citep{Doi2013,Doi2015}. Another observation with the VLA A-configuration at 8.5 GHz also displayed a slightly elongate structure and gave a flux density of $S_{8.5} \sim$ 14.6 mJy \citep{Orienti2010,Doi2013}. Furthermore, a radio spectrum was derived by \citet{Doi2013} with a VLA B-configuration quasi-simultaneous observation, finding spectral indexes of $\alpha_{\rm{1.4-5}} \sim$ 0.52 at 1.4-5 GHz, $\alpha_{\rm{5-8.5}} \sim$ 0.94 at 5-8.5 GHz, and $\alpha_{\rm{8.5-15}} \sim$ 1.64 at 8.5-15 GHz, which become steeper toward higher frequencies. Besides, this source was not detected at 22 GHz, which could be due to the steep spectral slope leading to the flux density at 22 GHz below the limit of the instrument \citep{Doi2016}.

\subsection{J1014$-$0418 (PG 1011$-$040)}

This target is in the PG quasar sample \citep{Boroson1992}. It is a RQ source having a slightly extended structure at kpc-scale (Fig. \ref{p} left panel) and a flux density of $S_{\rm{int}} \sim$ 0.38 mJy. The radio power is relatively weak with a luminosity of $\log L_{\rm{int}} \sim$ 38.23. The black hole mass is $\log (M_{\rm{BH}} / M_{\odot}) \sim$ 6.86 which is at a mean level and in agreement with previous study \citep{Davis2011}. The spectral slope is steep with $\alpha_{\rm{in-band}} \sim$ 0.79. It was not detected at 1.4 GHz. Additionally, this object was also observed with the A-array at 5 GHz giving a similar flux density and with the B-array at 8.5 GHz giving $S_{8.5} \sim$ 0.27 mJy, which result in a steep slope of $\alpha_{\rm{5-8.5}} \sim$ 0.57 between 5 and 8.5 GHz \citep{Kellermann1989,Laor2019}.

\subsection{J1511$-$2119}

This source has a very compact core at kpc-scale (Fig. \ref{t} left panel) with a flux density of $S_{\rm{int}} \sim$ 18.19 mJy. It is a radio bright object with $\log L_{\rm{int}} \sim$ 39.66 above the average, but has a small black hole with $\log (M_{\rm{BH}} / M_{\odot}) \sim$ 6.63 below the average. It was detected at 1.4 GHz with an associated counterpart of NVSS J151159$-$211900. The flux density and luminosity are $S_{1.4} \sim$ 46.9 mJy and $\log L_{1.4} \sim$ 39.48 respectively. The spectral slope is steep with $\alpha_{\rm{in-band}} \sim$ 0.81 and $\alpha_{\rm{1.4-5.5}} \sim$ 0.69. An observation with the VLA A-configuration at 8.4 GHz gave a flux density of $S_{8.4} \sim$ 7.8 mJy and a slightly extended morphology \citep{Thean2000}. We modeled the radio spectrum from 1.4 to 8.4 GHz with a power-law obtaining a steep spectral index of $\alpha_{\rm{1.4-8.4}} \sim$ 0.93, as shown in Fig. \ref{spectra}.

\subsection{J1615$-$0936}
\label{app_j1615}

This target is detected at radio frequencies for the first time, no previous detection with the 1.4 GHz surveys. The central core is slightly extended at kpc-scale (Fig. \ref{u} left panel) with a flux density of $S_{\rm{int}} \sim$ 0.68 mJy. The luminosity and black hole mass are below and above the averages having $\log L_{\rm{int}} \sim$ 38.59 and $\log (M_{\rm{BH}} / M_{\odot}) \sim$ 6.94 respectively. It has a steep radio spectrum with $\alpha_{\rm{in-band}} \sim$ 1.15.

A nearby component (centered at R.A. 16:15:18.66 and Dec. $-$09:36:12.75) to the west of the central core, has a separation of 5.9 arcsec corresponding to a projected linear size of 7.7 kpc at $z$ = 0.065. We overlaid the radio contours on the optical image from the Pan-STARRS (Fig. \ref{cc} left panel). No optical counterpart is associated with it. The flux density and luminosity are $S_{\rm{int}} \sim$ 0.15 mJy and $\log L_{\rm{int}} \sim$ 37.91 respectively, which are weaker than the central core. It shows slightly diffuse emission. The spectral slope is steeper than the core with $\alpha_{\rm{in-band}} \sim$ 1.6.

\subsection{J1937$-$0613}
\label{app_j1937}

This target exhibits diffuse emission around the central core at kpc-scale (Fig. \ref{w} right panel) with a flux density of $S_{\rm{int}} \sim$ 12.12 mJy. The luminosity and black hole mass are $\log L_{\rm{int}} \sim$ 38.20 and $\log (M_{\rm{BH}} / M_{\odot}) \sim$ 6.03 respectively, which are lower than the average values. It was also detected at 1.4 GHz associating with a counterpart of NVSS J193733$-$061304. The flux density and luminosity are $S_{1.4} \sim$ 42.2 mJy and $\log L_{1.4} \sim$ 38.15 respectively. The spectrum is steep with $\alpha_{\rm{in-band}} \sim$ 0.64 and $\alpha_{\rm{1.4-5.5}} \sim$ 0.91. It looks like a spiral galaxy on the optical image from the Pan-STARRS (Fig. \ref{cc} middle panel). We overlaid the radio contours on the optical image and found that the morphology of the diffuse emission is similar to that of the host galaxy.

\subsection{J2021$-$2235}
\label{app_j2021}

This target is one of the RL NLS1s in the sample with a flux density of $S_{\rm{int}} \sim$ 9.47 mJy. It is a radio bright source having $\log L_{\rm{int}} \sim$ 40.70, which is consistent with previous study \citep{Komossa2006}. The black hole mass is relatively small with $\log (M_{\rm{BH}} / M_{\odot}) \sim$ 6.65, which lies in the range of $\log (M_{\rm{BH}} / M_{\odot}) \sim$ 6.48-7.57 according to different methods \citep{Komossa2006,Foschini2015}. It has an associated counterpart of NVSS J202104$-$223520 at 1.4 GHz with a flux density of $S_{1.4} \sim$ 24.6 mJy and a luminosity of $\log L_{1.4} \sim$ 40.50. The spectral slope is steep with $\alpha_{\rm{in-band}} \sim$ 0.95 and $\alpha_{\rm{1.4-5.5}} \sim$ 0.70. This source exhibits a compact core at kpc-scale (Fig. \ref{x} left panel). An optical morphological study reveals that it is interacting with a nearby galaxy to the northeast of the target \citep{Berton2019}. We overlaid the radio contours on the optical image from the Pan-STARRS (Fig. \ref{cc} right panel). The nearby galaxy can not be resolved in the current VLA map.

\section{Radio maps}
\label{radiomap}
\counterwithin{figure}{section}

\begin{figure*}
\centering
\includegraphics[width=.41\textwidth, trim={1cm 13.5cm 4cm 2cm}, clip]{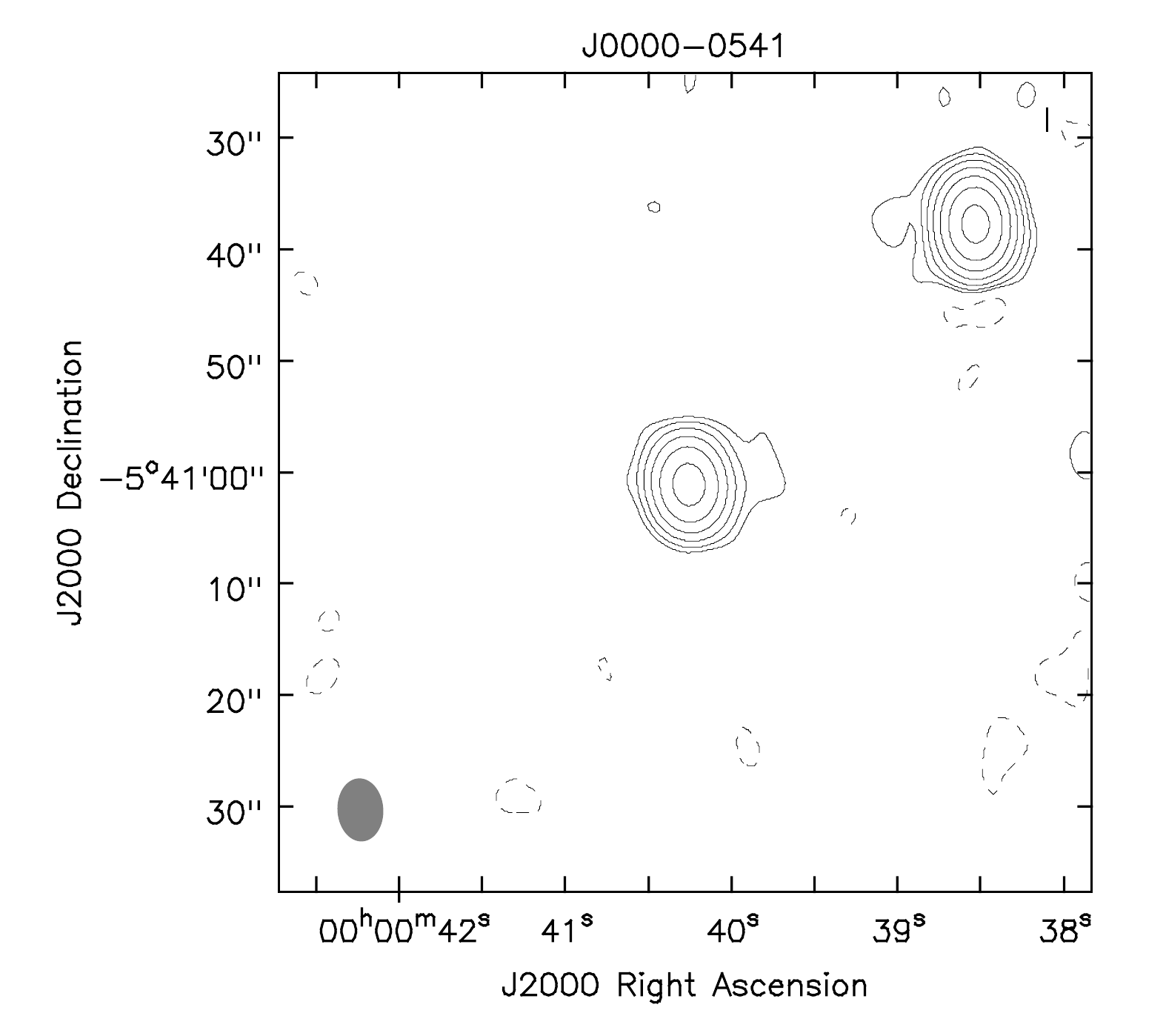}
\includegraphics[width=.41\textwidth, trim={1cm 13.5cm 4cm 2cm}, clip]{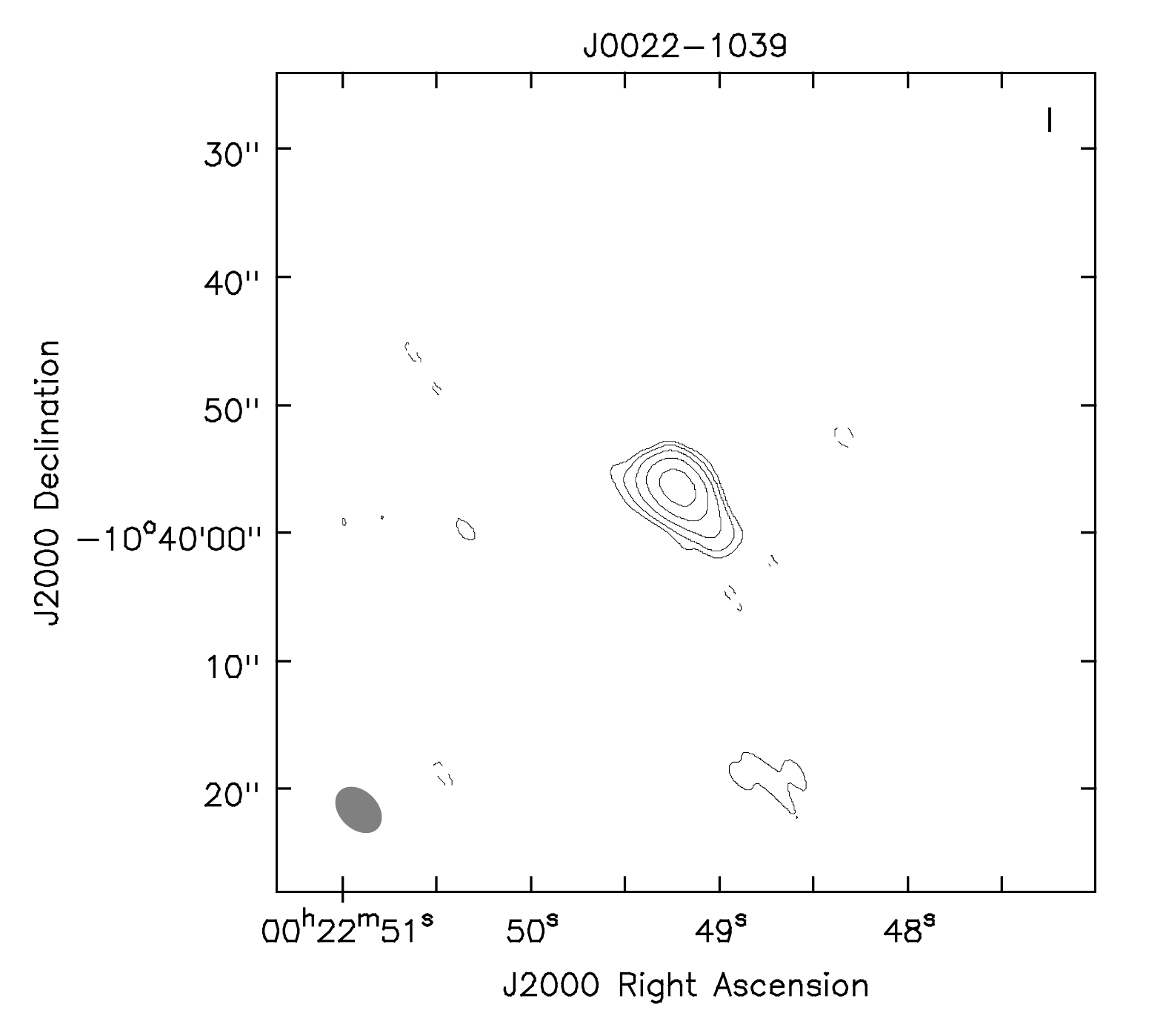}
\caption{\textit{left panel:} J0000$-$0541, rms = 8 $\mu$Jy beam$^{-1}$, contour levels at $-$3, 3 $\times$ 2$^n$, $n \in$ [0,6], beam size 11.58 $\times$ 8.37 kpc. \textit{right panel:} J0022$-$1039, rms = 6 $\mu$Jy beam$^{-1}$, contour levels at $-$3, 3 $\times$ 2$^n$, $n \in$ [0,4], beam size 44.67 $\times$ 31.72 kpc.}
\label{a}
\end{figure*}

\begin{figure*}
\centering
\includegraphics[width=.41\textwidth, trim={1cm 13.5cm 4cm 2cm}, clip]{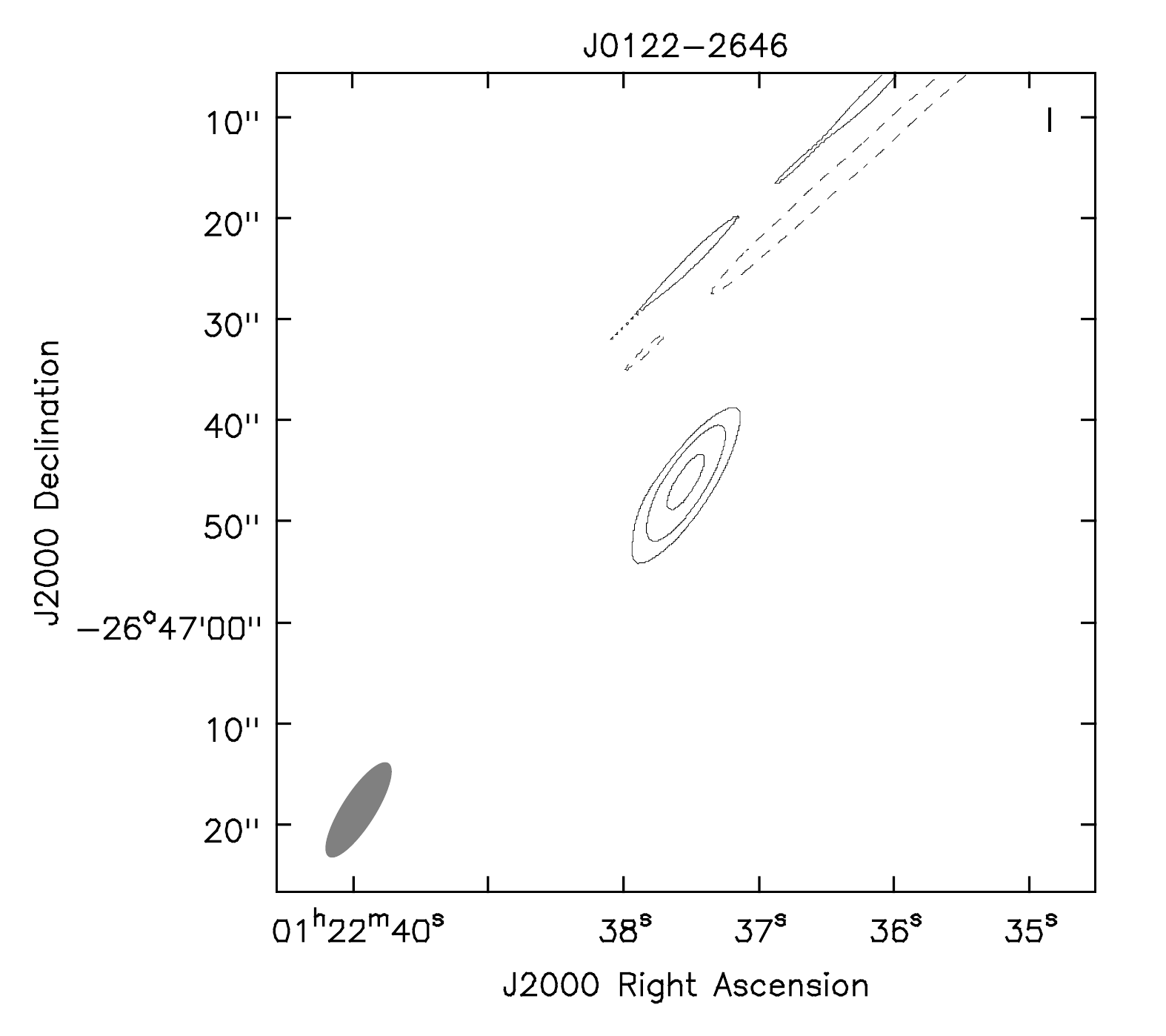}
\includegraphics[width=.41\textwidth, trim={1cm 13.5cm 4cm 2cm}, clip]{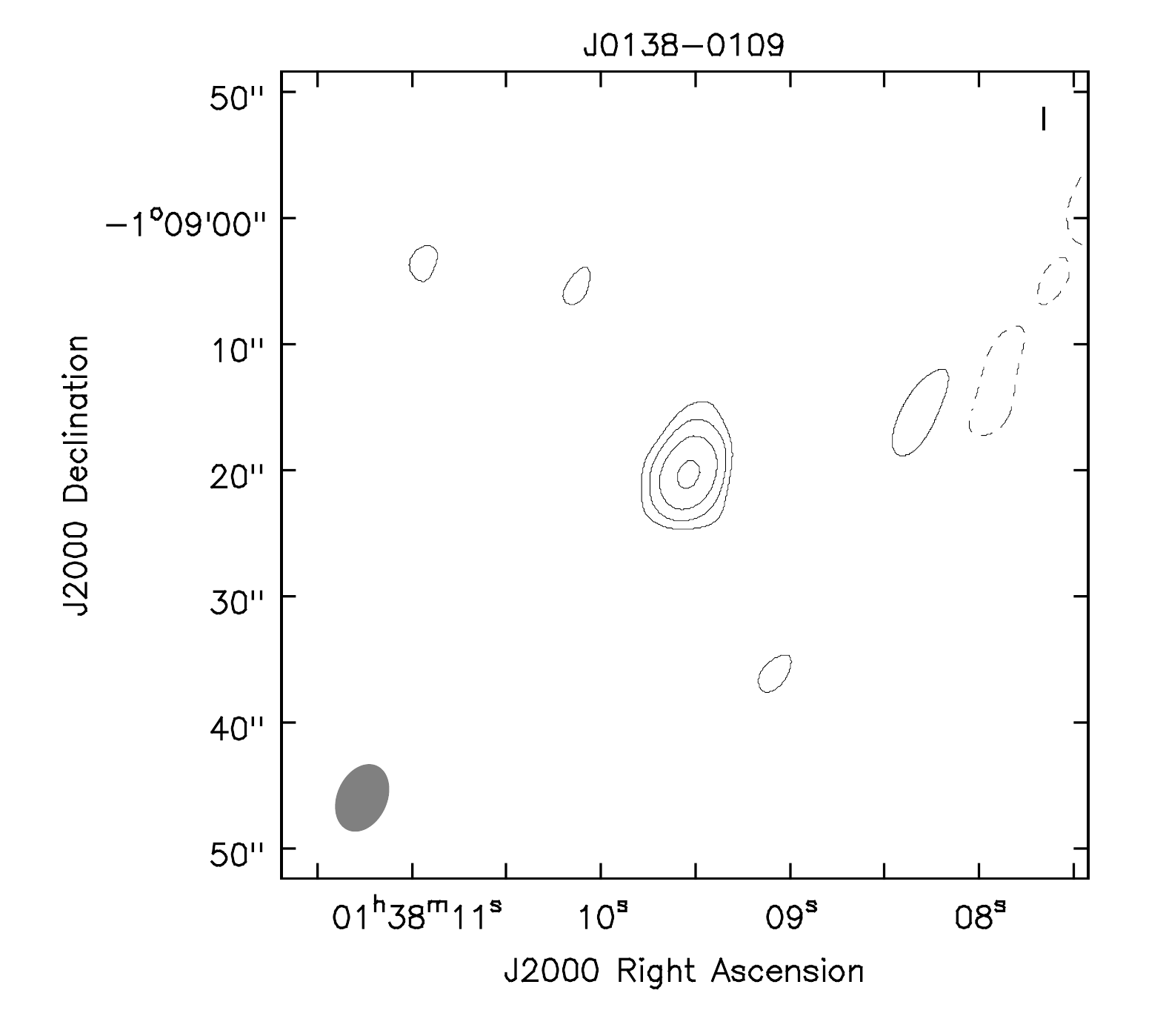}
\caption{\textit{left panel:} J0122$-$2646, rms = 55 $\mu$Jy beam$^{-1}$, contour levels at $-$3, 3 $\times$ 2$^n$, $n \in$ [0,2], beam size 120.29 $\times$ 35.55 kpc. \textit{right panel:} J0138$-$0109, rms = 8 $\mu$Jy beam$^{-1}$, contour levels at $-$3, 3 $\times$ 2$^n$, $n \in$ [0,3], beam size 37.10 $\times$ 26.15 kpc.}
\label{b}
\end{figure*}

\begin{figure*}
\centering
\includegraphics[width=.41\textwidth, trim={1cm 13.5cm 4cm 2cm}, clip]{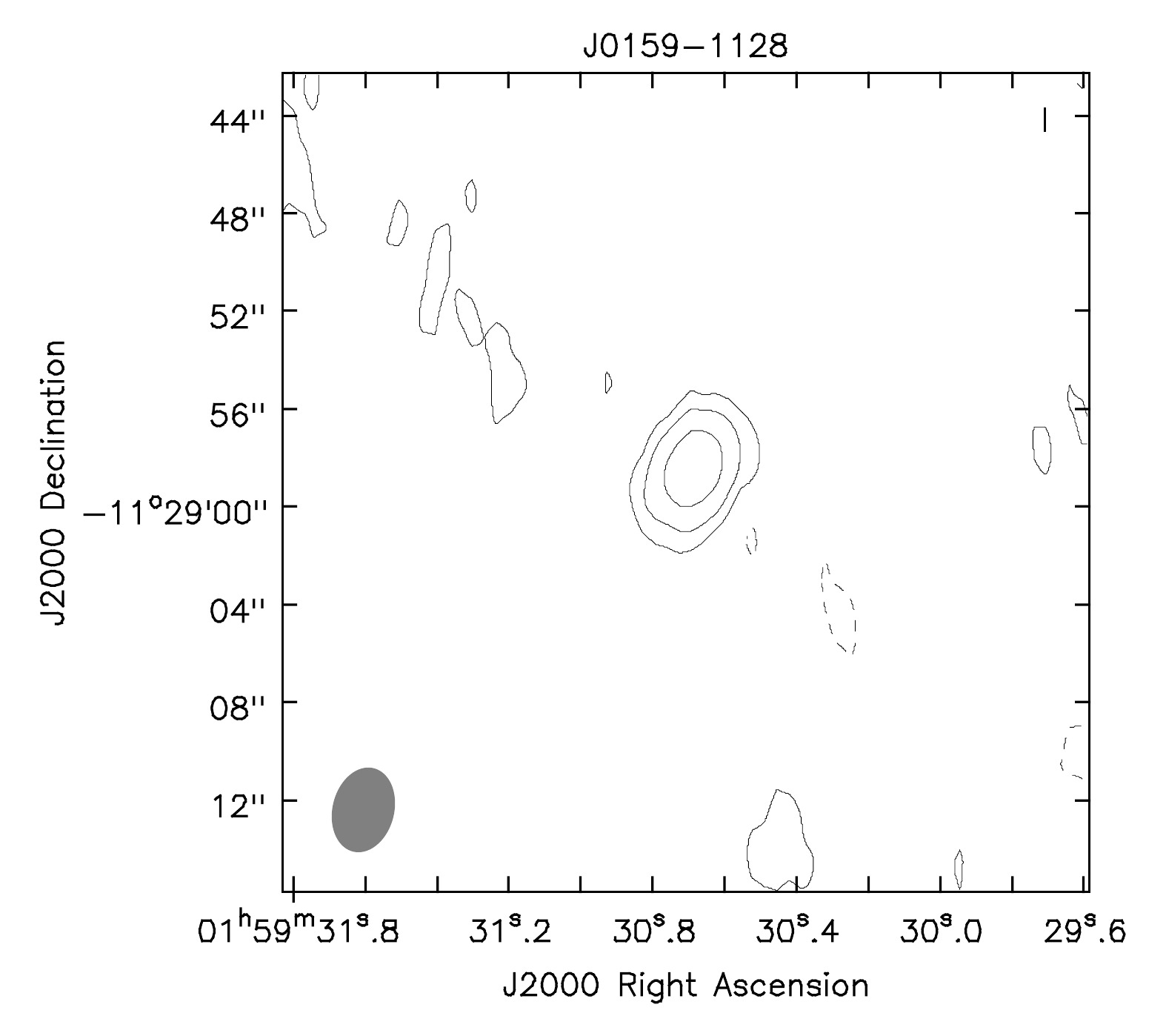}
\includegraphics[width=.41\textwidth, trim={1cm 13.5cm 4cm 2cm}, clip]{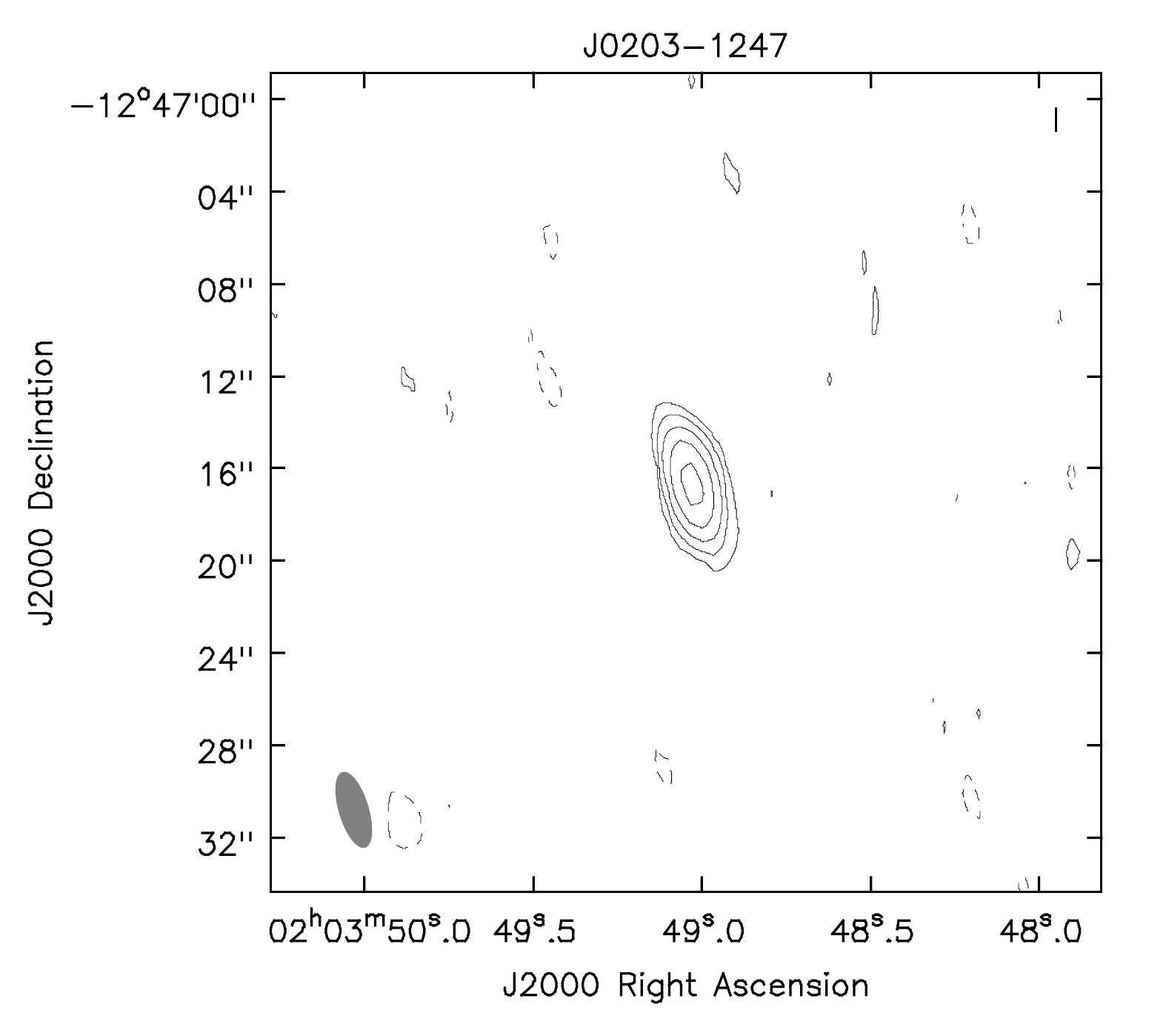}
\caption{\textit{left panel:} J0159$-$1128, rms = 5 $\mu$Jy beam$^{-1}$, contour levels at $-$3, 3 $\times$ 2$^n$, $n \in$ [0,2], beam size 12.92 $\times$ 9.22 kpc. \textit{right panel:} J0203$-$1247, rms = 5 $\mu$Jy beam$^{-1}$, contour levels at $-$3, 3 $\times$ 2$^n$, $n \in$ [0,4], beam size 3.82 $\times$ 1.42 kpc.}
\label{c}
\end{figure*}

\begin{figure*}
\centering
\includegraphics[width=.41\textwidth, trim={1cm 13.5cm 4cm 2cm}, clip]{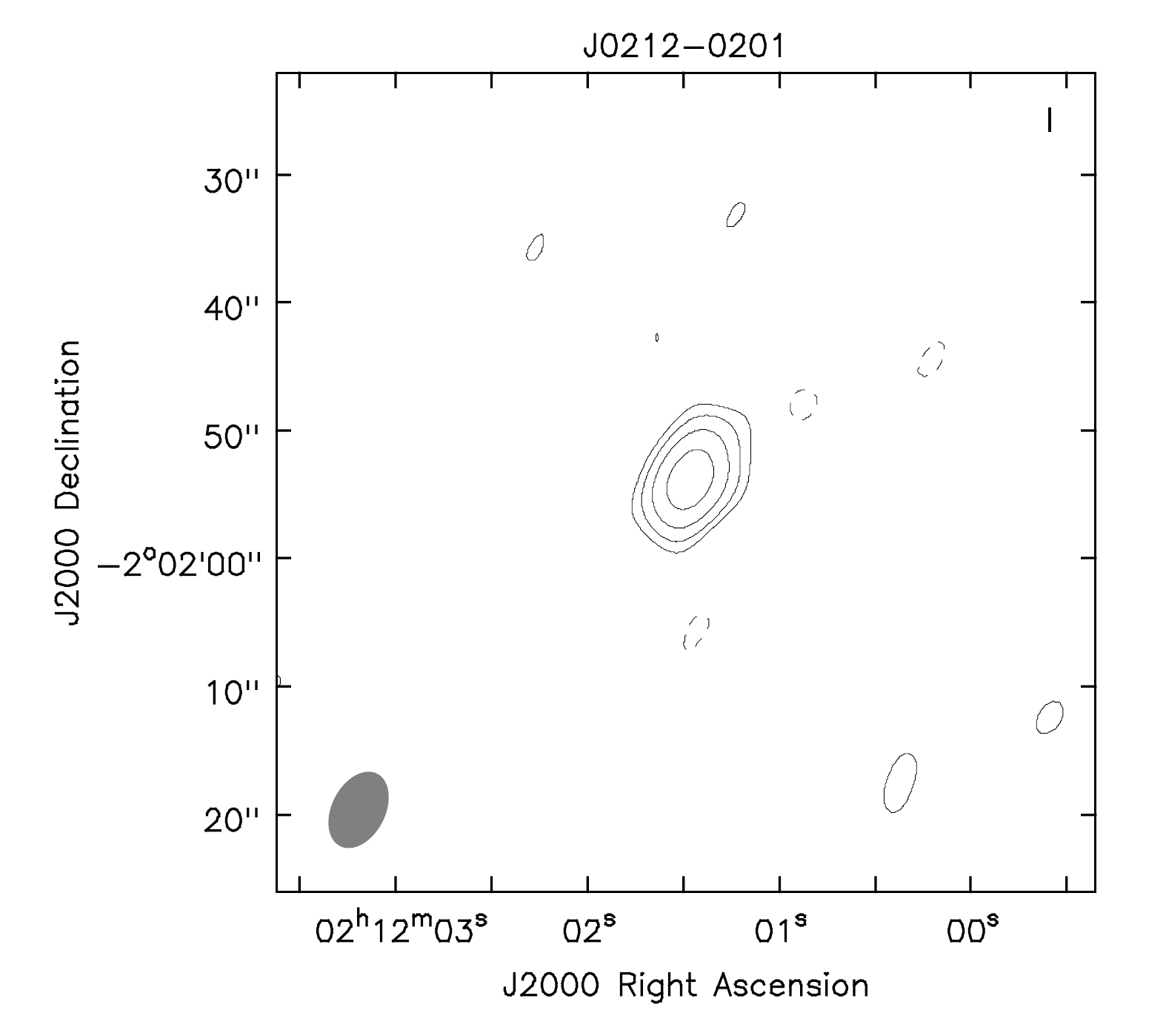}
\includegraphics[width=.41\textwidth, trim={1cm 13.5cm 4cm 2cm}, clip]{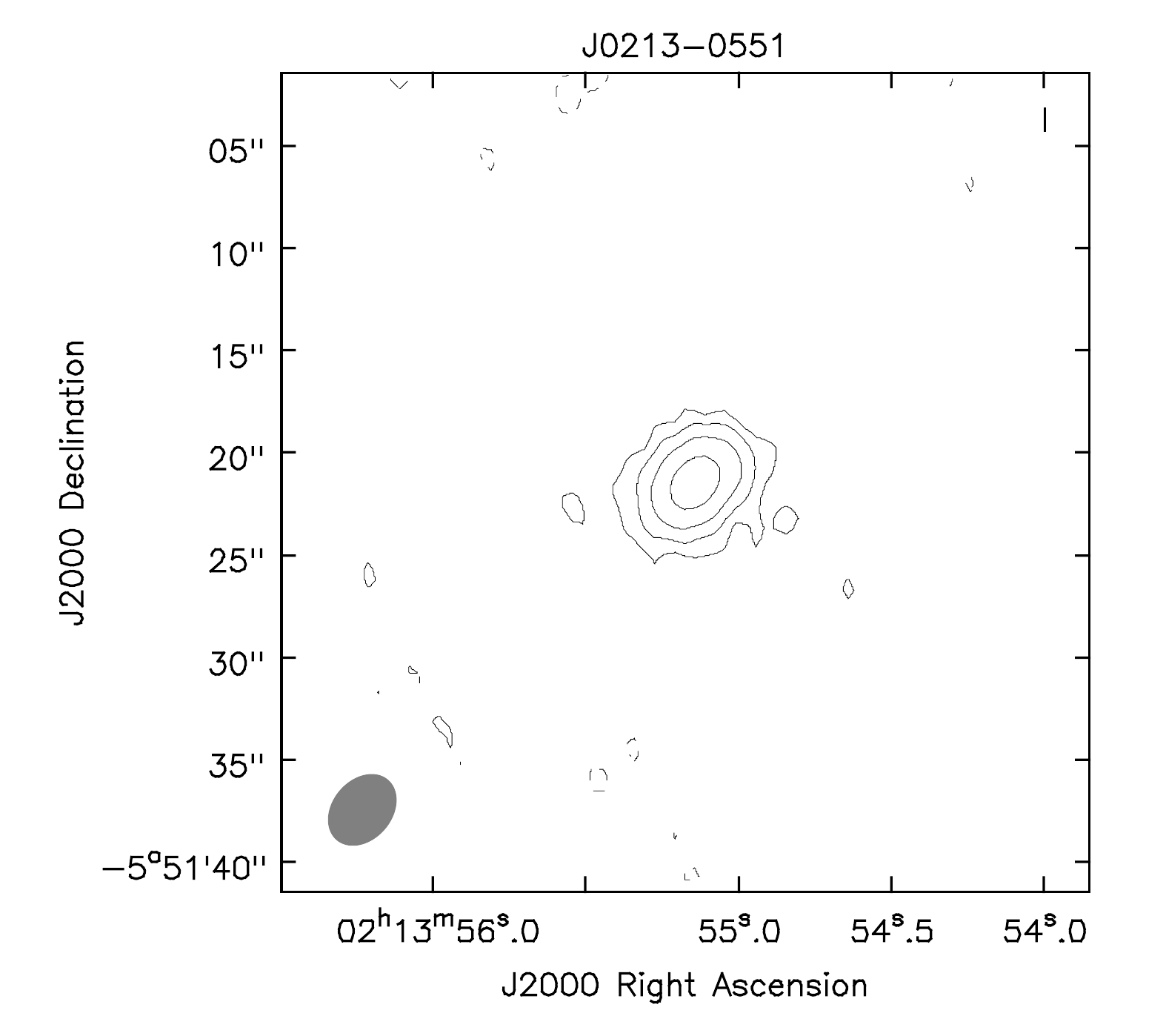}
\caption{\textit{left panel:} J0212$-$0201, rms = 6 $\mu$Jy beam$^{-1}$, contour levels at $-$3, 3 $\times$ 2$^n$, $n \in$ [0,3], beam size 73.33 $\times$ 47.79 kpc. \textit{right panel:} J0213$-$0551, rms = 5 $\mu$Jy beam$^{-1}$, contour levels at $-$3, 3 $\times$ 2$^n$, $n \in$ [0,3], beam size 12.24 $\times$ 9.13 kpc.}
\label{d}
\end{figure*}

\begin{figure*}
\centering
\includegraphics[width=.41\textwidth, trim={1cm 13.5cm 4cm 2cm}, clip]{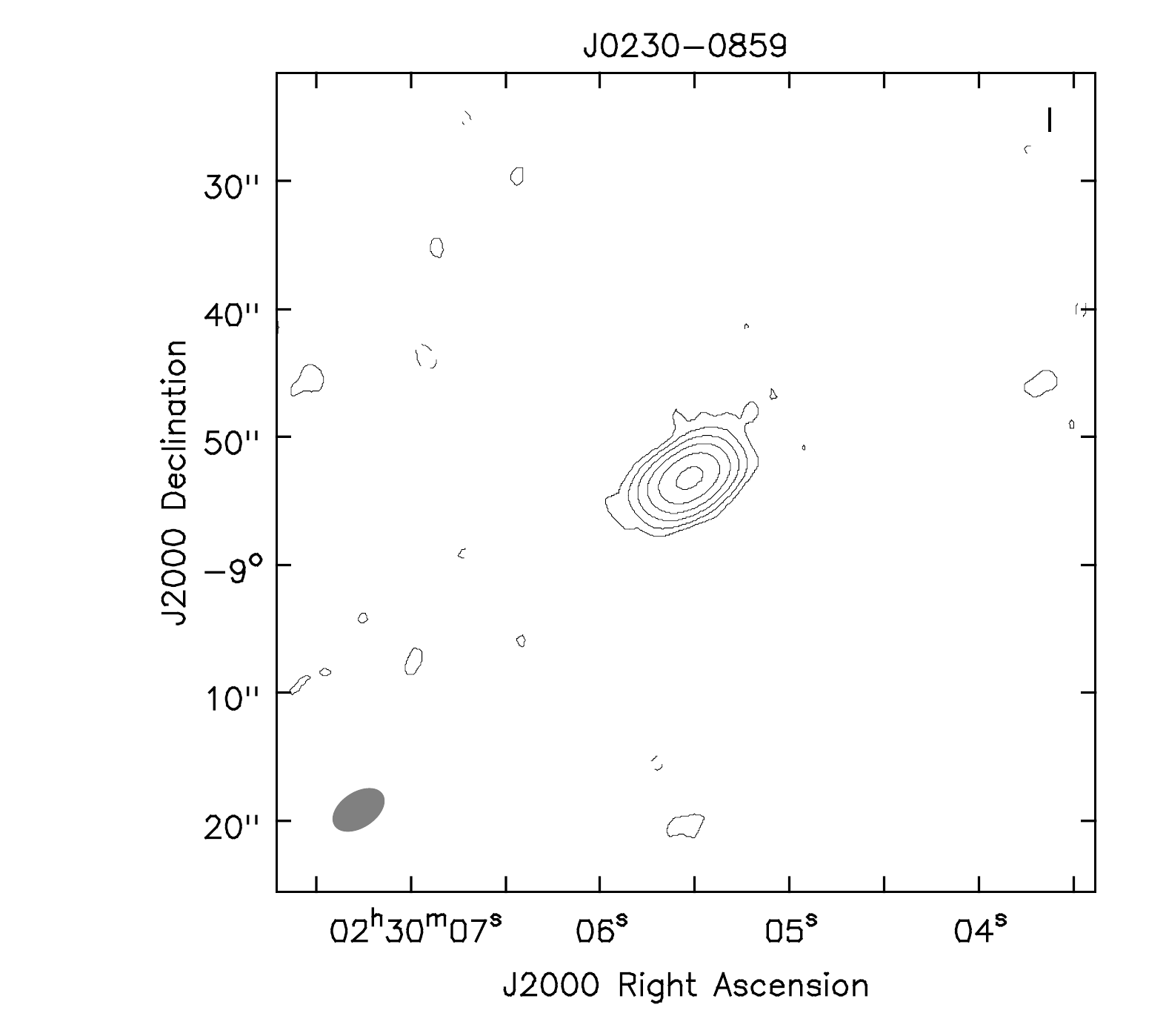}
\includegraphics[width=.41\textwidth, trim={1cm 13.5cm 4cm 2cm}, clip]{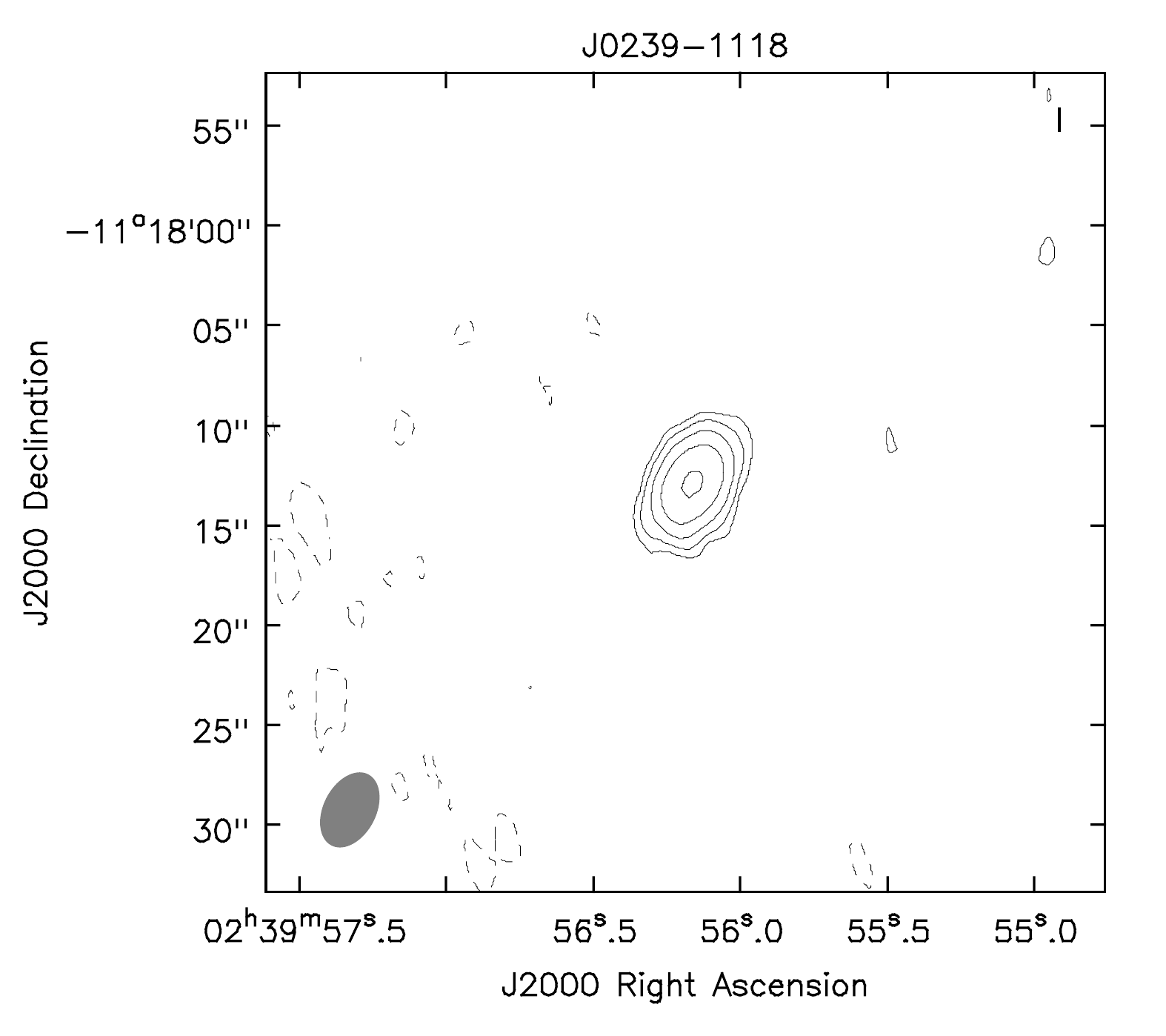}
\caption{\textit{left panel:} J0230$-$0859, rms = 9 $\mu$Jy beam$^{-1}$, contour levels at $-$3, 3 $\times$ 2$^n$, $n \in$ [0,5], beam size 1.50 $\times$ 0.92 kpc. \textit{right panel:} J0239$-$1118, rms = 6 $\mu$Jy beam$^{-1}$, contour levels at $-$3, 3 $\times$ 2$^n$, $n \in$ [0,4], beam size 19.13 $\times$ 12.45 kpc.}
\label{e}
\end{figure*}

\begin{figure*}
\centering
\includegraphics[width=.41\textwidth, trim={1cm 13.5cm 4cm 2cm}, clip]{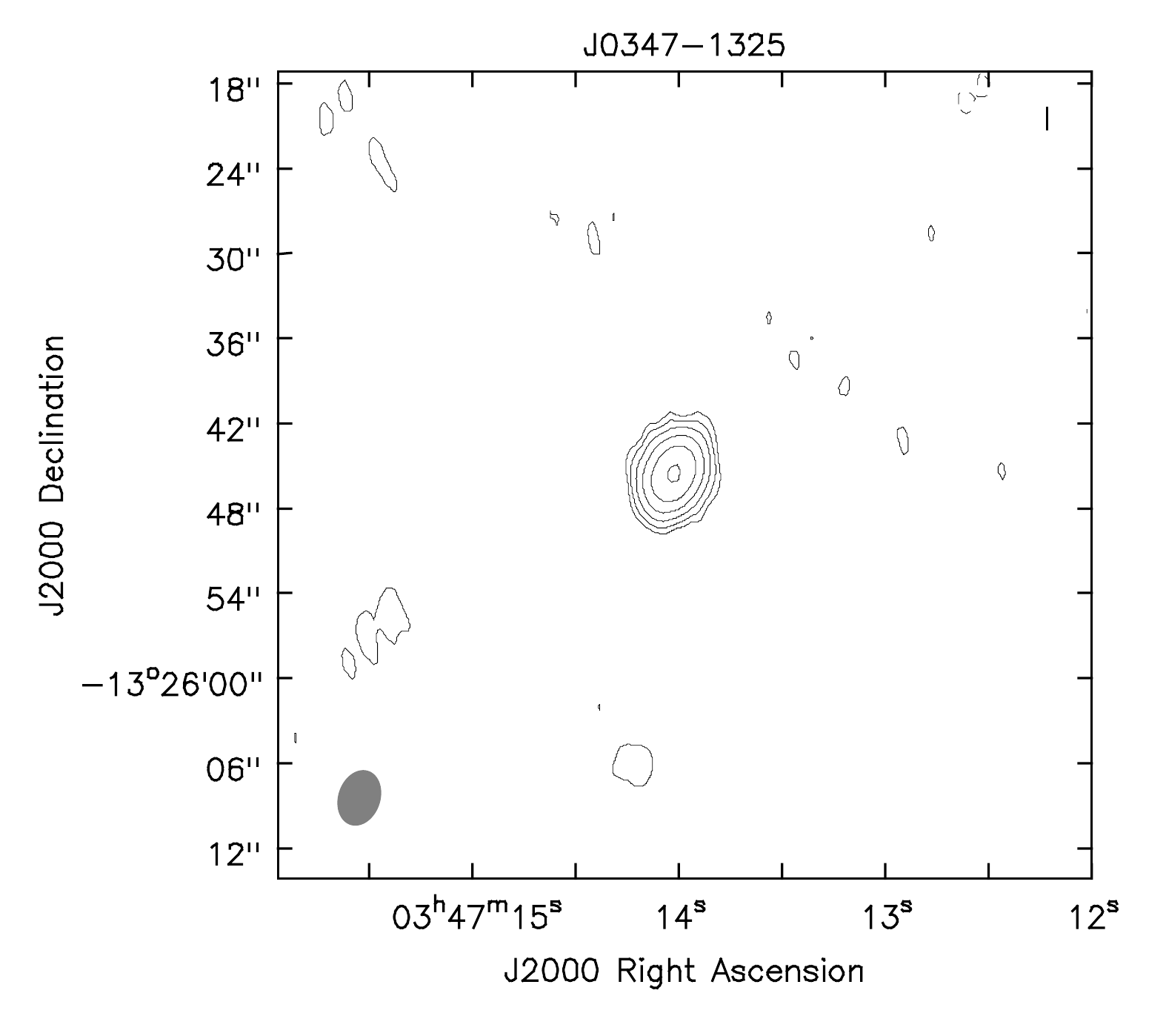}
\includegraphics[width=.41\textwidth, trim={1cm 13.5cm 4cm 2cm}, clip]{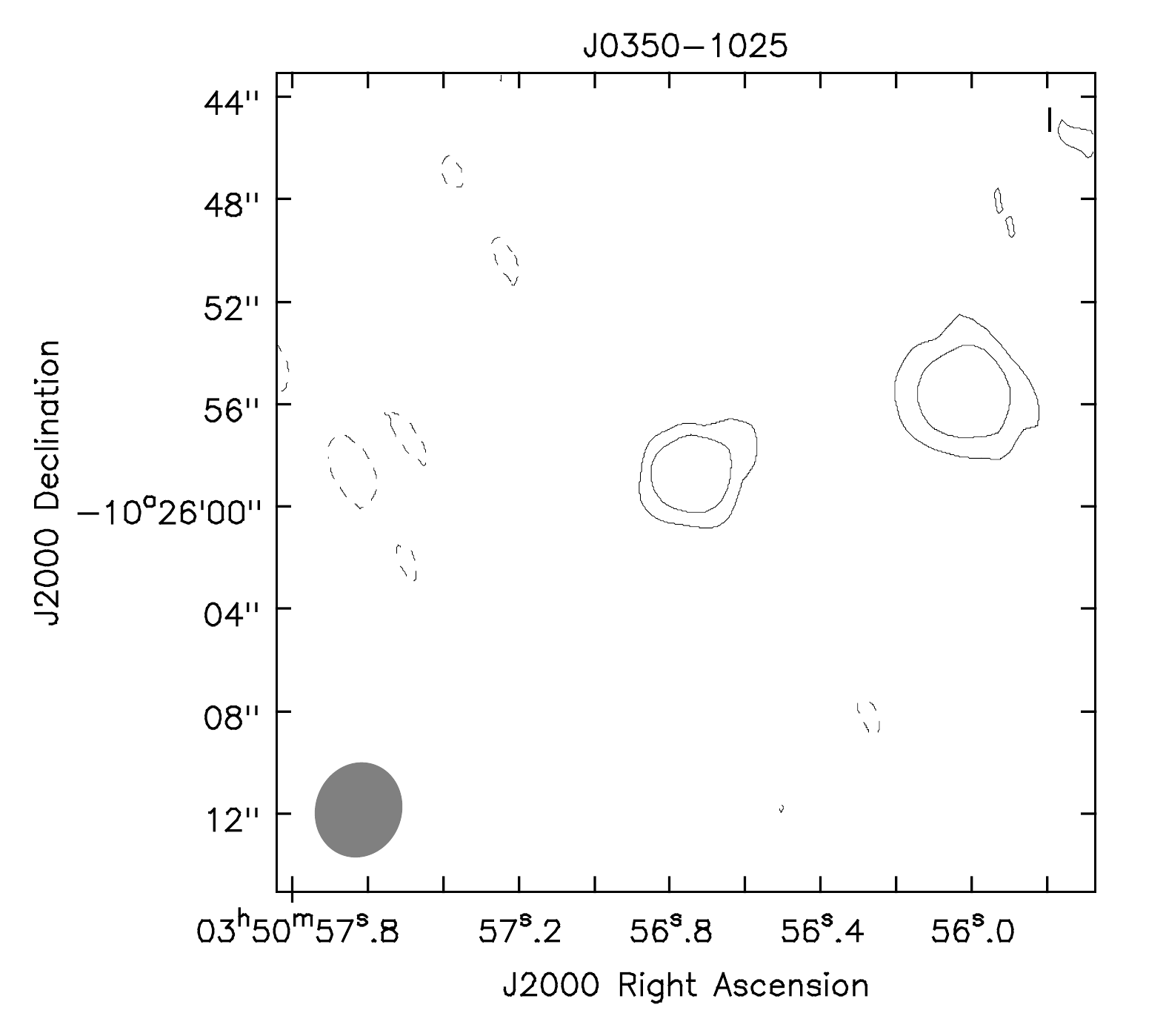}
\caption{\textit{left panel:} J0347$-$1325, rms = 6 $\mu$Jy beam$^{-1}$, contour levels at $-$3, 3 $\times$ 2$^n$, $n \in$ [0,5], beam size 17.82 $\times$ 13.10 kpc. \textit{right panel:} J0350$-$1025, rms = 5 $\mu$Jy beam$^{-1}$, contour levels at $-$3, 3 $\times$ 2$^n$, $n \in$ [0,1], beam size 10.83 $\times$ 9.72 kpc.}
\label{f}
\end{figure*}

\begin{figure*}
\centering
\includegraphics[width=.41\textwidth, trim={1cm 13.5cm 4cm 2cm}, clip]{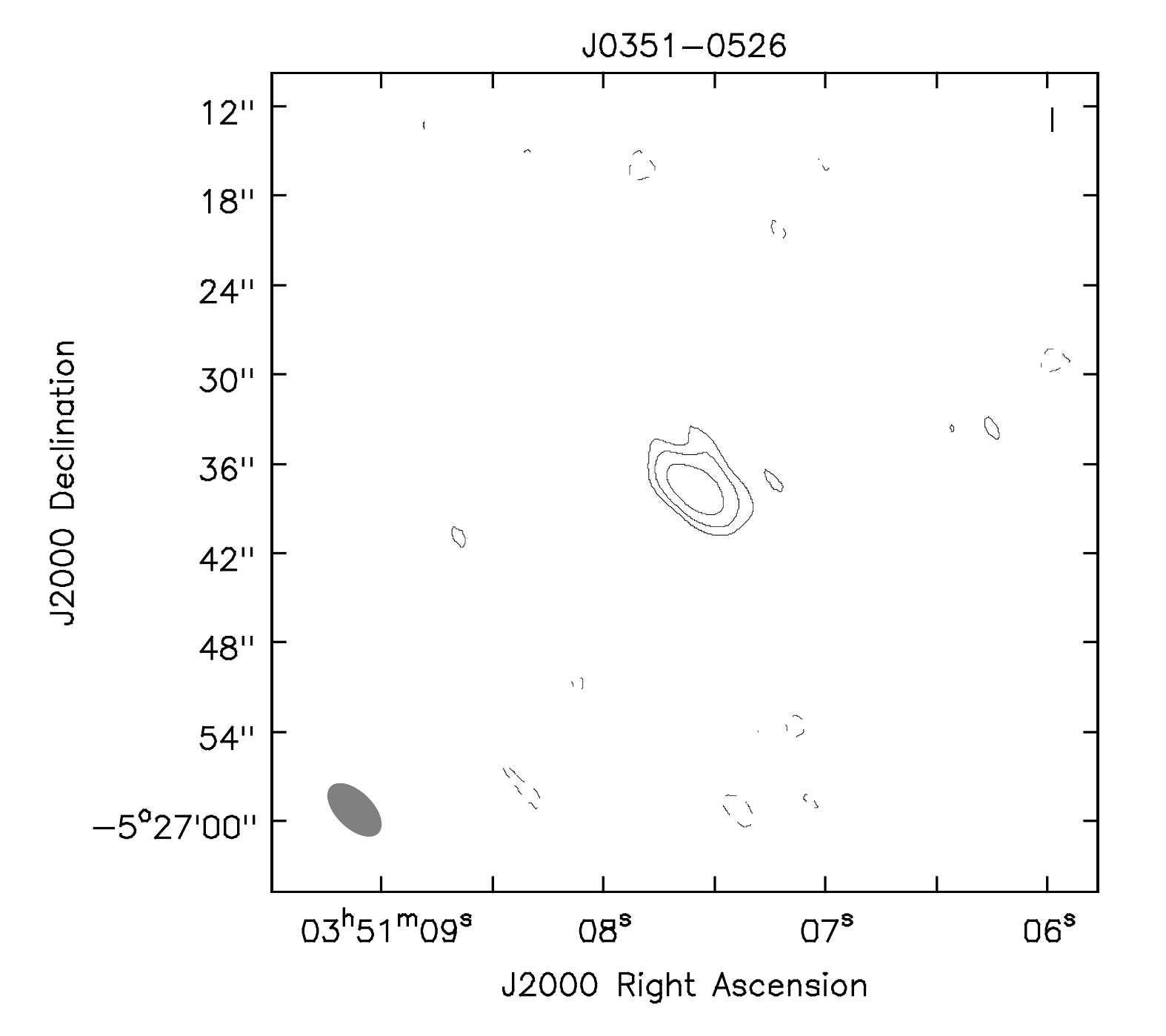}
\includegraphics[width=.41\textwidth, trim={1cm 13.5cm 4cm 2cm}, clip]{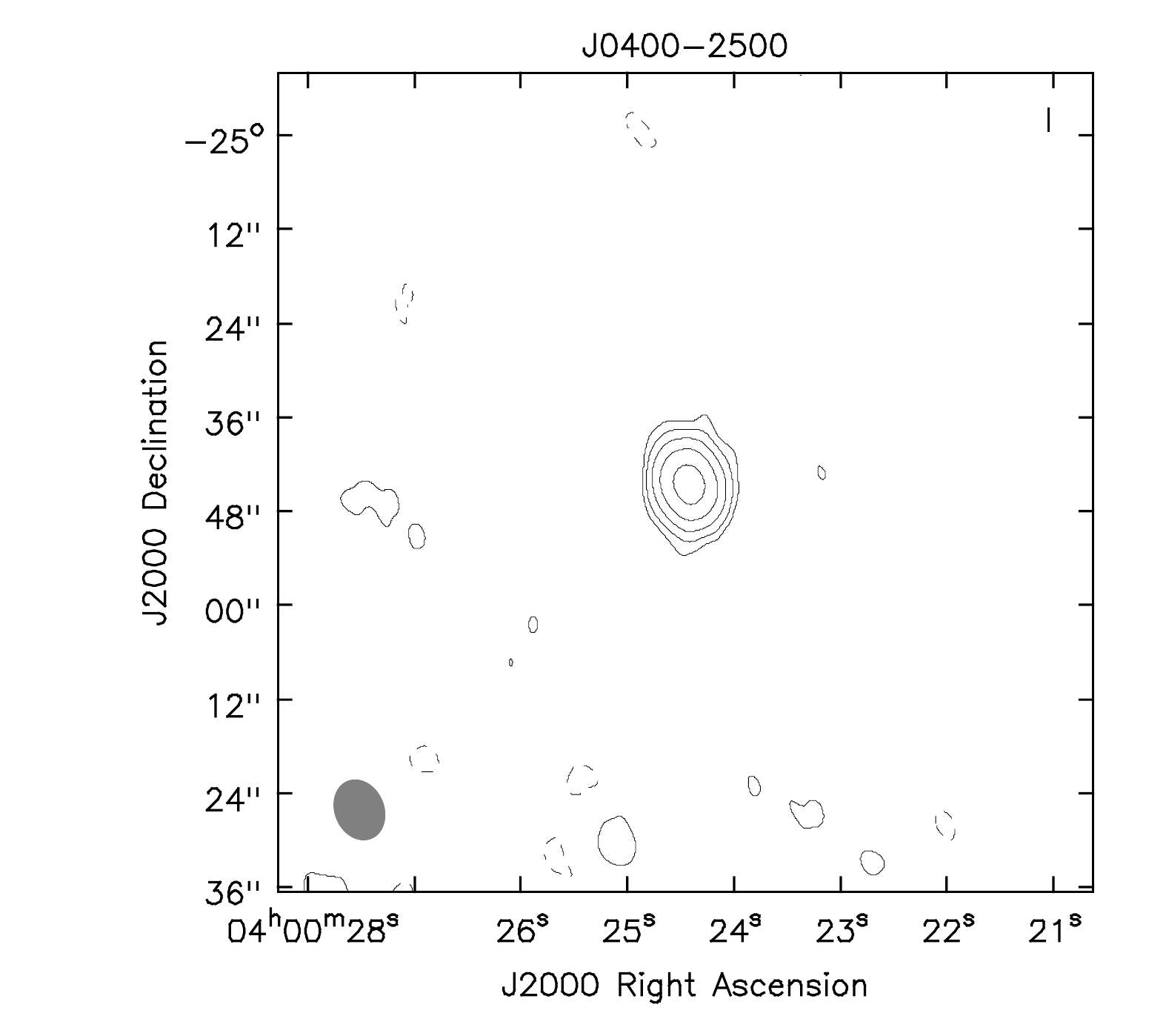}
\caption{\textit{left panel:} J0351$-$0526, rms = 5 $\mu$Jy beam$^{-1}$, contour levels at $-$3, 3 $\times$ 2$^n$, $n \in$ [0,2], beam size 6.51 $\times$ 3.52 kpc. \textit{right panel:} J0400$-$2500, rms = 19 $\mu$Jy beam$^{-1}$, contour levels at $-$3, 3 $\times$ 2$^n$, $n \in$ [0,4], beam size 17.04 $\times$ 13.46 kpc.}
\label{g}
\end{figure*}

\begin{figure*}
\centering
\includegraphics[width=.41\textwidth, trim={1cm 13.5cm 4cm 2cm}, clip]{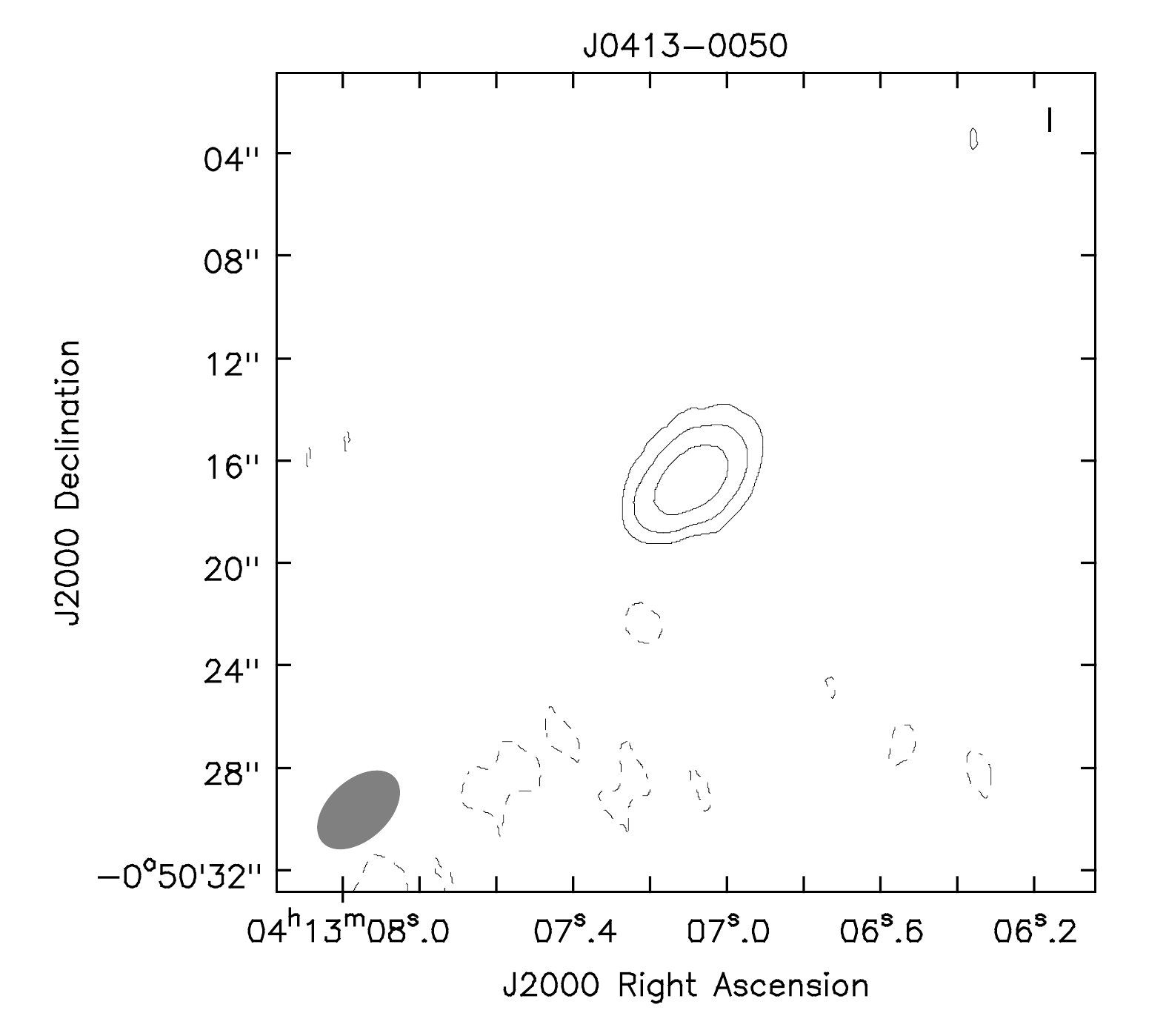}
\includegraphics[width=.41\textwidth, trim={1cm 13.5cm 4cm 2cm}, clip]{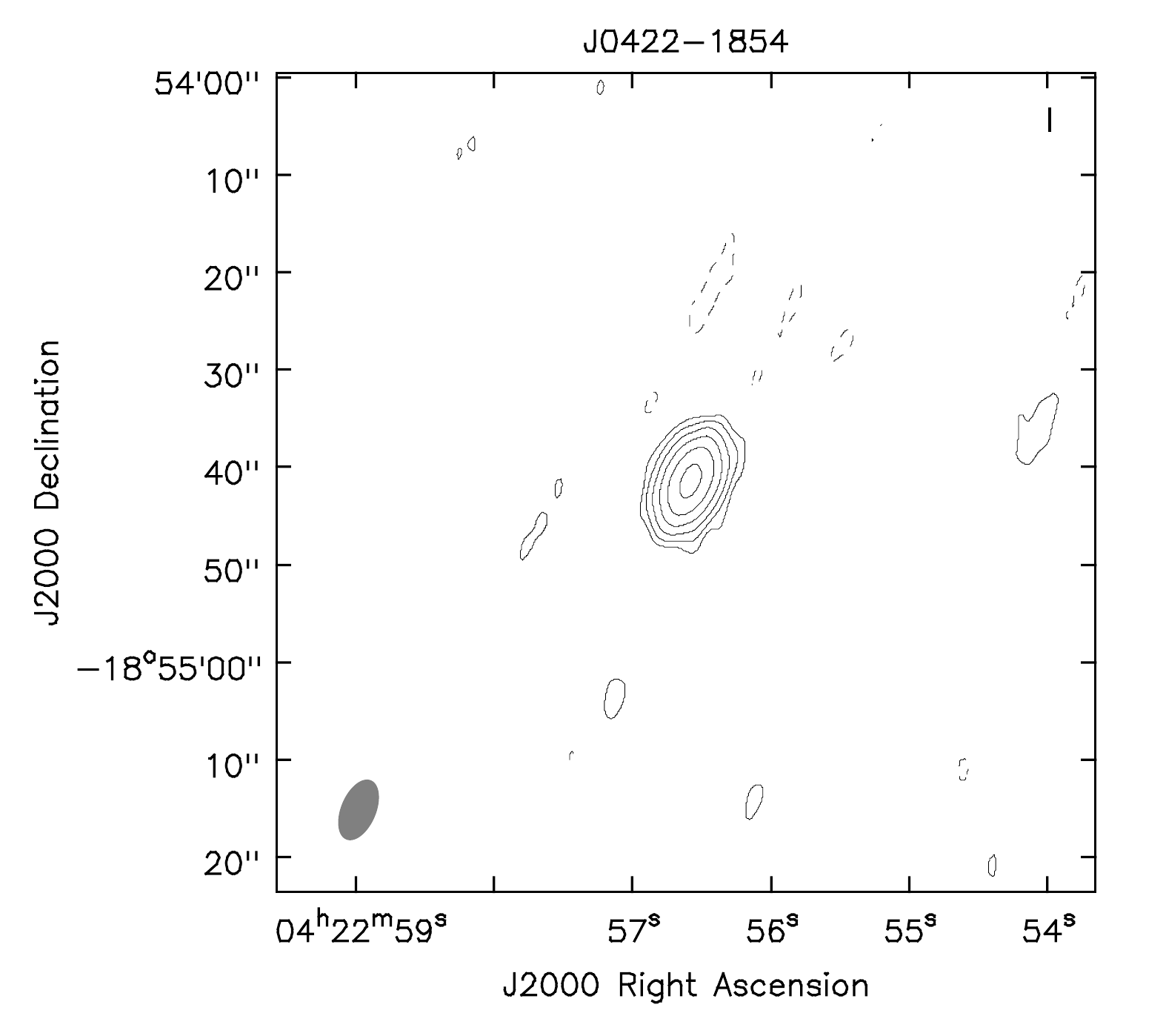}
\caption{\textit{left panel:} J0413$-$0050, rms = 7 $\mu$Jy beam$^{-1}$, contour levels at $-$3, 3 $\times$ 2$^n$, $n \in$ [0,2], beam size 3.22 $\times$ 1.98 kpc. \textit{right panel:} J0422$-$1854, rms = 9 $\mu$Jy beam$^{-1}$, contour levels at $-$3, 3 $\times$ 2$^n$, $n \in$ [0,5], beam size 9.00 $\times$ 4.84 kpc.}
\label{h}
\end{figure*}

\begin{figure*}
\centering
\includegraphics[width=.41\textwidth, trim={1cm 13.5cm 4cm 2cm}, clip]{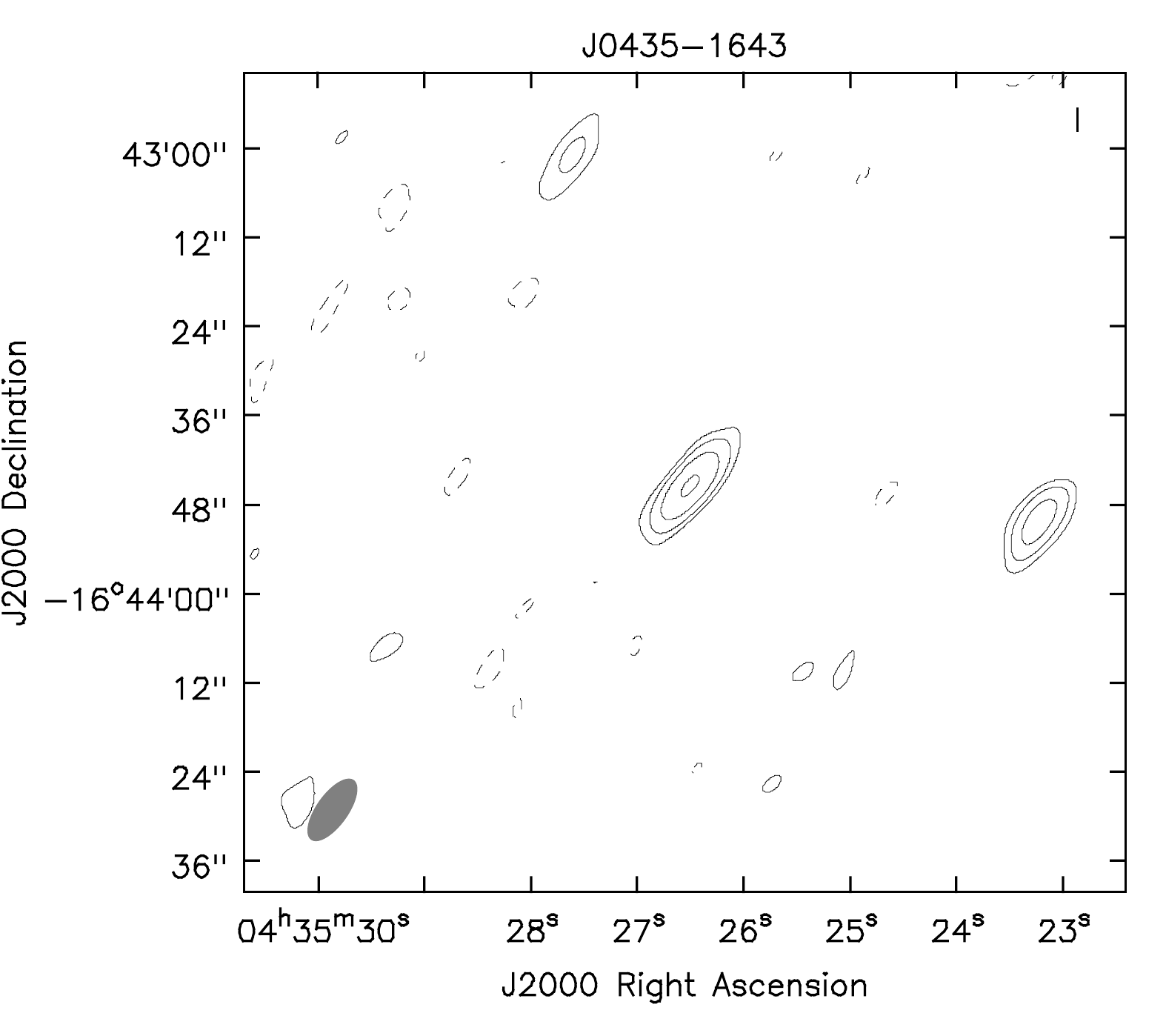}
\includegraphics[width=.41\textwidth, trim={1cm 13.5cm 4cm 2cm}, clip]{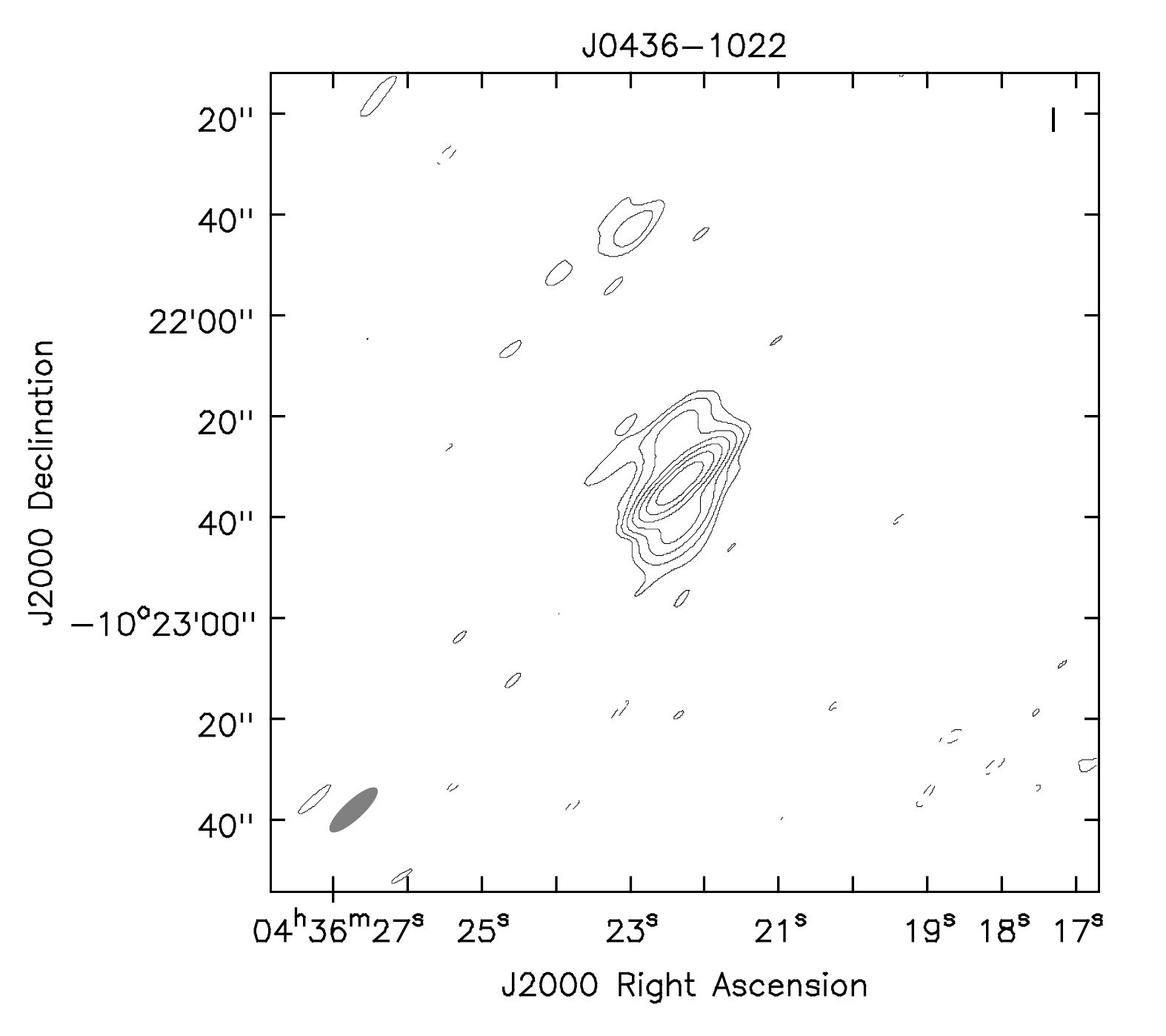}
\caption{\textit{left panel:} J0435$-$1643, rms = 5 $\mu$Jy beam$^{-1}$, contour levels at $-$3, 3 $\times$ 2$^n$, $n \in$ [0,3], beam size 21.33 $\times$ 9.02 kpc. \textit{right panel:} J0436$-$1022, rms = 11 $\mu$Jy beam$^{-1}$, contour levels at $-$3, 3 $\times$ 2$^n$, $n \in$ [0,6], beam size 9.22 $\times$ 2.65 kpc.}
\label{i}
\end{figure*}

\begin{figure*}
\centering
\includegraphics[width=.41\textwidth, trim={1cm 13.5cm 4cm 2cm}, clip]{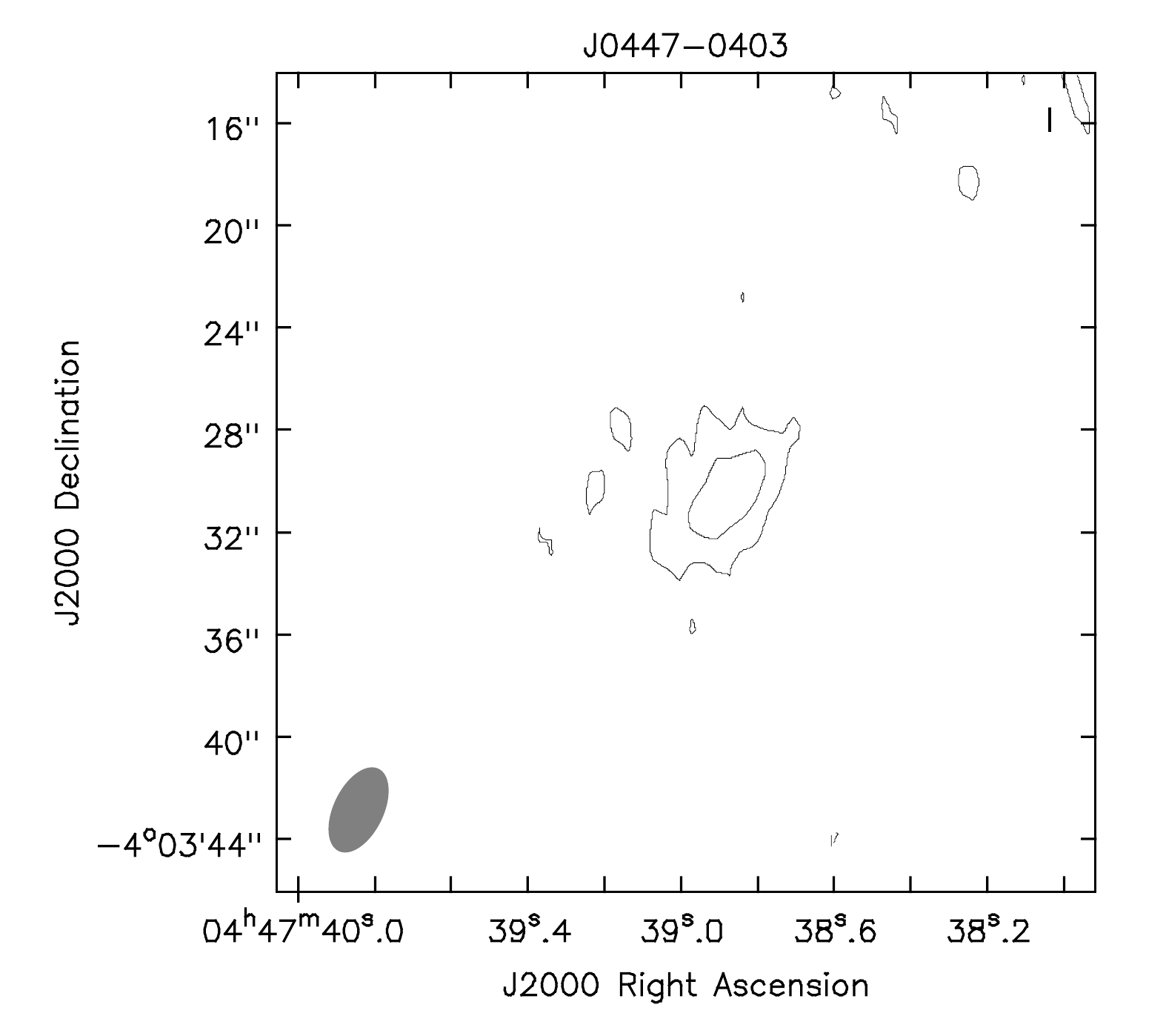}
\includegraphics[width=.41\textwidth, trim={1cm 13.5cm 4cm 2cm}, clip]{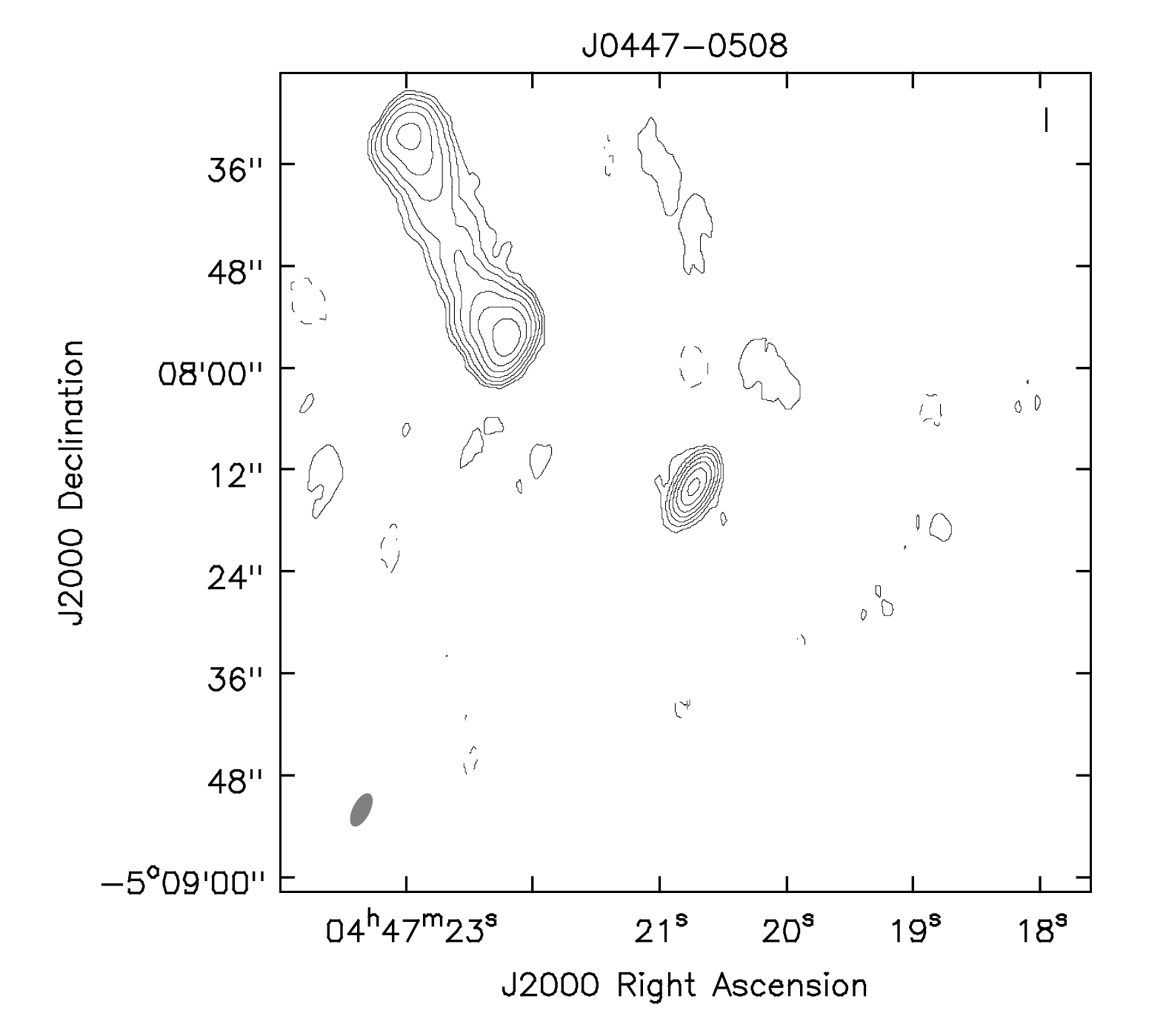}
\caption{\textit{left panel:} J0447$-$0403, rms = 5 $\mu$Jy beam$^{-1}$, contour levels at $-$3, 3 $\times$ 2$^n$, $n \in$ [0,1], beam size 3.39 $\times$ 1.84 kpc. \textit{right panel:} J0447$-$0508, rms = 15 $\mu$Jy beam$^{-1}$, contour levels at $-$3, 3 $\times$ 2$^n$, $n \in$ [0,6], beam size 7.44 $\times$ 3.39 kpc.}
\label{j}
\end{figure*}

\begin{figure*}
\centering
\includegraphics[width=.41\textwidth, trim={1cm 13.5cm 4cm 2cm}, clip]{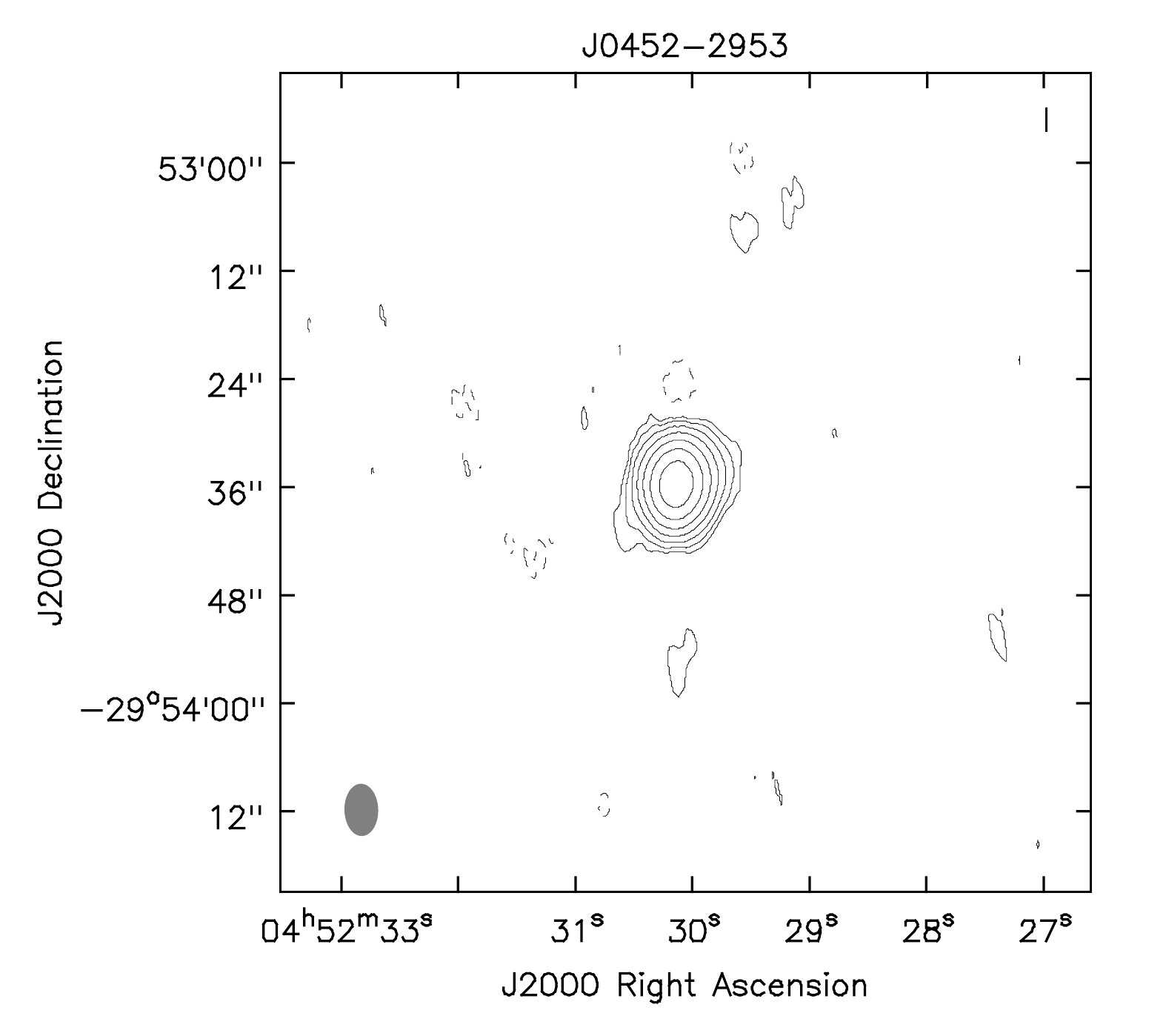}
\includegraphics[width=.41\textwidth, trim={1cm 13.5cm 4cm 2cm}, clip]{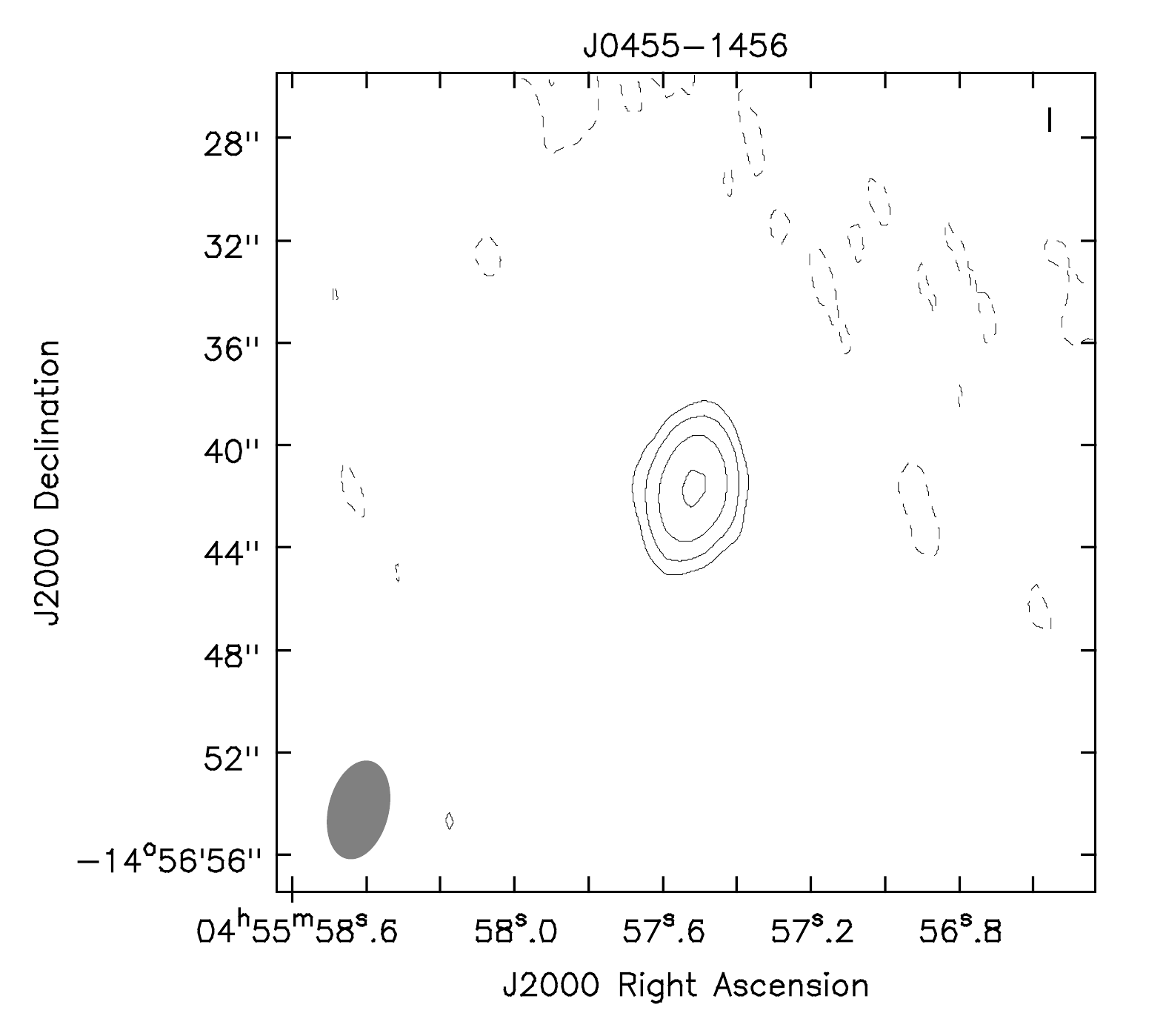}
\caption{\textit{left panel:} J0452$-$2953, rms = 8 $\mu$Jy beam$^{-1}$, contour levels at $-$3, 3 $\times$ 2$^n$, $n \in$ [0,6], beam size 40.37 $\times$ 25.70 kpc. \textit{right panel:} J0455$-$1456, rms = 5 $\mu$Jy beam$^{-1}$, contour levels at $-$3, 3 $\times$ 2$^n$, $n \in$ [0,3], beam size 12.13 $\times$ 7.22 kpc.}
\label{k}
\end{figure*}

\begin{figure*}
\centering
\includegraphics[width=.41\textwidth, trim={1cm 13.5cm 4cm 2cm}, clip]{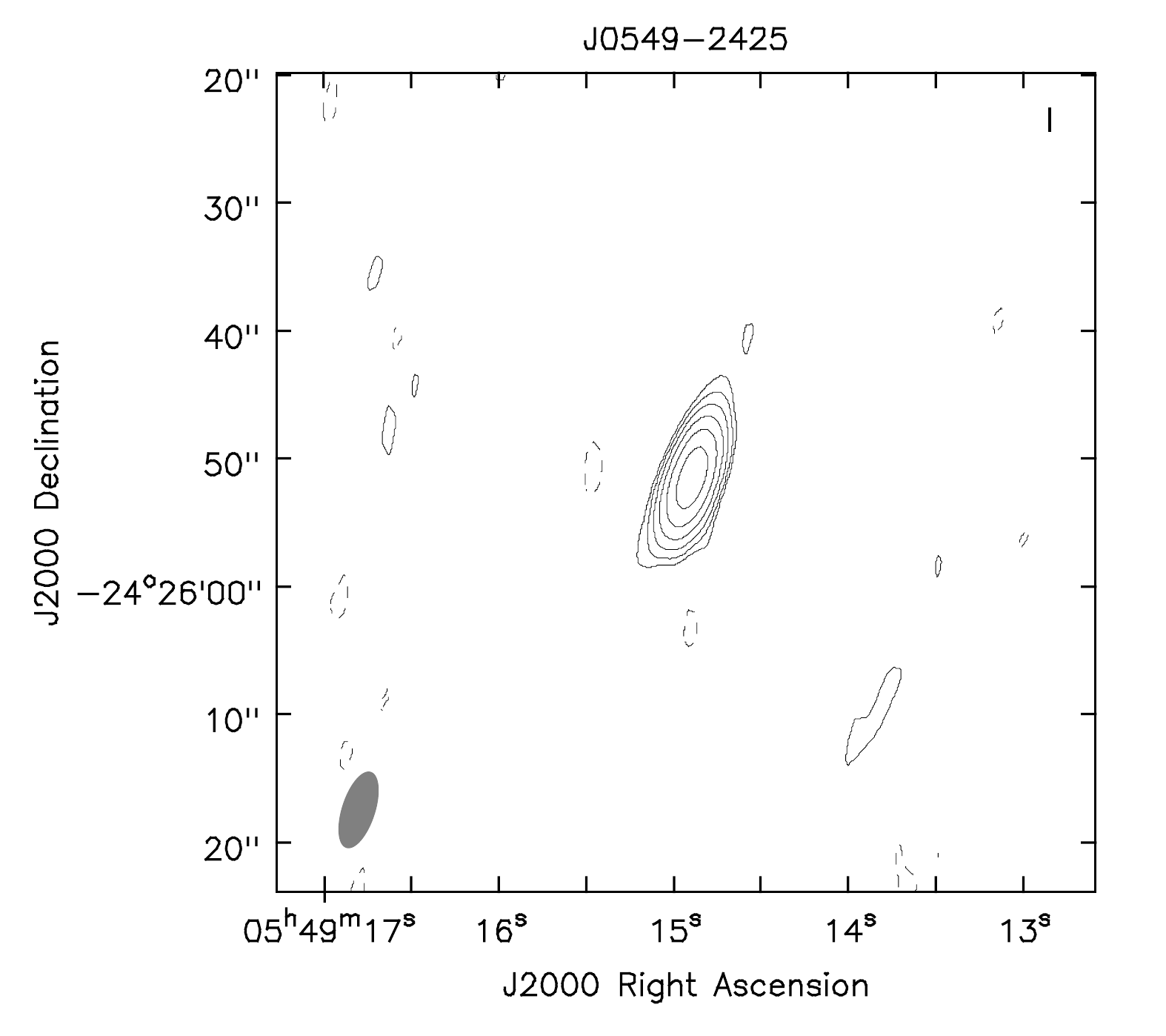}
\includegraphics[width=.41\textwidth, trim={1cm 13.5cm 4cm 2cm}, clip]{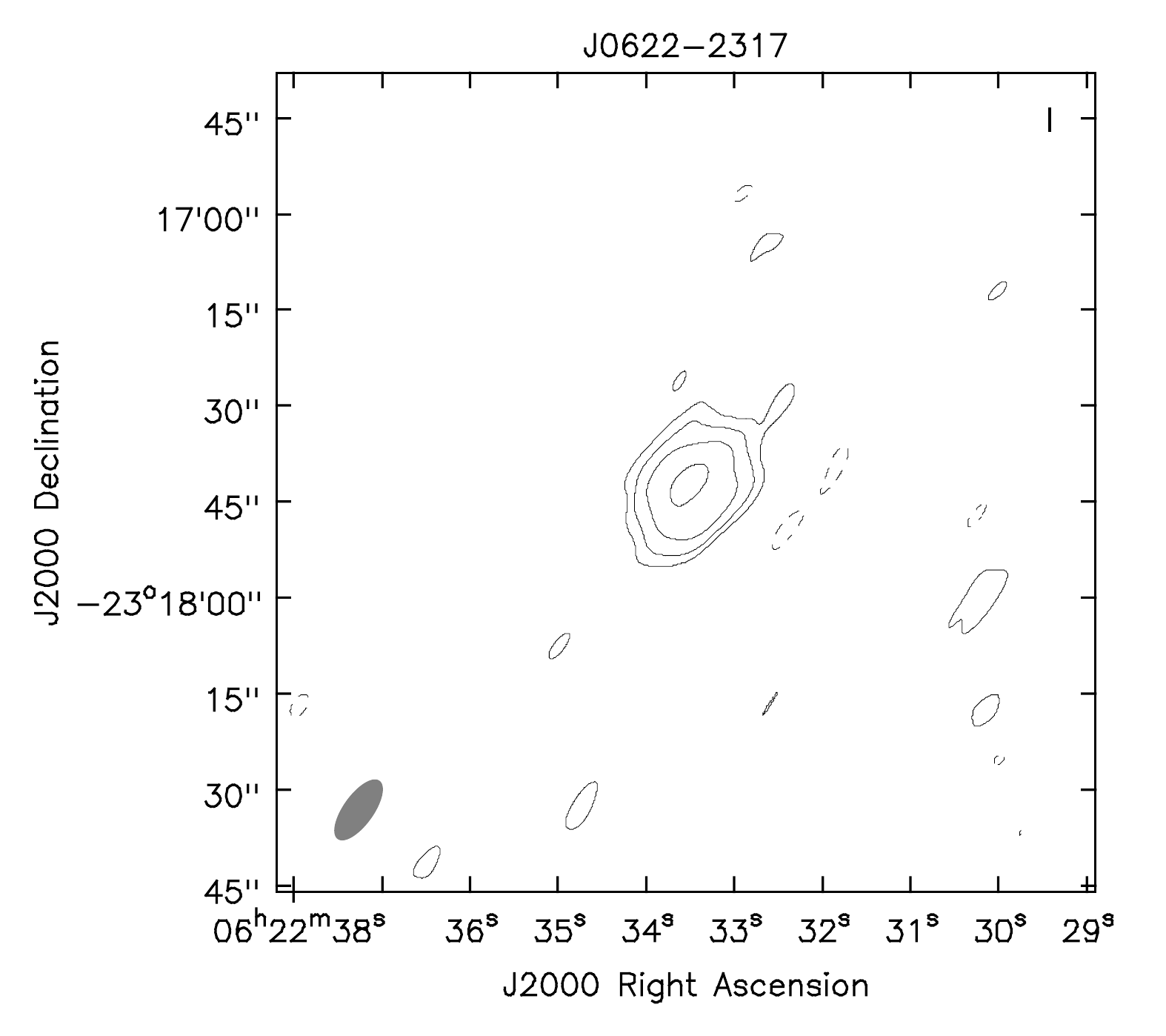}
\caption{\textit{left panel:} J0549$-$2425, rms = 10 $\mu$Jy beam$^{-1}$, contour levels at $-$3, 3 $\times$ 2$^n$, $n \in$ [0,5], beam size 5.95 $\times$ 2.43 kpc. \textit{right panel:} J0622$-$2317, rms = 10 $\mu$Jy beam$^{-1}$, contour levels at $-$3, 3 $\times$ 2$^n$, $n \in$ [0,3], beam size 8.89 $\times$ 3.67 kpc.}
\label{l}
\end{figure*}

\begin{figure*}
\centering
\includegraphics[width=.41\textwidth, trim={1cm 13.5cm 4cm 2cm}, clip]{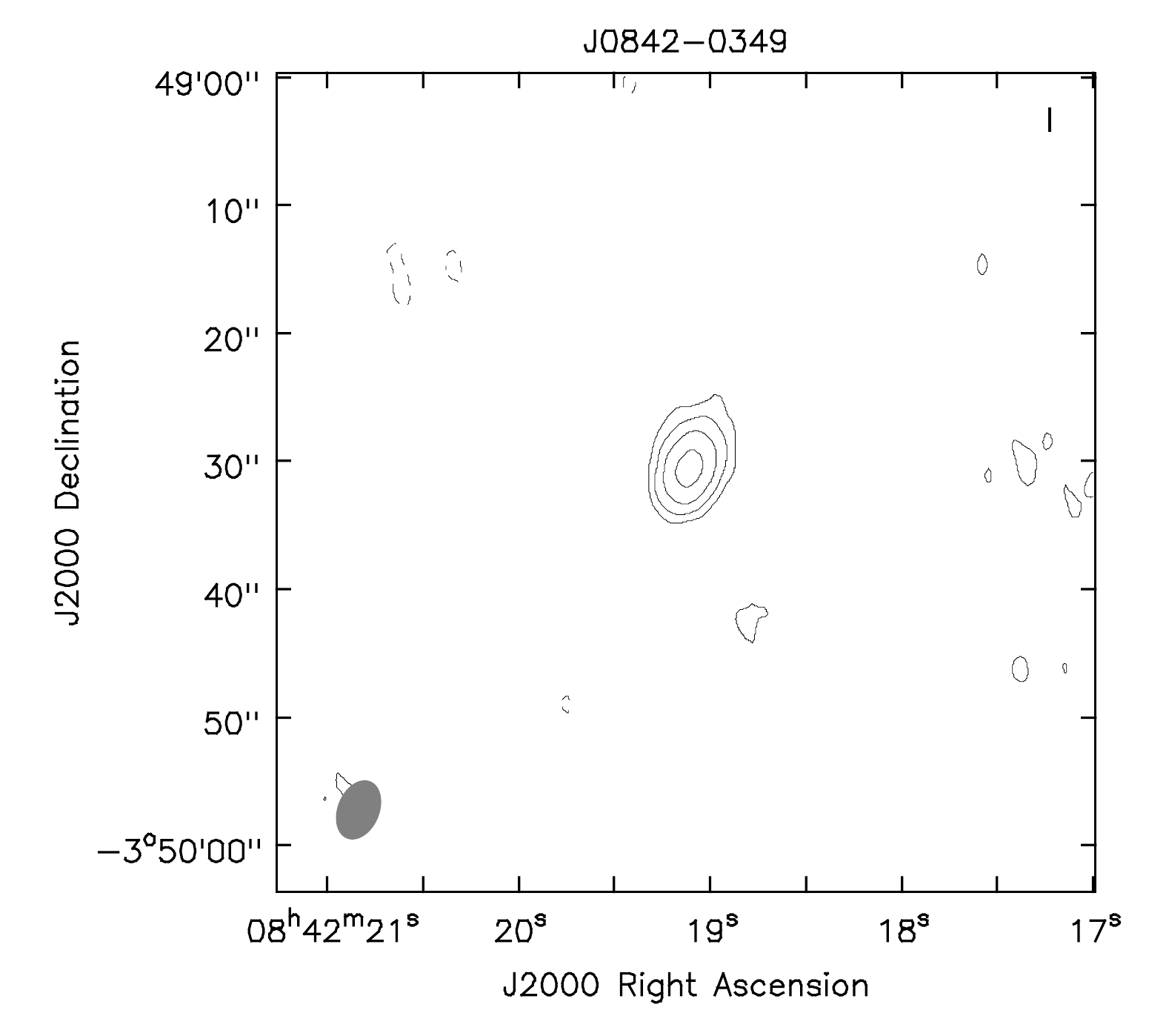}
\includegraphics[width=.41\textwidth, trim={1cm 13.5cm 4cm 2cm}, clip]{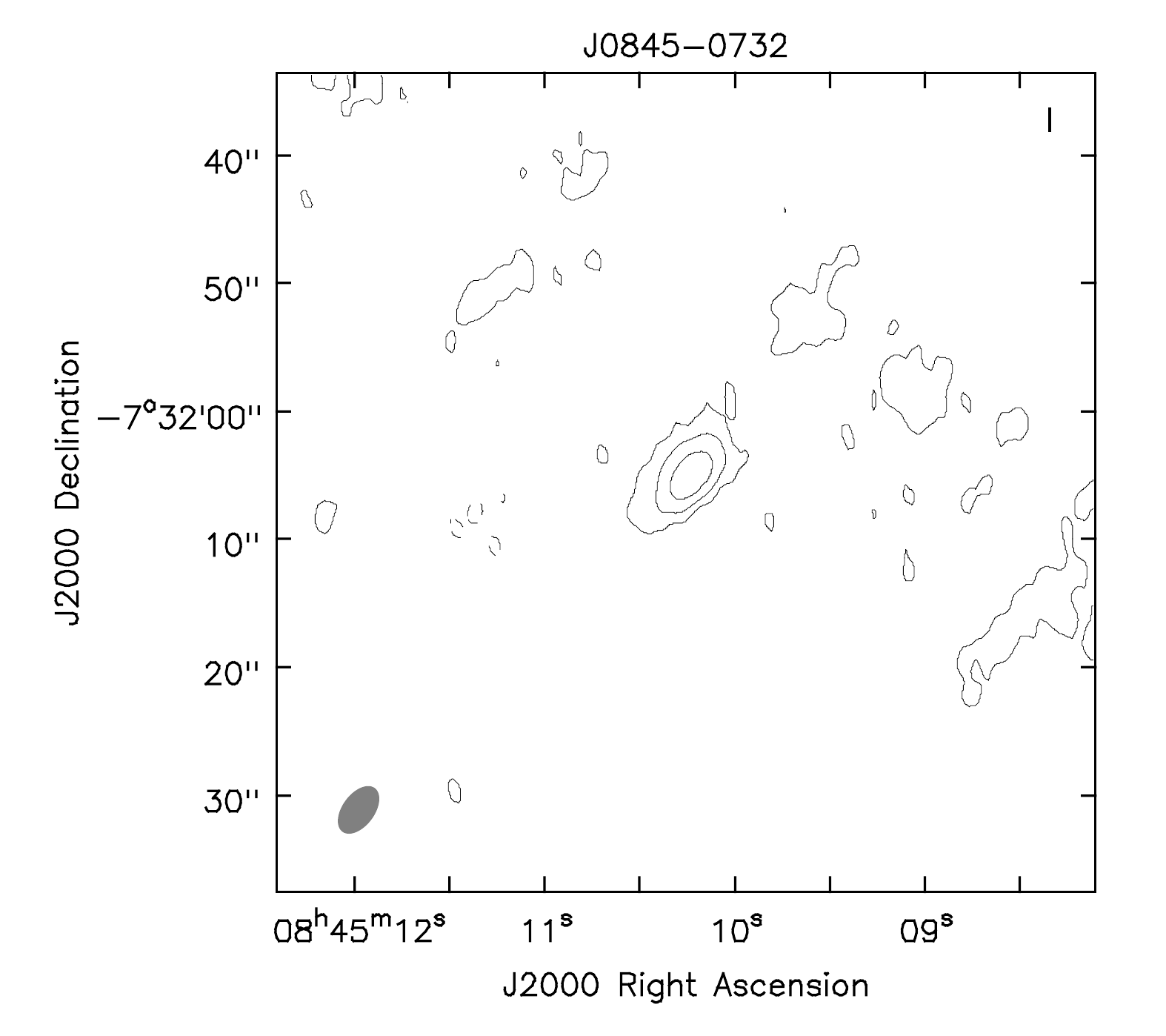}
\caption{\textit{left panel:} J0842$-$0349, rms = 9 $\mu$Jy beam$^{-1}$, contour levels at $-$3, 3 $\times$ 2$^n$, $n \in$ [0,3], beam size 43.52 $\times$ 29.16 kpc. \textit{right panel:} J0845$-$0732, rms = 7 $\mu$Jy beam$^{-1}$, contour levels at $-$3, 3 $\times$ 2$^n$, $n \in$ [0,2], beam size 9.62 $\times$ 5.58 kpc.}
\label{m}
\end{figure*}

\begin{figure*}
\centering
\includegraphics[width=.41\textwidth, trim={1cm 13.5cm 4cm 2cm}, clip]{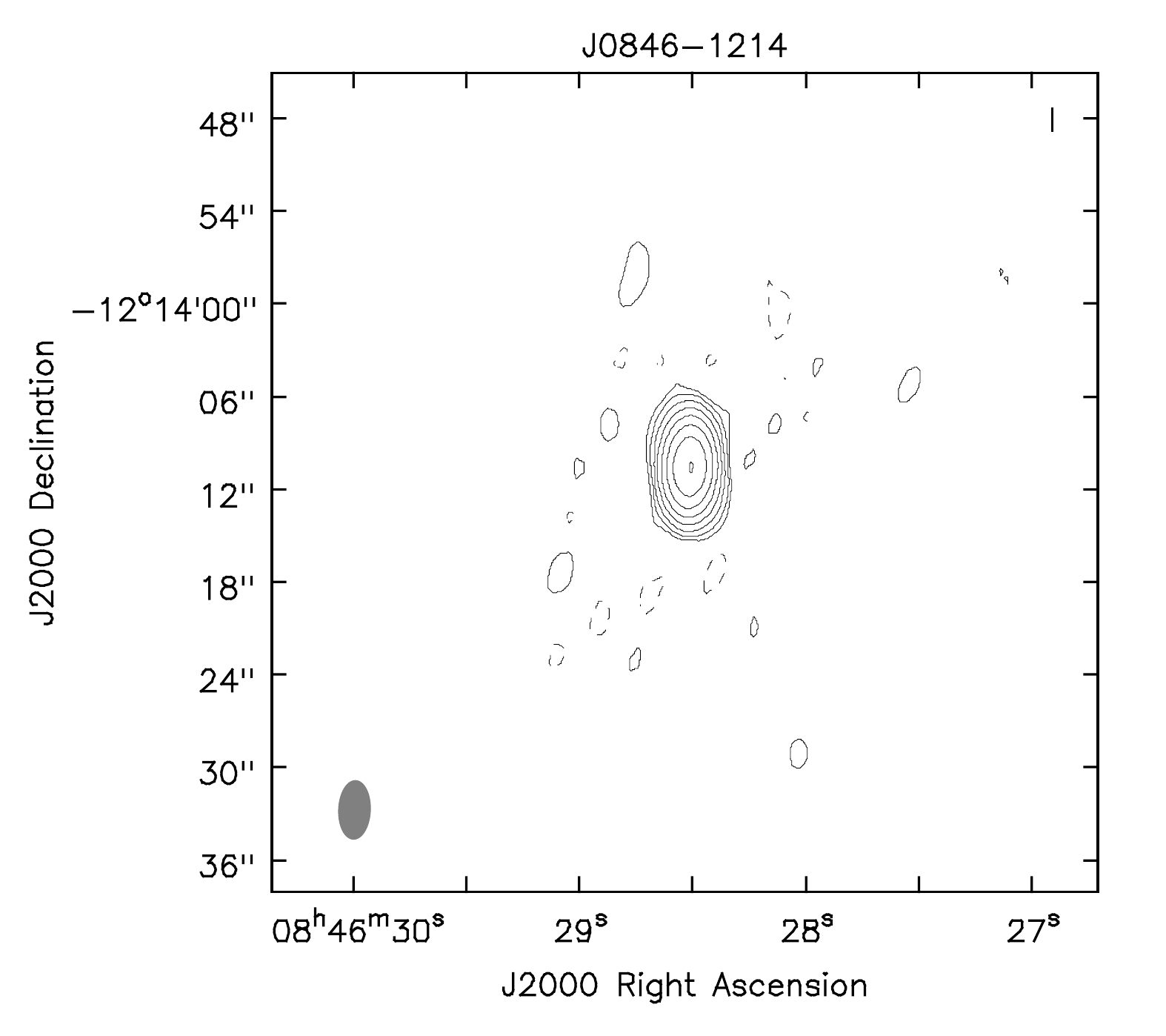}
\includegraphics[width=.41\textwidth, trim={1cm 13.5cm 4cm 2cm}, clip]{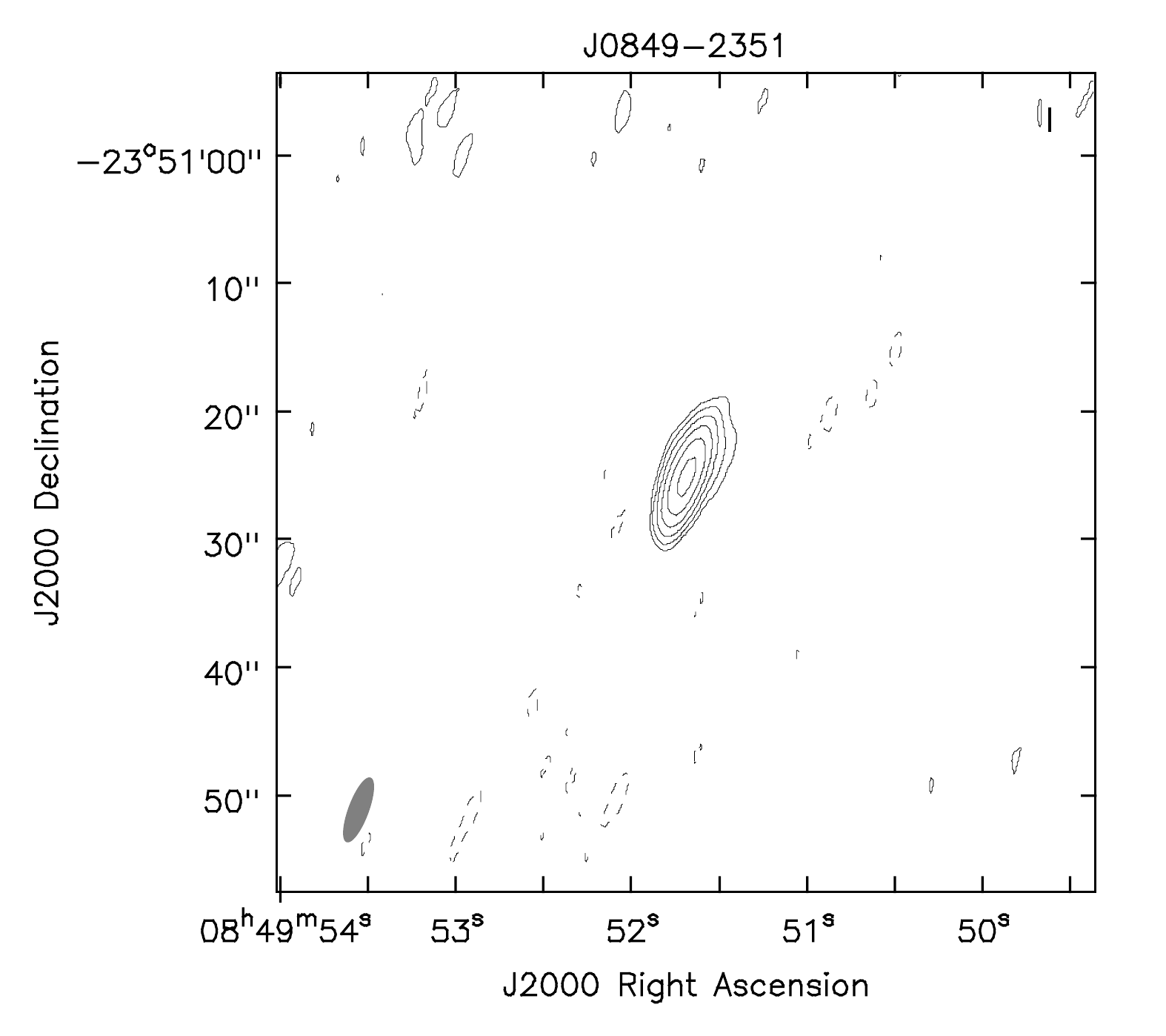}
\caption{\textit{left panel:} J0846$-$1214, rms = 13 $\mu$Jy beam$^{-1}$, contour levels at $-$3, 3 $\times$ 2$^n$, $n \in$ [0,7], beam size 9.14 $\times$ 4.93 kpc. \textit{right panel:} J0849$-$2351, rms = 6 $\mu$Jy beam$^{-1}$, contour levels at $-$3, 3 $\times$ 2$^n$, $n \in$ [0,5], beam size 15.29 $\times$ 4.36 kpc.}
\label{n}
\end{figure*}

\begin{figure*}
\centering
\includegraphics[width=.41\textwidth, trim={1cm 13.5cm 4cm 2cm}, clip]{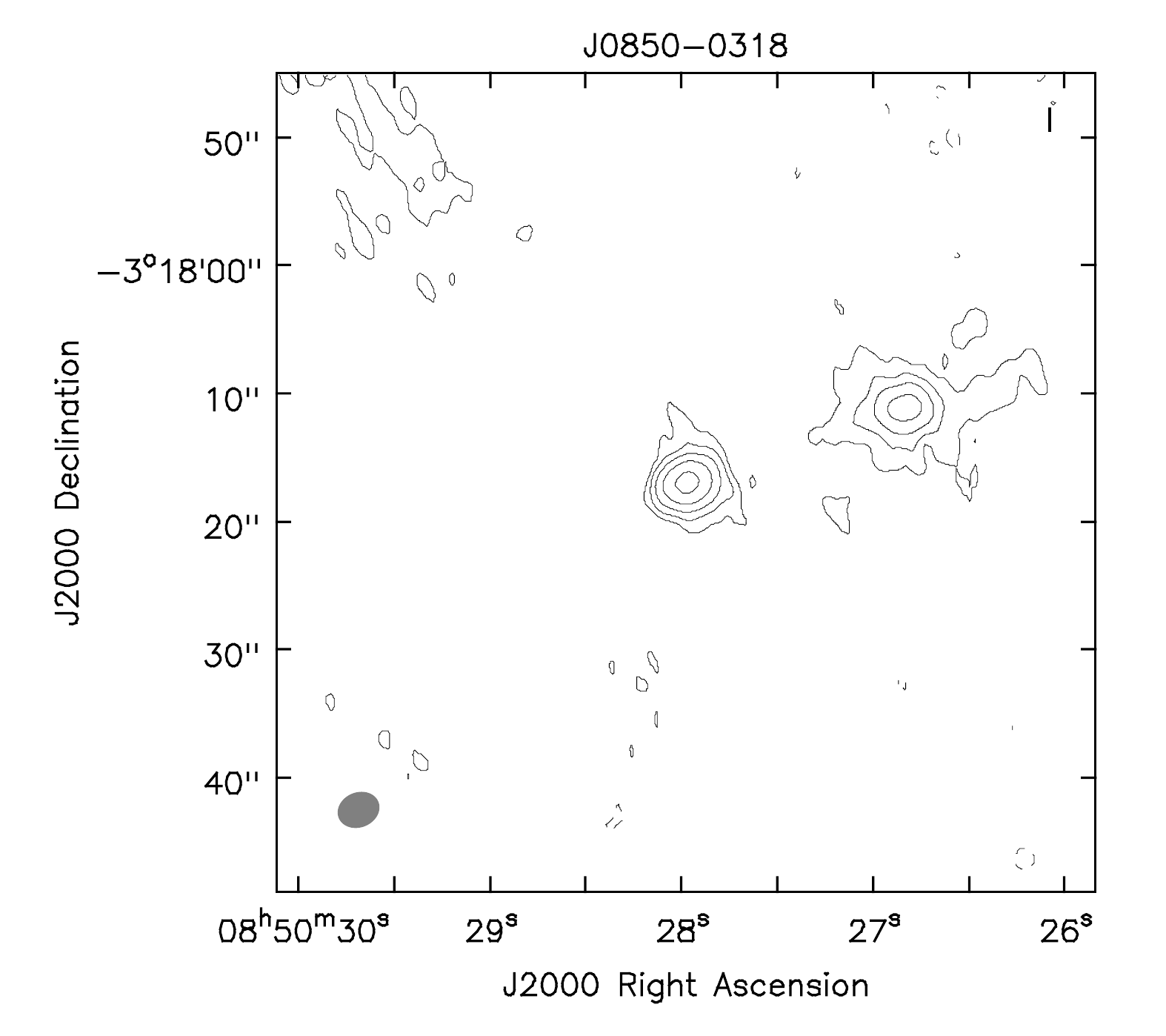}
\includegraphics[width=.41\textwidth, trim={1cm 13.5cm 4cm 2cm}, clip]{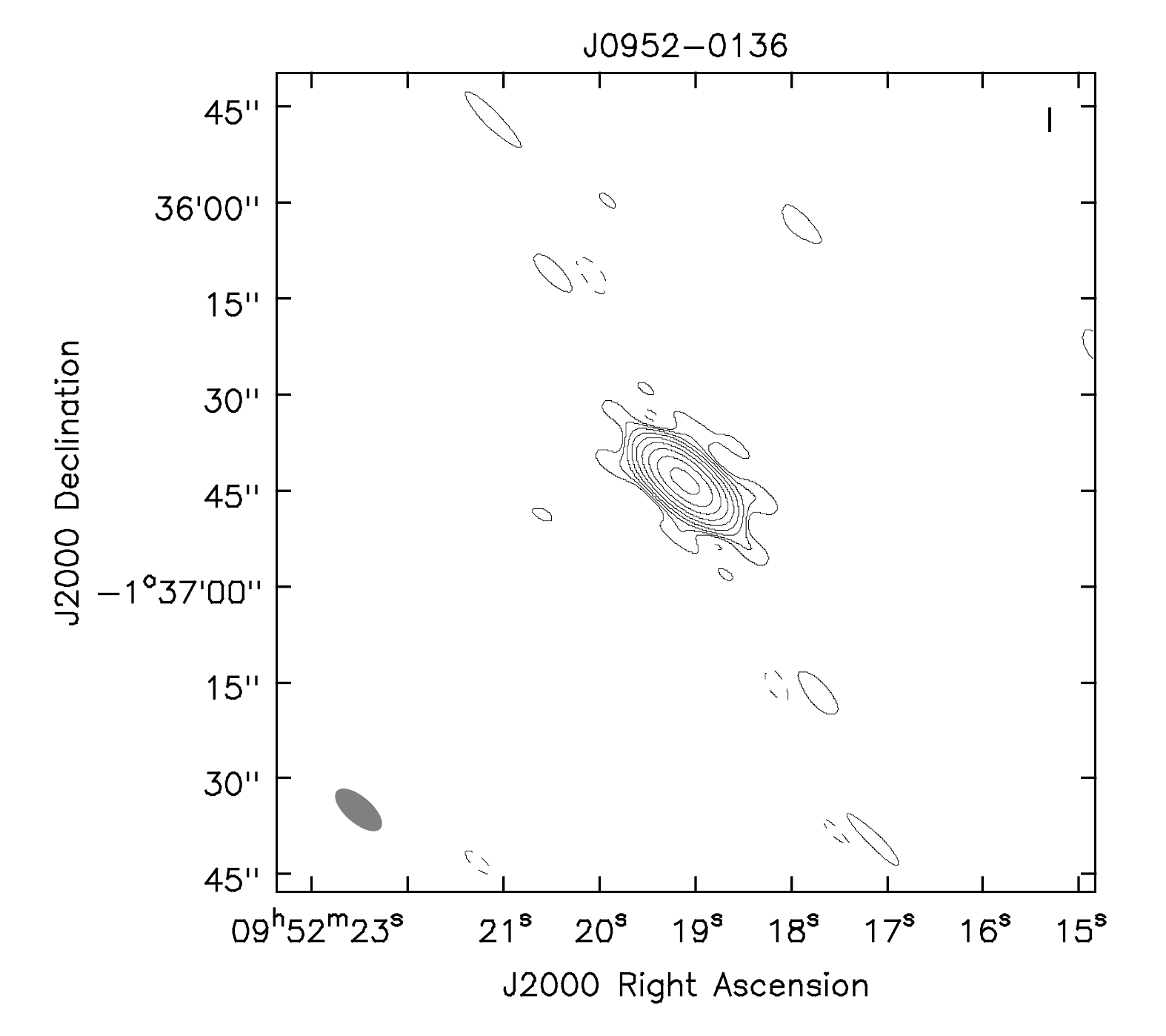}
\caption{\textit{left panel:} J0850$-$0318, rms = 6 $\mu$Jy beam$^{-1}$, contour levels at $-$3, 3 $\times$ 2$^n$, $n \in$ [0,4], beam size 12.25 $\times$ 9.91 kpc. \textit{right panel:} J0952$-$0136, rms = 20 $\mu$Jy beam$^{-1}$, contour levels at $-$3, 3 $\times$ 2$^n$, $n \in$ [0,8], beam size 3.63 $\times$ 1.63 kpc.}
\label{o}
\end{figure*}

\begin{figure*}
\centering
\includegraphics[width=.41\textwidth, trim={1cm 13.5cm 4cm 2cm}, clip]{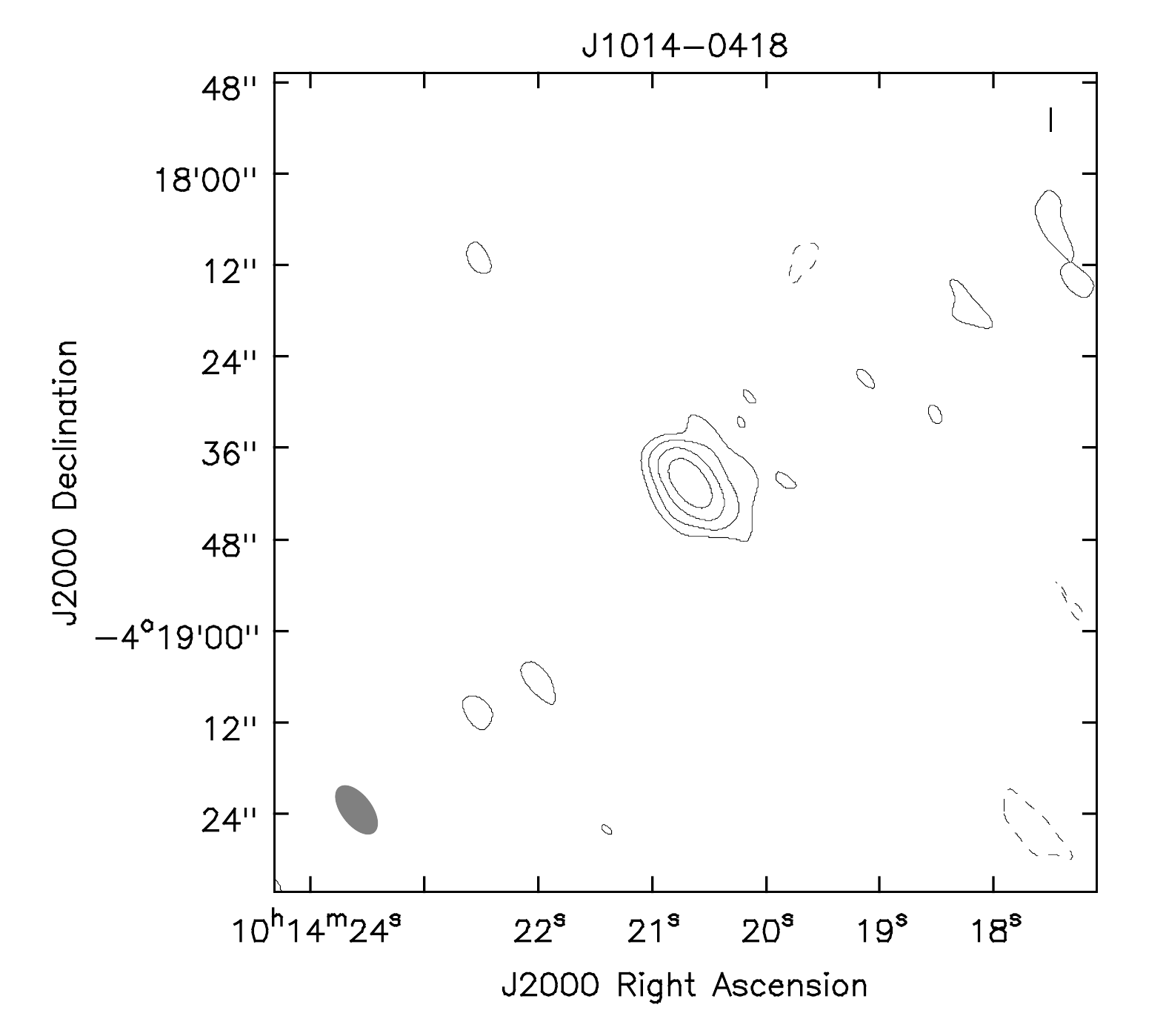}
\includegraphics[width=.41\textwidth, trim={1cm 13.5cm 4cm 2cm}, clip]{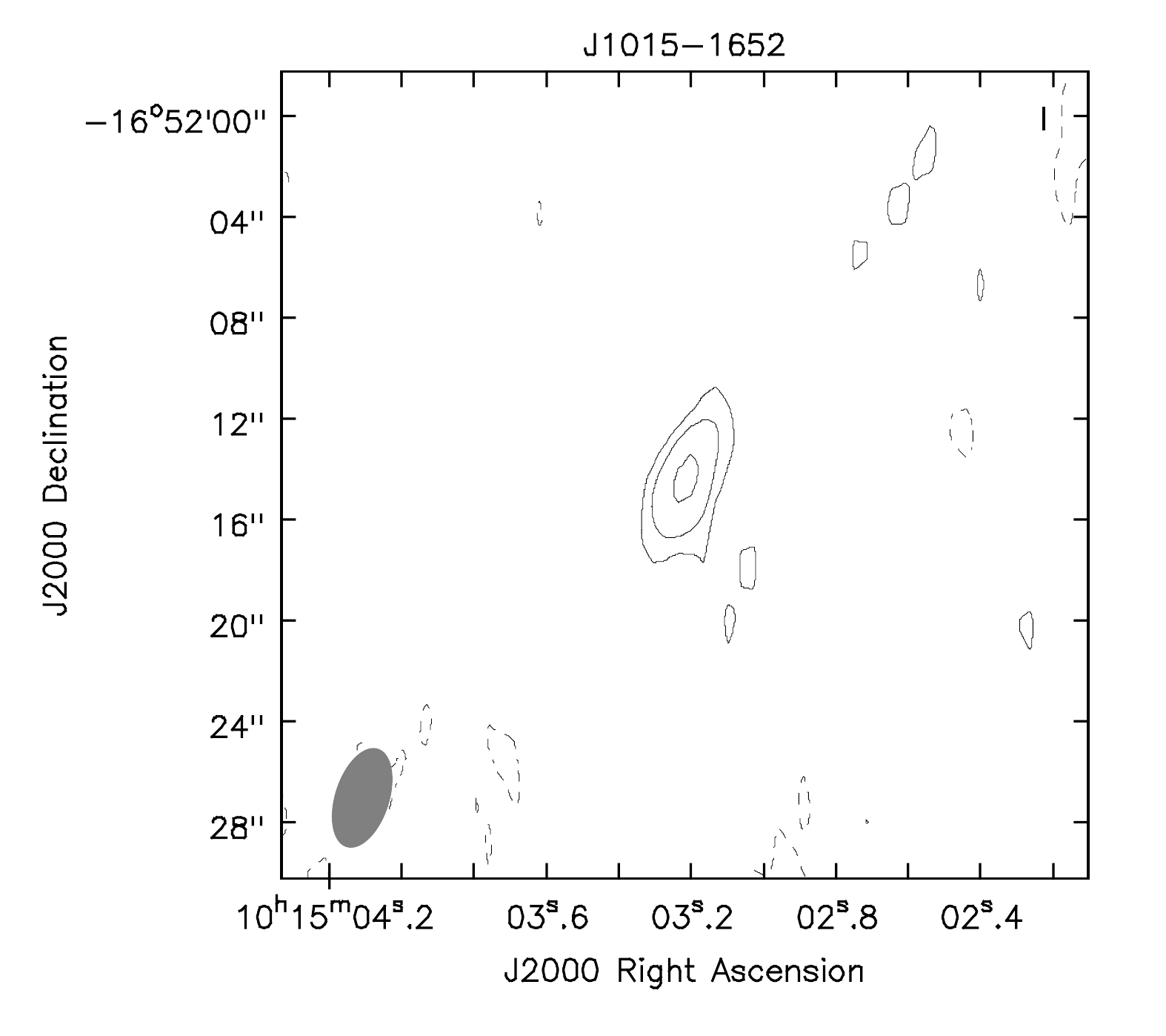}
\caption{\textit{left panel:} J1014$-$0418, rms = 6 $\mu$Jy beam$^{-1}$, contour levels at $-$3, 3 $\times$ 2$^n$, $n \in$ [0,3], beam size 9.34 $\times$ 4.91 kpc. \textit{right panel:} J1015$-$1652, rms = 5 $\mu$Jy beam$^{-1}$, contour levels at $-$3, 3 $\times$ 2$^n$, $n \in$ [0,2], beam size 46.56 $\times$ 24.32 kpc.}
\label{p}
\end{figure*}

\begin{figure*}
\centering
\includegraphics[width=.41\textwidth, trim={1cm 13.5cm 4cm 2cm}, clip]{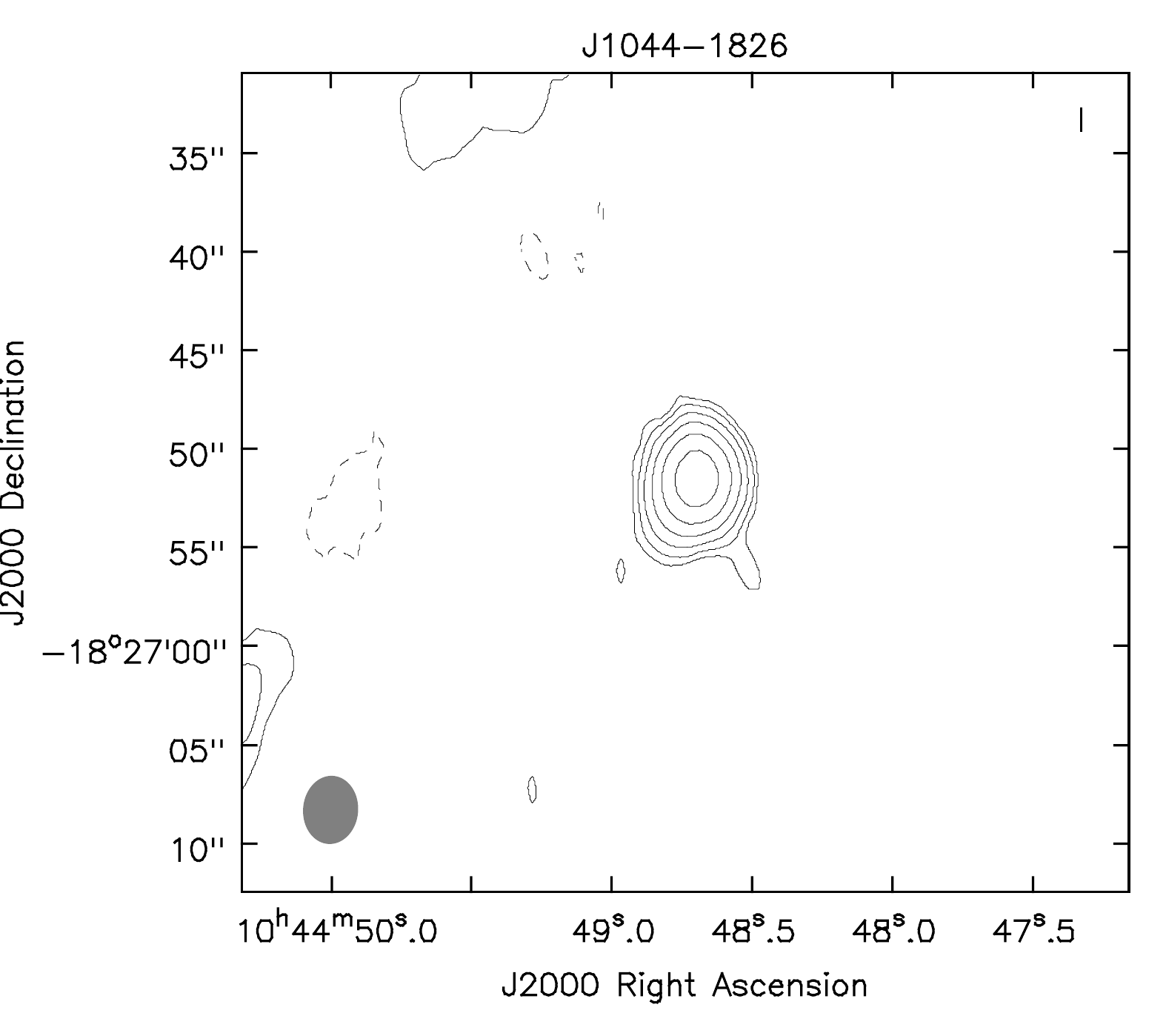}
\includegraphics[width=.41\textwidth, trim={1cm 13.5cm 4cm 2cm}, clip]{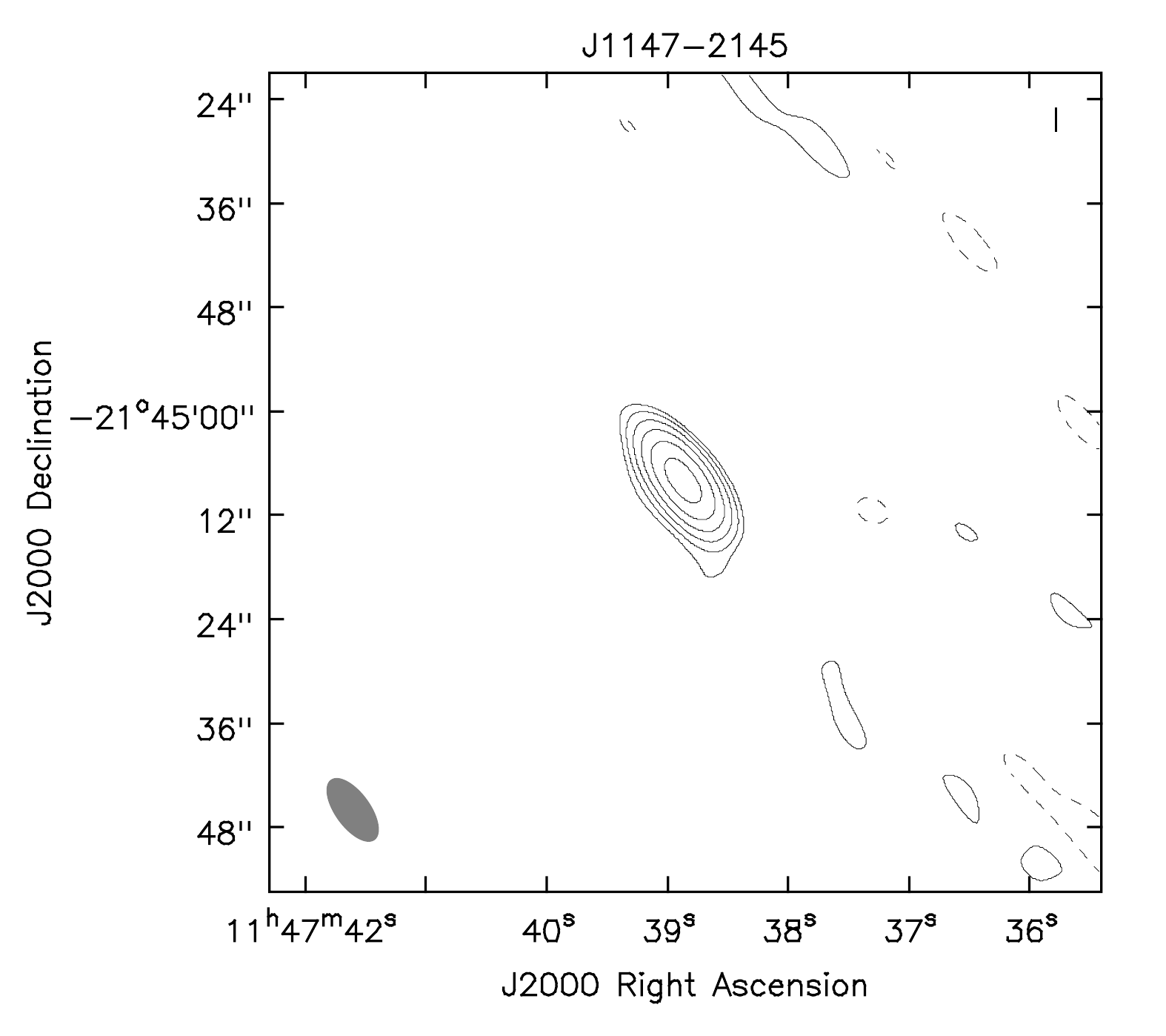}
\caption{\textit{left panel:} J1044$-$1826, rms = 10 $\mu$Jy beam$^{-1}$, contour levels at $-$3, 3 $\times$ 2$^n$, $n \in$ [0,5], beam size 8.66 $\times$ 6.96 kpc. \textit{right panel:} J1147$-$2145, rms = 15 $\mu$Jy beam$^{-1}$, contour levels at $-$3, 3 $\times$ 2$^n$, $n \in$ [0,5], beam size 44.47 $\times$ 20.74 kpc.}
\label{q}
\end{figure*}

\begin{figure*}
\centering
\includegraphics[width=.41\textwidth, trim={1cm 13.5cm 4cm 2cm}, clip]{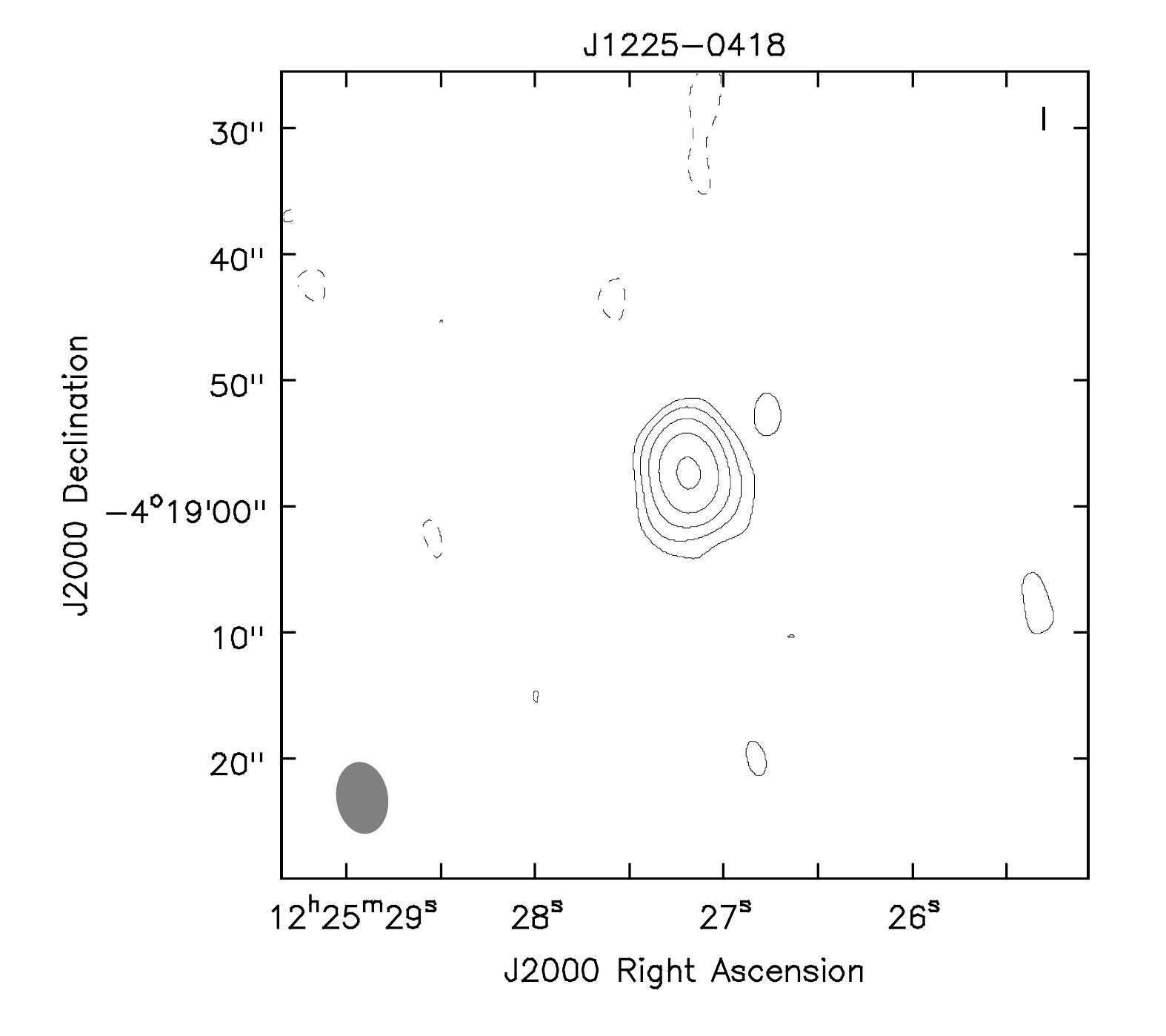}
\includegraphics[width=.41\textwidth, trim={1cm 13.5cm 4cm 2cm}, clip]{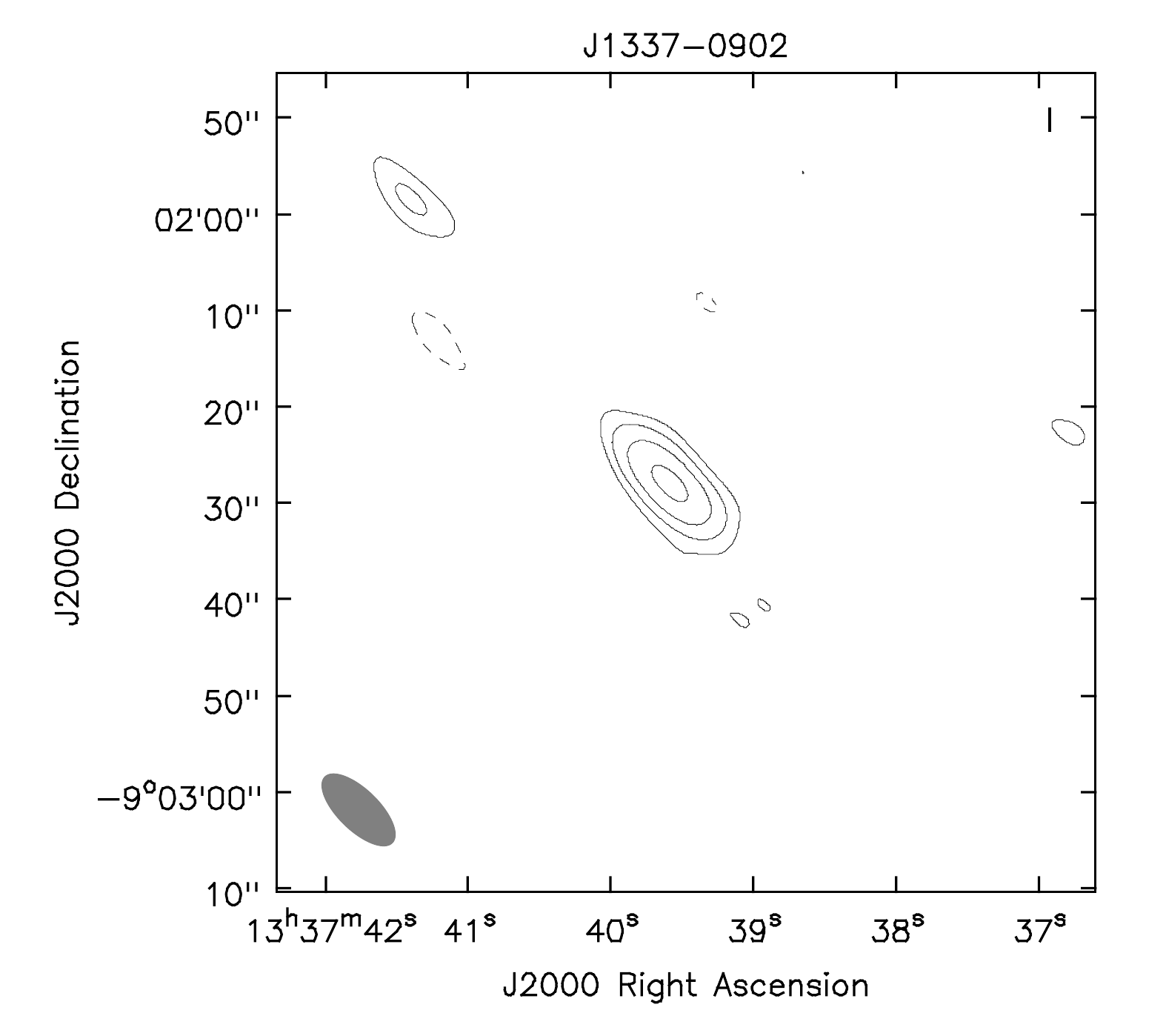}
\caption{\textit{left panel:} J1225$-$0418, rms = 5 $\mu$Jy beam$^{-1}$, contour levels at $-$3, 3 $\times$ 2$^n$, $n \in$ [0,4], beam size 17.65 $\times$ 12.54 kpc. \textit{right panel:} J1337$-$0902, rms = 6 $\mu$Jy beam$^{-1}$, contour levels at $-$3, 3 $\times$ 2$^n$, $n \in$ [0,3], beam size 17.23 $\times$ 7.44 kpc.}
\label{r}
\end{figure*}

\begin{figure*}
\centering
\includegraphics[width=.41\textwidth, trim={1cm 13.5cm 4cm 2cm}, clip]{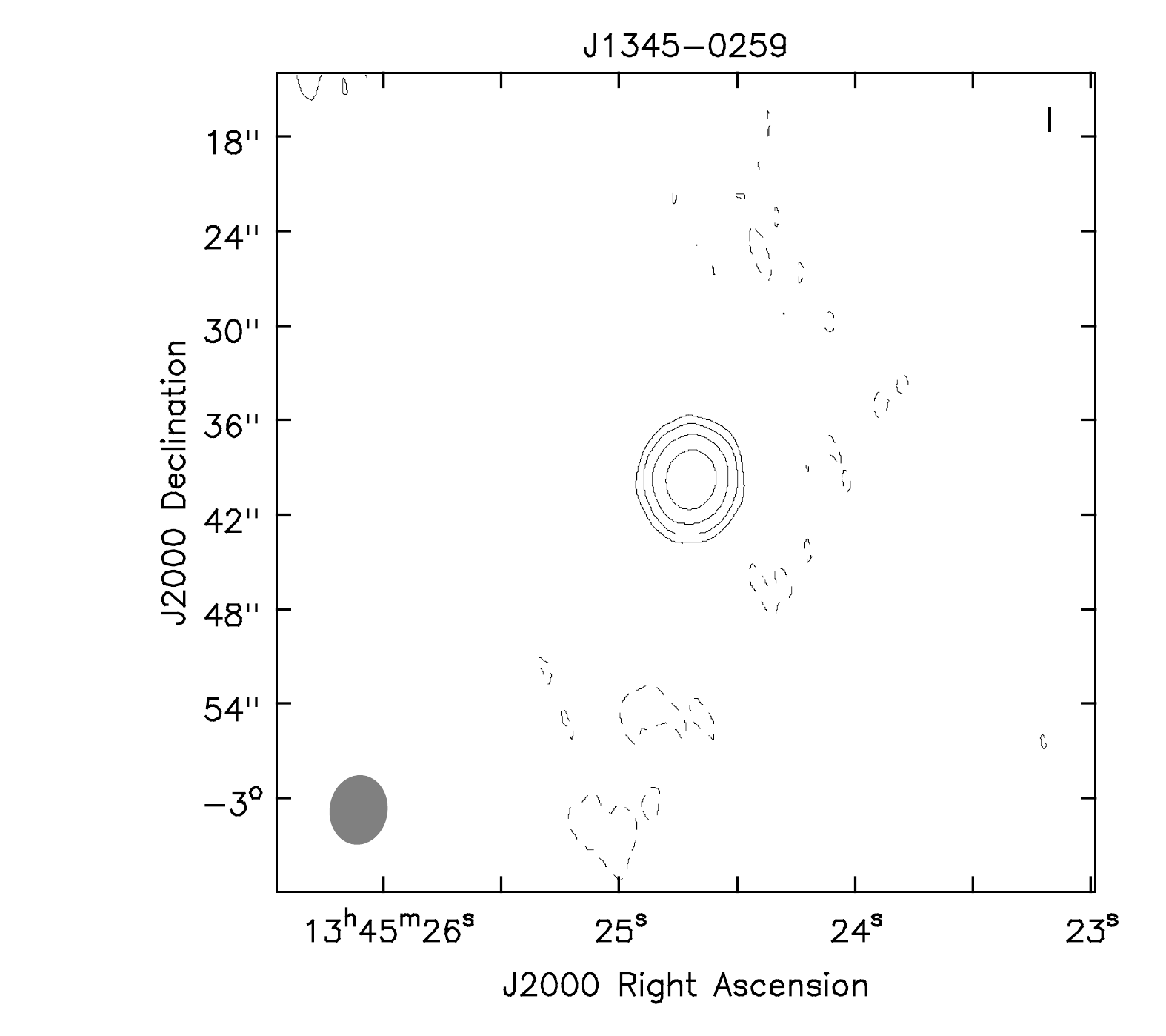}
\includegraphics[width=.41\textwidth, trim={1cm 13.5cm 4cm 2cm}, clip]{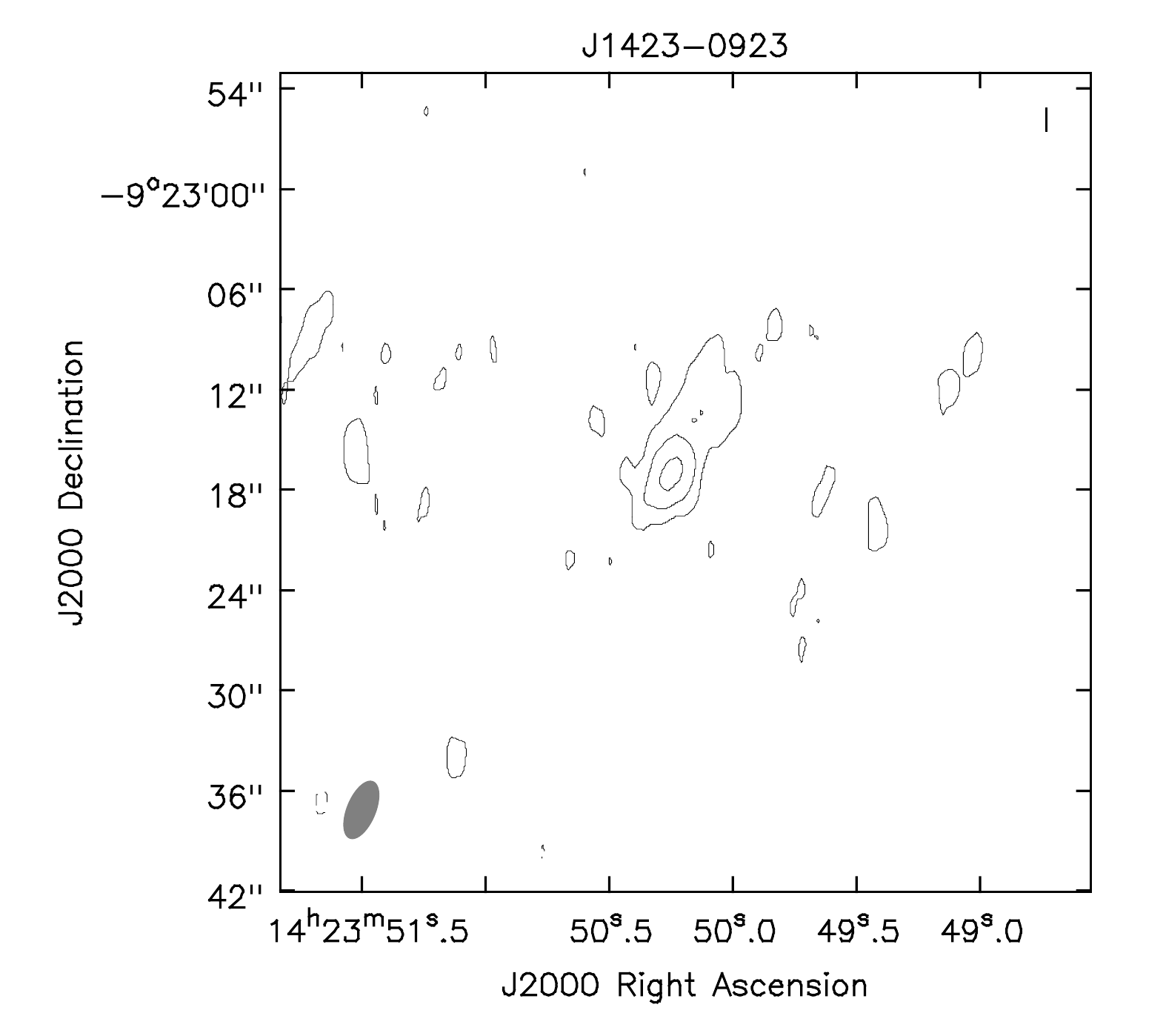}
\caption{\textit{left panel:} J1345$-$0259, rms = 9 $\mu$Jy beam$^{-1}$, contour levels at $-$3, 3 $\times$ 2$^n$, $n \in$ [0,3], beam size 8.20 $\times$ 6.78 kpc. \textit{right panel:} J1423$-$0923, rms = 6 $\mu$Jy beam$^{-1}$, contour levels at $-$3, 3 $\times$ 2$^n$, $n \in$ [0,2], beam size 5.42 $\times$ 2.50 kpc.}
\label{s}
\end{figure*}

\begin{figure*}
\centering
\includegraphics[width=.41\textwidth, trim={1cm 13.5cm 4cm 2cm}, clip]{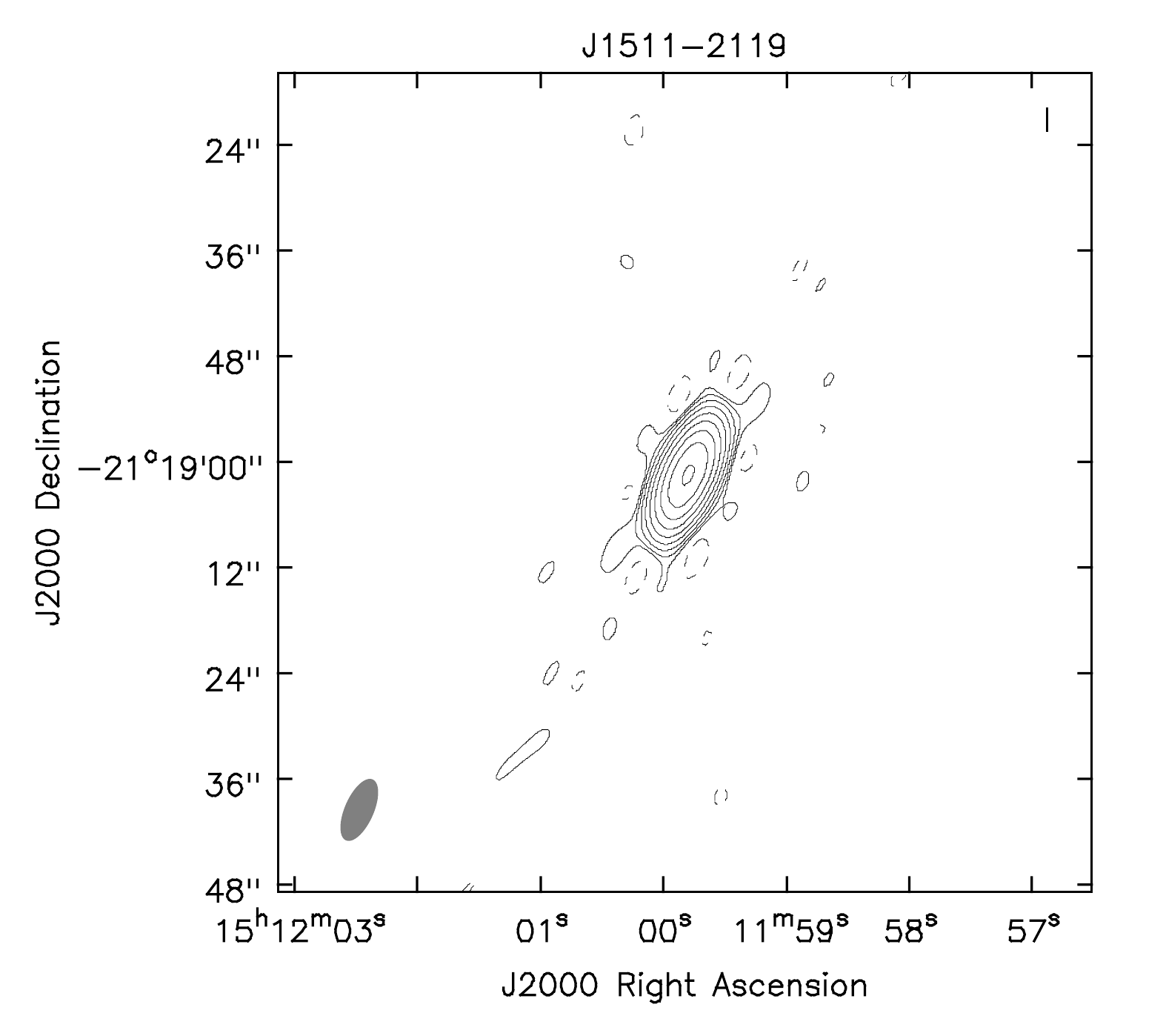}
\includegraphics[width=.41\textwidth, trim={1cm 13.5cm 4cm 2cm}, clip]{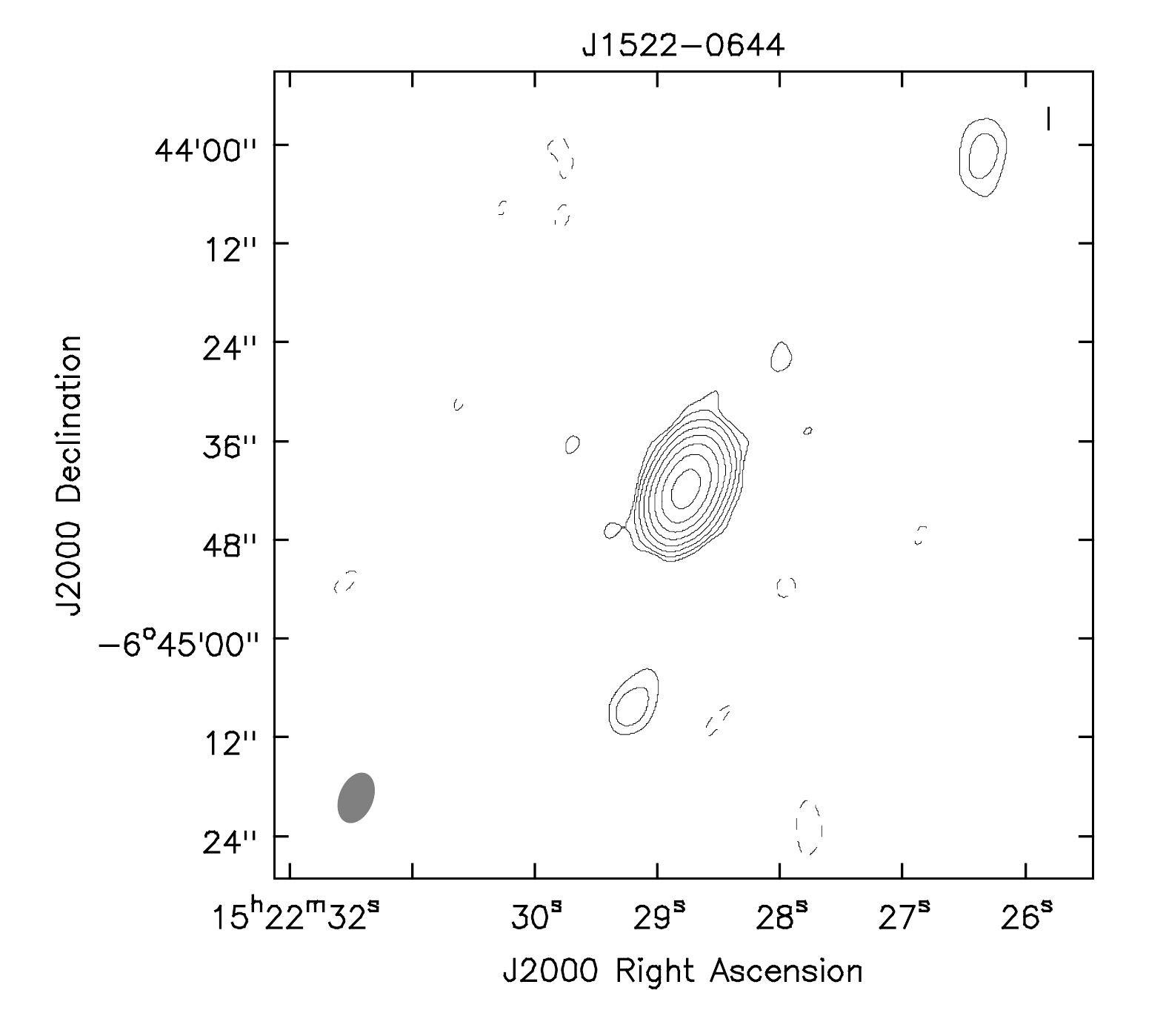}
\caption{\textit{left panel:} J1511$-$2119, rms = 21 $\mu$Jy beam$^{-1}$, contour levels at $-$3, 3 $\times$ 2$^n$, $n \in$ [0,8], beam size 7.07 $\times$ 3.02 kpc. \textit{right panel:} J1522$-$0644, rms = 8 $\mu$Jy beam$^{-1}$, contour levels at $-$3, 3 $\times$ 2$^n$, $n \in$ [0,7], beam size 11.46 $\times$ 7.32 kpc.}
\label{t}
\end{figure*}

\begin{figure*}
\centering
\includegraphics[width=.41\textwidth, trim={1cm 13.5cm 4cm 2cm}, clip]{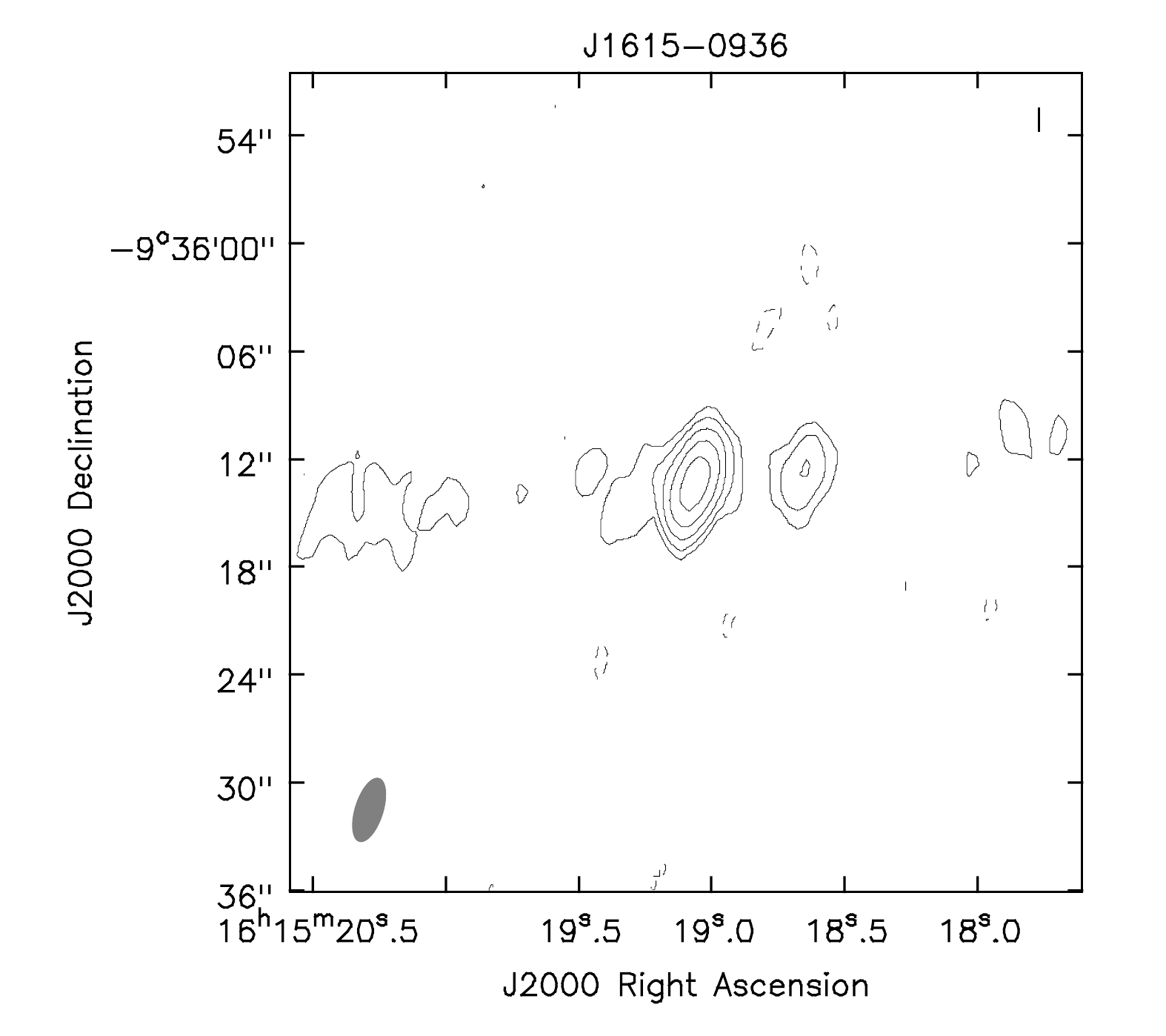}
\includegraphics[width=.41\textwidth, trim={1cm 13.5cm 4cm 2cm}, clip]{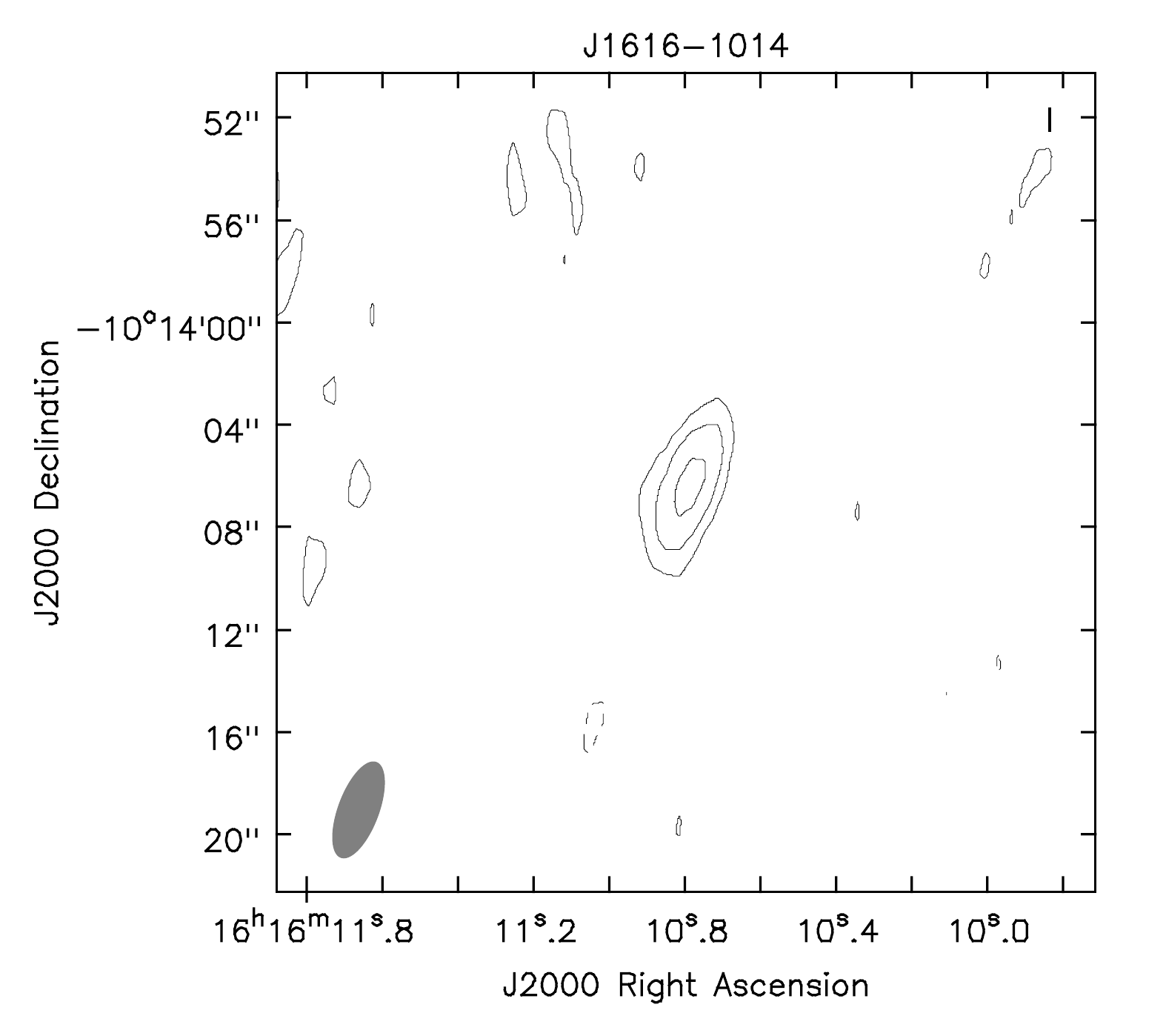}
\caption{\textit{left panel:} J1615$-$0936, rms = 7 $\mu$Jy beam$^{-1}$, contour levels at $-$3, 3 $\times$ 2$^n$, $n \in$ [0,4], beam size 5.17 $\times$ 2.17 kpc. \textit{right panel:} J1616$-$1014, rms = 5 $\mu$Jy beam$^{-1}$, contour levels at $-$3, 3 $\times$ 2$^n$, $n \in$ [0,2], beam size 6.73 $\times$ 2.66 kpc.}
\label{u}
\end{figure*}

\begin{figure*}
\centering
\includegraphics[width=.41\textwidth, trim={1cm 13.5cm 4cm 2cm}, clip]{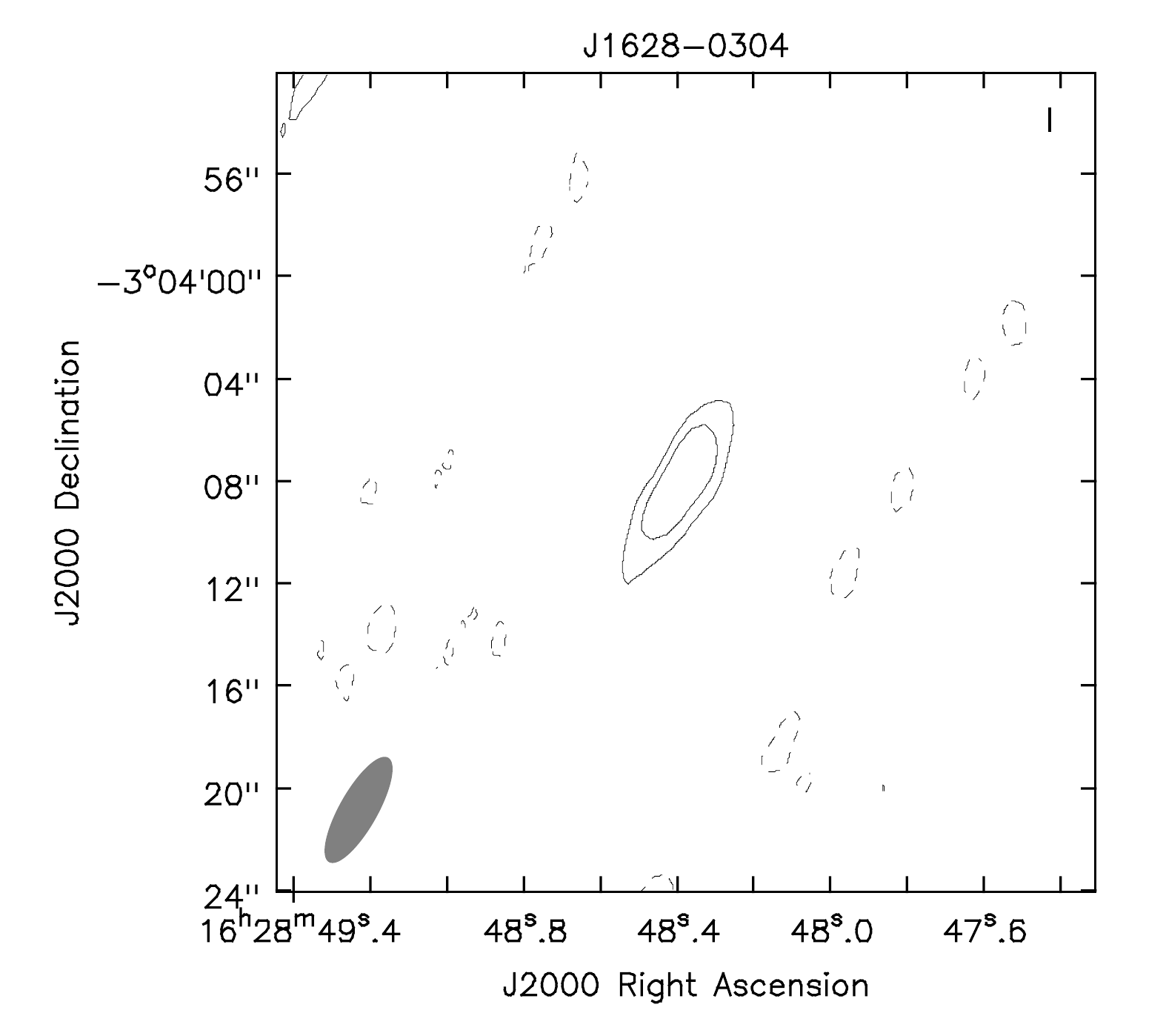}
\includegraphics[width=.41\textwidth, trim={1cm 13.5cm 4cm 2cm}, clip]{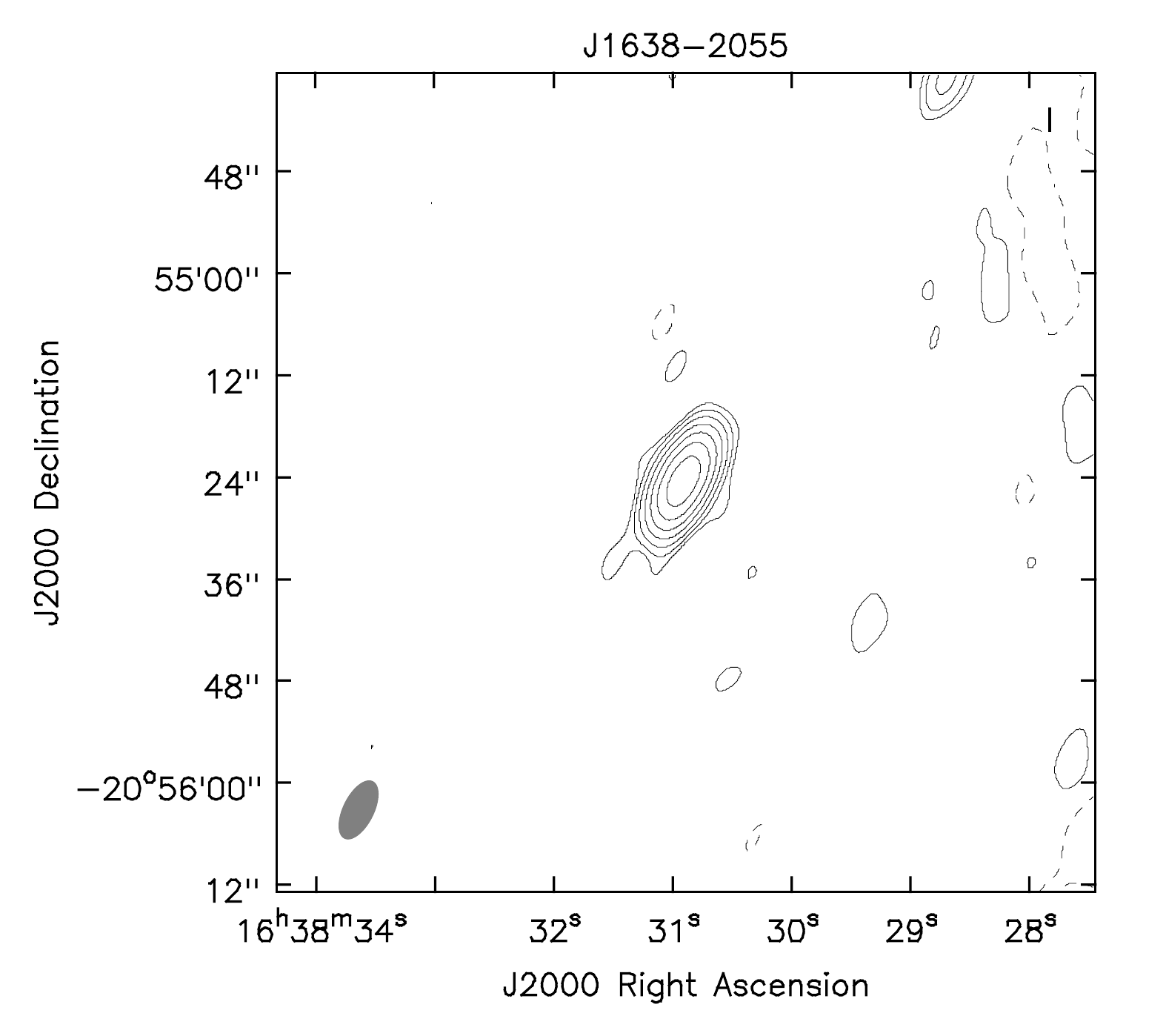}
\caption{\textit{left panel:} J1628$-$0304, rms = 6 $\mu$Jy beam$^{-1}$, contour levels at $-$3, 3 $\times$ 2$^n$, $n \in$ [0,1], beam size 9.58 $\times$ 2.99 kpc. \textit{right panel:} J1638$-$2055, rms = 11 $\mu$Jy beam$^{-1}$, contour levels at $-$3, 3 $\times$ 2$^n$, $n \in$ [0,6], beam size 4.22 $\times$ 2.01 kpc.}
\label{v}
\end{figure*}

\begin{figure*}
\centering
\includegraphics[width=.41\textwidth, trim={1cm 13.5cm 4cm 2cm}, clip]{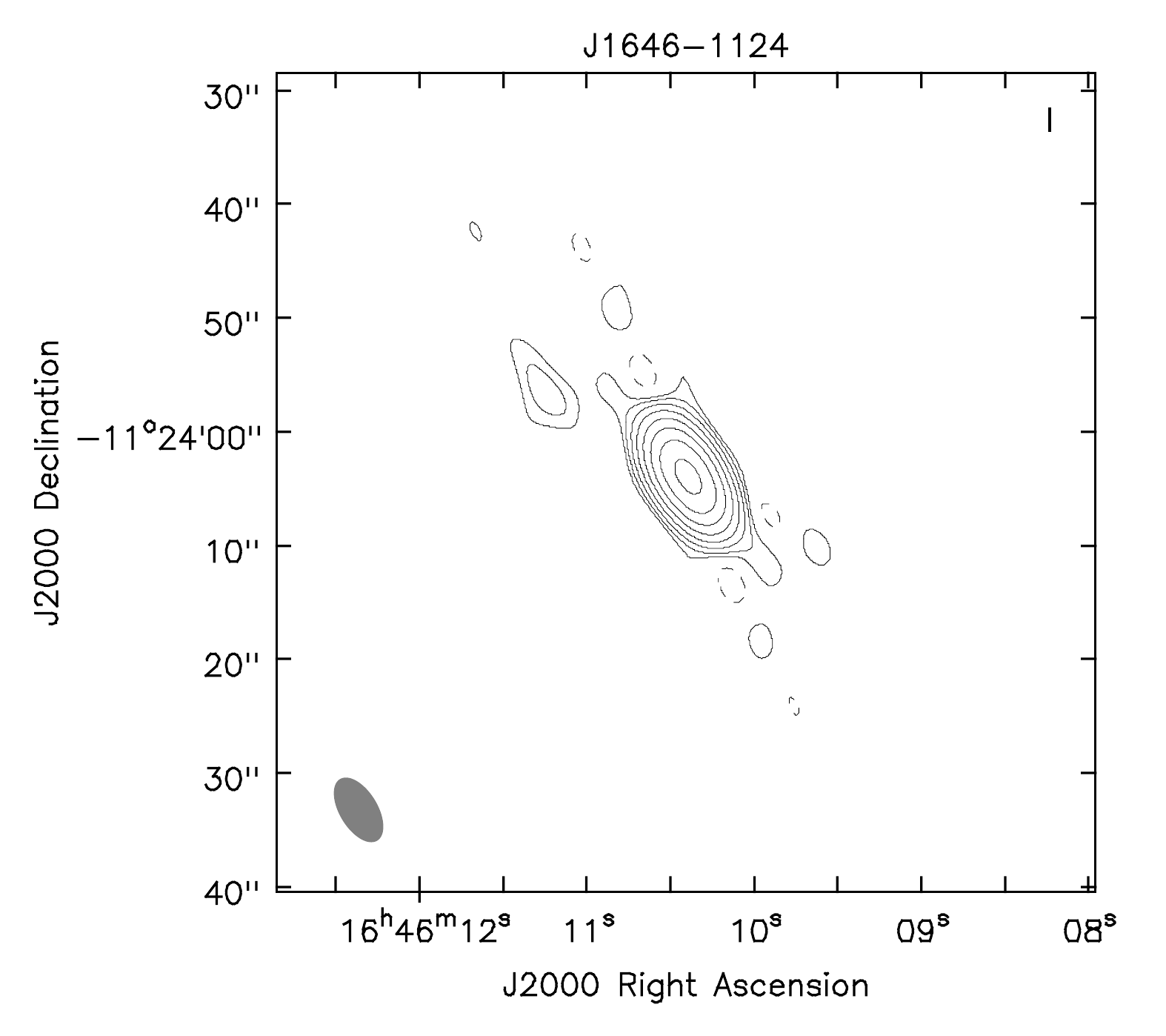}
\includegraphics[width=.41\textwidth, trim={1cm 13.5cm 4cm 2cm}, clip]{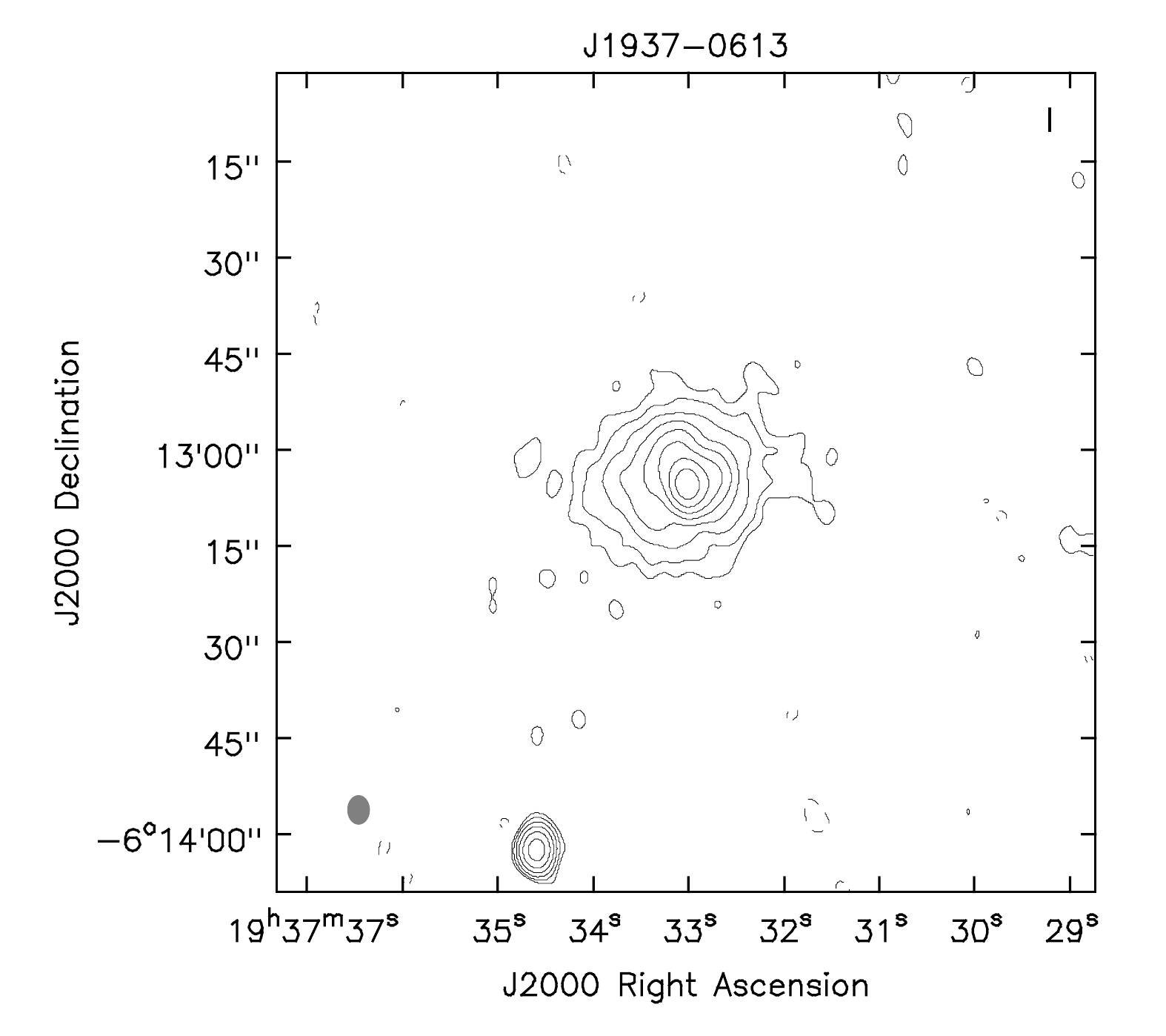}
\caption{\textit{left panel:} J1646$-$1124, rms = 22 $\mu$Jy beam$^{-1}$, contour levels at $-$3, 3 $\times$ 2$^n$, $n \in$ [0,7], beam size 10.09 $\times$ 5.26 kpc. \textit{right panel:} J1937$-$0613, rms = 9 $\mu$Jy beam$^{-1}$, contour levels at $-$3, 3 $\times$ 2$^n$, $n \in$ [0,7], beam size 0.95 $\times$ 0.72 kpc.}
\label{w}
\end{figure*}

\begin{figure*}
\centering
\includegraphics[width=.41\textwidth, trim={1cm 13.5cm 4cm 2cm}, clip]{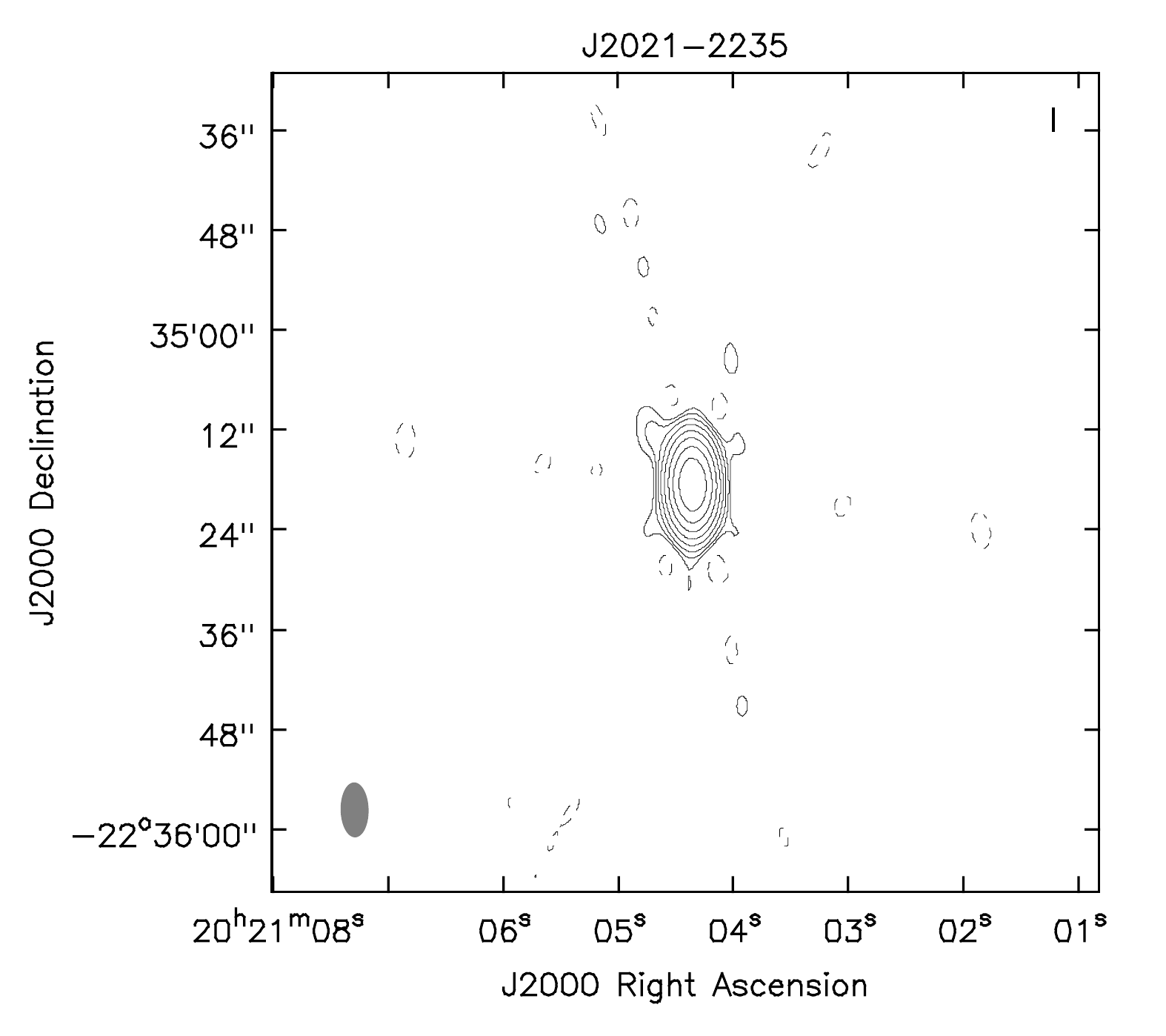}
\includegraphics[width=.41\textwidth, trim={1cm 13.5cm 4cm 2cm}, clip]{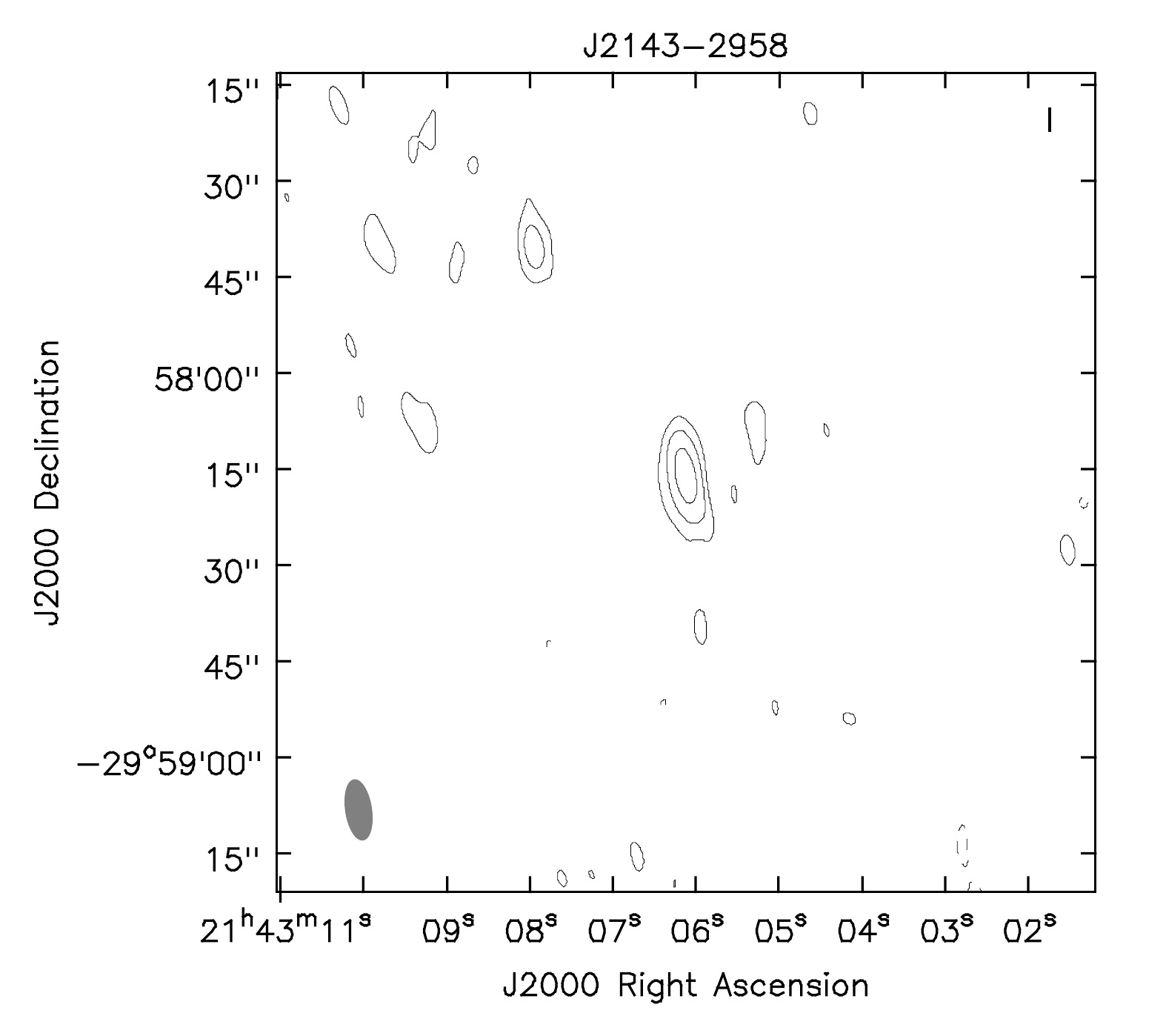}
\caption{\textit{left panel:} J2021$-$2235, rms = 12 $\mu$Jy beam$^{-1}$, contour levels at $-$3, 3 $\times$ 2$^n$, $n \in$ [0,7], beam size 28.12 $\times$ 14.02 kpc. \textit{right panel:} J2143$-$2958, rms = 8 $\mu$Jy beam$^{-1}$, contour levels at $-$3, 3 $\times$ 2$^n$, $n \in$ [0,2], beam size 25.77 $\times$ 10.94 kpc.}
\label{x}
\end{figure*}

We present the radio maps of the southern NLS1s with the VLA C-configuration observations at 5.5 GHz in Figs. \ref{a} - \ref{x}.

\section{Optical images overlaid with radio contours}
\label{overlay}
\counterwithin{figure}{section}

\begin{figure*}
\centering
\includegraphics[width=0.33\textwidth, trim={5cm 3cm 5cm 1cm}, clip]{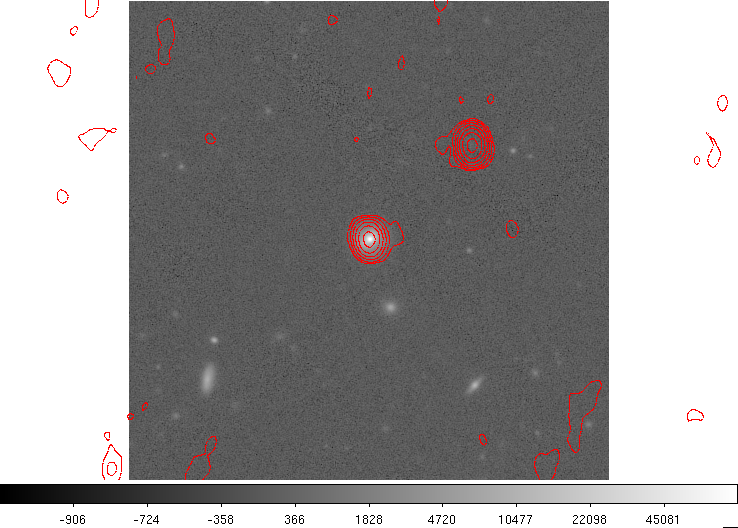}
\includegraphics[width=0.33\textwidth, trim={5cm 3cm 5cm 1cm}, clip]{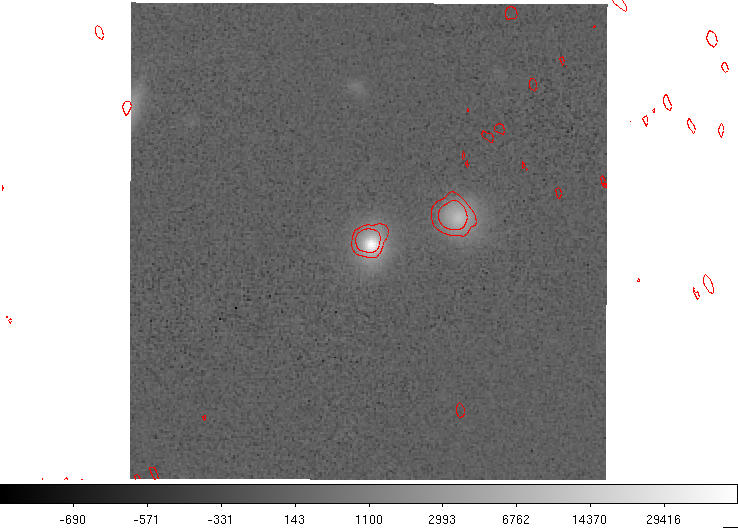}
\includegraphics[width=0.33\textwidth, trim={5cm 3cm 5cm 1cm}, clip]{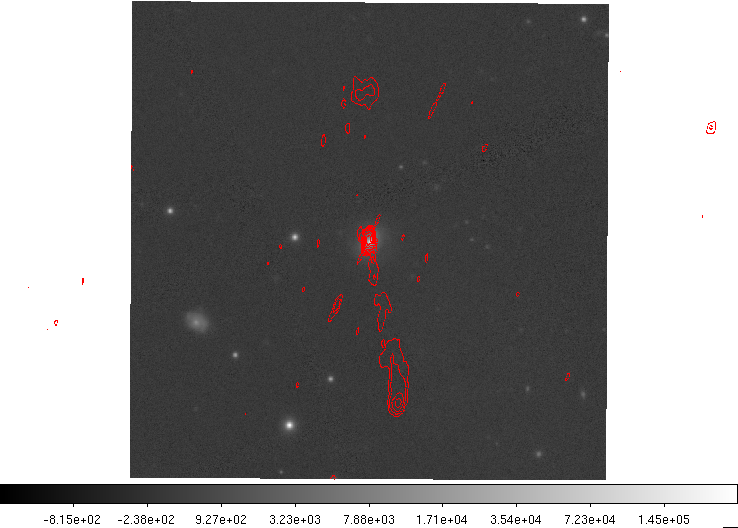}
\caption{\textit{Left:} J0000$-$0541, \textit{middle:} J0350$-$1025, \textit{right:} J0354$-$1340.}
\label{aa}
\end{figure*}

\begin{figure*}
\centering
\includegraphics[width=0.33\textwidth, trim={5cm 3cm 5cm 1cm}, clip]{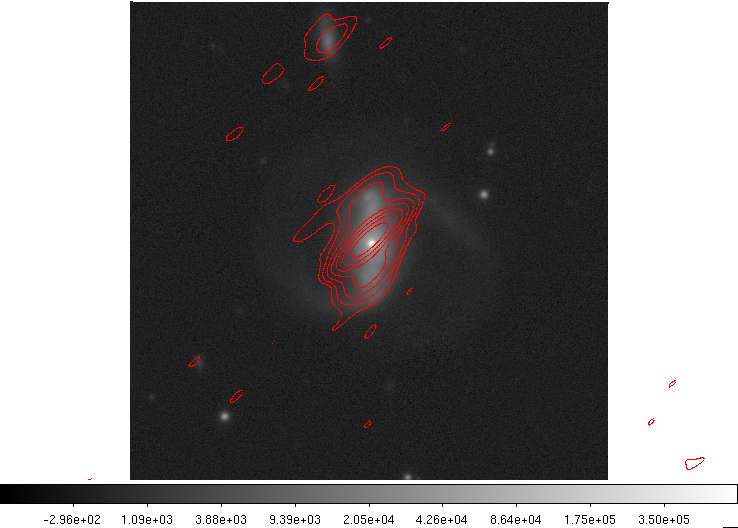}
\includegraphics[width=0.33\textwidth, trim={5cm 3cm 5cm 1cm}, clip]{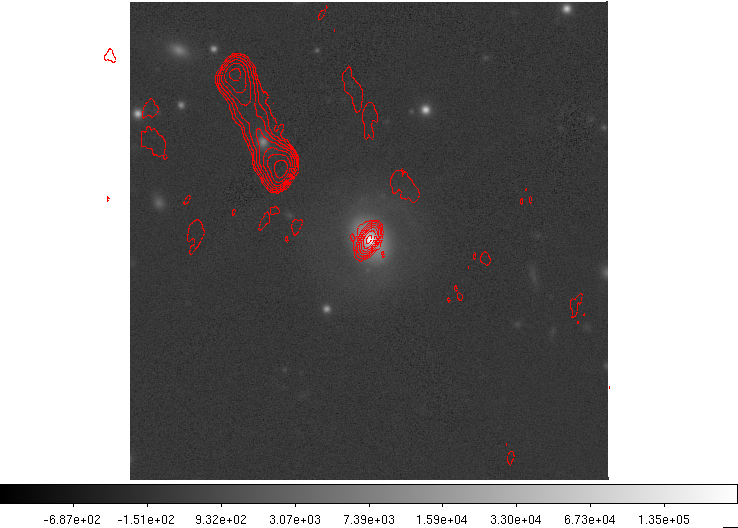}
\includegraphics[width=0.33\textwidth, trim={5cm 3cm 5cm 1cm}, clip]{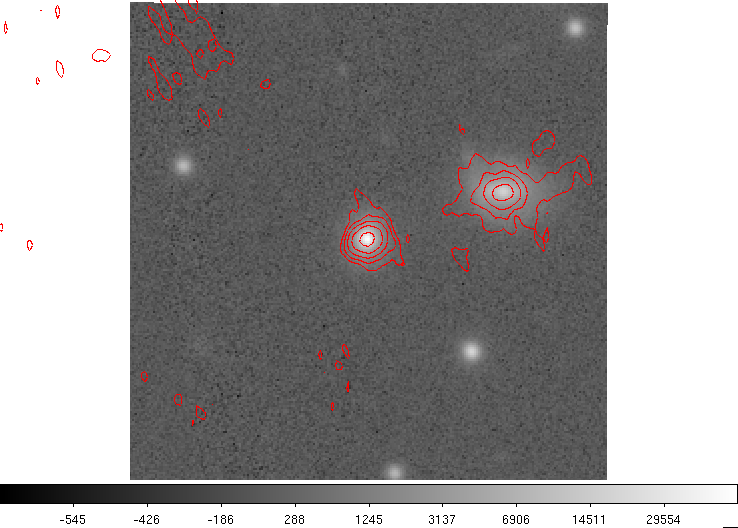}
\caption{\textit{Left:} J0436$-$1022, \textit{middle:} J0447$-$0508, \textit{right:} J0850$-$0318.}
\label{bb}
\end{figure*}

\begin{figure*}
\centering
\includegraphics[width=0.33\textwidth, trim={5cm 3cm 5cm 1cm}, clip]{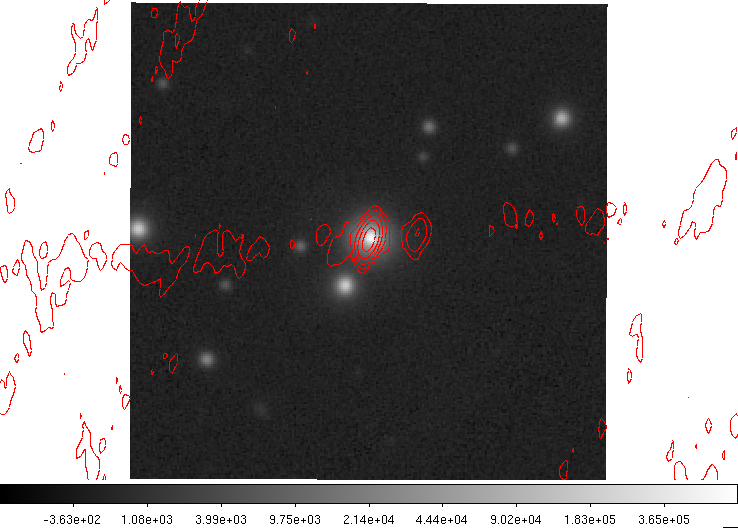}
\includegraphics[width=0.33\textwidth, trim={5cm 3cm 5cm 1cm}, clip]{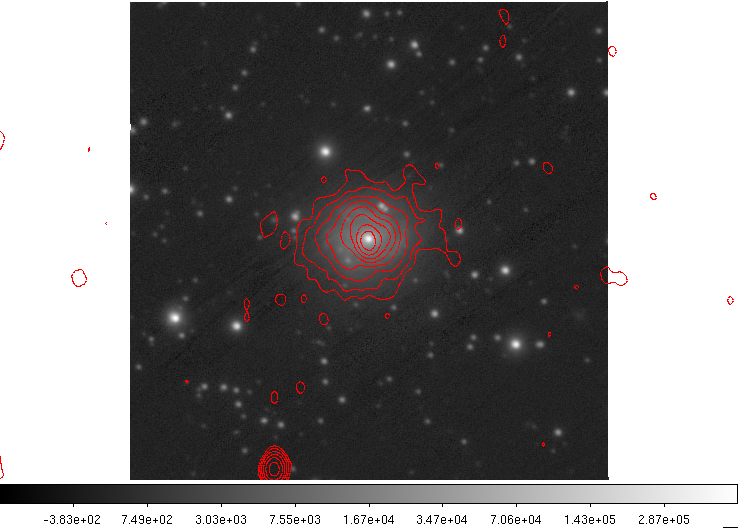}
\includegraphics[width=0.33\textwidth, trim={5cm 3cm 5cm 1cm}, clip]{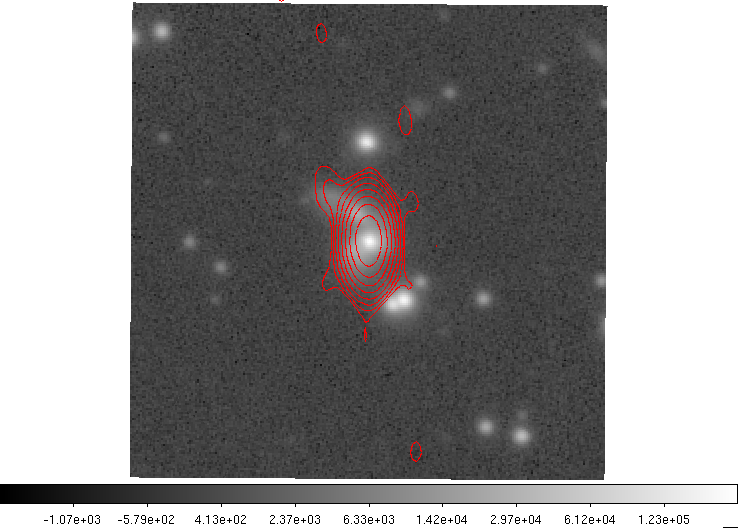}
\caption{\textit{Left:} J1615$-$0936, \textit{middle:} J1937$-$0613, \textit{right:} J2021$-$2235.}
\label{cc}
\end{figure*}

We present the optical images from the Pan-STARRS overlaid with the radio contours for nine NLS1s in Figs. \ref{aa} - \ref{cc}. The radio contours are the same as Figs. \ref{J0354} and \ref{radiomap}.

\section{Tables}
\counterwithin{table}{section}

We present the radio measurements of the southern NLS1s observed with the VLA C-configuration at 5.5 GHz. The coordinate, redshift, scale, core size, position angle, and integrated and peak flux densities and luminosities are reported in Tab. \ref{radio_measurement}. The radio loudness, flux concentration, in-band spectral index between 5 and 6 GHz, different classifications, flux density in optical B-band, and black hole mass are reported in Tab. \ref{classify+optical}. We additionally present the radio measurements of 21 NLS1s detected with the 1.4 GHz surveys, including flux density, luminosity, radio loudness, spectral index between 1.4 and 5.5 GHz, and different types, as listed in Tab. \ref{first+nvss}.

\begin{landscape}
\begin{table}
\centering
\caption{The radio measurements at 5.5 GHz of the southern NLS1 sample}
\begin{scriptsize}
\begin{tabular}{cccccccccccc}
\hline
\hline
Name & R.A. & Dec. & $z$ & Scale & Core maj & Core min & Core PA & $S_{\rm{int}}$ & $S_{\rm{p}}$ & $\log L_{\rm{int}}$ & $\log L_{\rm{p}}$ \\
- & - & - & - & (kpc \, arcsec$^{-1}$) & (kpc) & (kpc) & (deg) & ($\mu$Jy) & ($\mu$Jy \, beam$^{-1}$) & (erg \, s$^{-1}$) & (erg \, s$^{-1}$) \\
\hline
J0000$-$0541 & 00:00:40.25 & $-$05:41:01.05 & 0.094 & 2.09 & 3.71 & 1.29 & 85 & 1164.0 $\pm$ 16.0 & 1058.5 $\pm$ 8.5 & 39.16 $\pm$ 0.01 & 39.12 $\pm$ 0.01 \\
J0022$-$1039 & 00:22:49.22 & $-$10:39:56.51 & 0.414 & 10.97 & 22.94 & 12.62 & 41 & 514.0 $\pm$ 27.0 & 425.0 $\pm$ 14.0 & 40.25 $\pm$ 0.06 & 40.17 $\pm$ 0.06 \\
J0122$-$2646 & 01:22:37.54 & $-$26:46:46.30 & 0.417 & 11.08 & 50.40 & 7.86 & 136 & 941.0 $\pm$ 102.0 & 825.0 $\pm$ 44.0 & 40.52 $\pm$ 0.09 & 40.47 $\pm$ 0.08 \\
J0138$-$0109 & 01:38:09.54 & $-$01:09:20.20 & 0.273 & 6.76 & < 19.60 & < 8.11 & - & 235.0 $\pm$ 14.0 & 214.4 $\pm$ 7.6 & 39.38 $\pm$ 0.06 & 39.34 $\pm$ 0.05 \\
J0159$-$1128 & 01:59:30.69 & $-$11:28:58.44 & 0.161 & 3.73 & 8.06 & 3.02 & 140 & 122.4 $\pm$ 9.0 & 96.7 $\pm$ 4.3 & 38.57 $\pm$ 0.05 & 38.47 $\pm$ 0.04 \\
J0203$-$1247 & 02:03:49.03 & $-$12:47:16.66 & 0.053 & 1.13 & < 1.13 & < 0.67 & - & 360.0 $\pm$ 10.0 & 314.9 $\pm$ 4.9 & 38.11 $\pm$ 0.01 & 38.05 $\pm$ 0.01 \\
J0212$-$0201 & 02:12:01.46 & $-$02:01:53.75 & 0.438 & 11.71 & 13.94 & 6.21 & 141 & 229.4 $\pm$ 8.8 & 223.3 $\pm$ 4.8 & 39.84 $\pm$ 0.05 & 39.83 $\pm$ 0.05 \\
J0213$-$0551 & 02:13:55.14 & $-$05:51:21.44 & 0.140 & 3.20 & 5.51 & 4.26 & 123 & 202.0 $\pm$ 9.6 & 166.5 $\pm$ 4.8 & 38.74 $\pm$ 0.03 & 38.66 $\pm$ 0.03 \\
J0230$-$0859 & 02:30:05.53 & $-$08:59:53.20 & 0.016 & 0.34 & 0.60 & 0.44 & 115 & 1232.0 $\pm$ 25.0 & 1029.0 $\pm$ 13.0 & 37.58 $\pm$ 0.01 & 37.50 $\pm$ 0.01 \\
J0239$-$1118 & 02:39:56.16 & $-$11:18:12.91 & 0.203 & 4.84 & < 9.57 & < 6.22 & - & < 327.7 $\pm$ 5.1 & 327.7 $\pm$ 5.1 & < 39.31 $\pm$ 0.02 & 39.31 $\pm$ 0.02 \\
J0347$-$1325 & 03:47:14.02 & $-$13:25:45.51 & 0.192 & 4.53 & < 8.91 & < 6.55 & - & < 625.2 $\pm$ 6.0 & 625.2 $\pm$ 6.0 & < 39.54 $\pm$ 0.01 & 39.54 $\pm$ 0.01 \\
J0350$-$1025 & 03:50:56.74 & $-$10:25:58.69 & 0.128 & 2.92 & < 5.42 & < 4.86 & - & < 60.9 $\pm$ 1.9 & 60.9 $\pm$ 1.9 & < 38.27 $\pm$ 0.03 & 38.27 $\pm$ 0.03 \\
J0351$-$0526 & 03:51:07.58 & $-$05:26:37.58 & 0.068 & 1.48 & 3.25 & 1.92 & 77 & 141.0 $\pm$ 12.0 & 107.4 $\pm$ 5.5 & 37.91 $\pm$ 0.04 & 37.79 $\pm$ 0.03 \\
J0354$-$1340 & 03:54:32.85 & $-$13:40:07.29 & 0.076 & 1.68 & < 3.69 & < 1.74 & - & < 5090.0 $\pm$ 25.0 & 5090.0 $\pm$ 25.0 & < 39.58 $\pm$ 0.01 & 39.58 $\pm$ 0.01 \\
J0400$-$2500 & 04:00:24.42 & $-$25:00:44.48 & 0.097 & 2.17 & < 8.52 & < 6.73 & - & < 1245.0 $\pm$ 14.0 & 1245.0 $\pm$ 14.0 & < 39.14 $\pm$ 0.01 & 39.14 $\pm$ 0.01 \\
J0413$-$0050 & 04:13:07.09 & $-$00:50:16.68 & 0.040 & 0.86 & 0.95 & 0.91 & 118 & 161.0 $\pm$ 11.0 & 140.7 $\pm$ 5.9 & 37.51 $\pm$ 0.03 & 37.45 $\pm$ 0.02 \\
J0422$-$1854 & 04:22:56.58 & $-$18:54:41.39 & 0.064 & 1.40 & 1.65 & 0.32 & 85 & 1126.0 $\pm$ 21.0 & 1069.0 $\pm$ 11.0 & 38.78 $\pm$ 0.01 & 38.76 $\pm$ 0.01 \\
J0435$-$1643 & 04:35:26.50 & $-$16:43:45.37 & 0.098 & 2.19 & < 10.67 & < 4.51 & - & 141.1 $\pm$ 5.9 & 131.9 $\pm$ 2.8 & 38.18 $\pm$ 0.04 & 38.15 $\pm$ 0.03 \\
J0436$-$1022 & 04:36:22.32 & $-$10:22:33.23 & 0.035 & 0.75 & 2.50 & 1.15 & 146 & 4620.0 $\pm$ 220.0 & 4031.0 $\pm$ 96.0 & 38.86 $\pm$ 0.02 & 38.80 $\pm$ 0.01 \\
J0447$-$0403 & 04:47:38.89 & $-$04:03:30.49 & 0.045 & 0.96 & 4.03 & 2.73 & 139 & 118.0 $\pm$ 12.0 & 42.3 $\pm$ 3.4 & 37.48 $\pm$ 0.05 & 37.03 $\pm$ 0.04 \\
J0447$-$0508 & 04:47:20.72 & $-$05:08:13.99 & 0.082 & 1.80 & 1.86 & 1.46 & 149 & 4040.0 $\pm$ 50.0 & 3600.0 $\pm$ 25.0 & 39.56 $\pm$ 0.01 & 39.51 $\pm$ 0.01 \\
J0452$-$2953 & 04:52:30.13 & $-$29:53:35.59 & 0.286 & 7.12 & 20.50 & 11.75 & 116 & 3373.0 $\pm$ 47.0 & 2564.0 $\pm$ 22.0 & 40.67 $\pm$ 0.01 & 40.56 $\pm$ 0.01 \\
J0455$-$1456 & 04:55:57.52 & $-$14:56:41.69 & 0.137 & 3.13 & < 3.06 & < 0.30 & - & 138.8 $\pm$ 5.2 & 136.1 $\pm$ 2.9 & 38.54 $\pm$ 0.03 & 38.53 $\pm$ 0.02 \\
J0549$-$2425 & 05:49:14.89 & $-$24:25:51.50 & 0.045 & 0.97 & < 2.97 & < 1.21 & - & < 1560.0 $\pm$ 11.0 & 1560.0 $\pm$ 11.0 & < 38.60 $\pm$ 0.01 & 38.60 $\pm$ 0.01 \\
J0622$-$2317 & 06:22:33.48 & $-$23:17:42.72 & 0.038 & 0.81 & 8.47 & 7.91 & 133 & 926.0 $\pm$ 50.0 & 281.0 $\pm$ 12.0 & 38.22 $\pm$ 0.02 & 37.70 $\pm$ 0.02 \\
J0842$-$0349 & 08:42:19.10 & $-$03:49:30.38 & 0.357 & 9.20 & 21.44 & 11.41 & 173 & 324.0 $\pm$ 18.0 & 269.4 $\pm$ 9.1 & 39.95 $\pm$ 0.08 & 39.87 $\pm$ 0.08 \\
J0845$-$0732 & 08:45:10.23 & $-$07:32:04.90 & 0.104 & 2.32 & 9.31 & 5.91 & 136 & 259.0 $\pm$ 26.0 & 127.7 $\pm$ 9.0 & 38.62 $\pm$ 0.06 & 38.31 $\pm$ 0.05 \\
J0846$-$1214 & 08:46:28.51 & $-$12:14:10.50 & 0.108 & 2.42 & < 1.43 & < 0.58 & - & 5310.0 $\pm$ 34.0 & 5237.0 $\pm$ 18.0 & 39.93 $\pm$ 0.01 & 39.93 $\pm$ 0.01 \\
J0849$-$2351 & 08:49:51.68 & $-$23:51:25.07 & 0.127 & 2.89 & 4.03 & 2.82 & 156 & 921.0 $\pm$ 17.0 & 748.5 $\pm$ 7.4 & 39.33 $\pm$ 0.01 & 39.24 $\pm$ 0.01 \\
J0850$-$0318 & 08:50:27.96 & $-$03:18:16.95 & 0.162 & 3.77 & 5.13 & 4.33 & 7 & 426.0 $\pm$ 19.0 & 357.2 $\pm$ 9.7 & 39.30 $\pm$ 0.03 & 39.23 $\pm$ 0.02 \\
J0952$-$0136 & 09:52:19.10 & $-$01:36:43.52 & 0.020 & 0.41 & 0.66 & 0.37 & 46 & 21279.0 $\pm$ 79.0 & 20429.0 $\pm$ 40.0 & 39.01 $\pm$ 0.01 & 38.99 $\pm$ 0.01 \\
J1014$-$0418 & 10:14:20.65 & $-$04:18:40.62 & 0.058 & 1.26 & 5.29 & 4.30 & 52 & 380.0 $\pm$ 23.0 & 247.9 $\pm$ 9.8 & 38.23 $\pm$ 0.03 & 38.04 $\pm$ 0.02 \\
J1015$-$1652 & 10:15:03.21 & $-$16:52:14.45 & 0.432 & 11.53 & < 23.28 & < 12.16 & - & 73.3 $\pm$ 9.7 & 69.9 $\pm$ 4.9 & 39.35 $\pm$ 0.17 & 39.33 $\pm$ 0.16 \\
J1044$-$1826 & 10:44:48.70 & $-$18:26:51.53 & 0.113 & 2.55 & < 3.06 & < 0.94 & - & 1594.0 $\pm$ 36.0 & 1527.0 $\pm$ 20.0 & 39.45 $\pm$ 0.01 & 39.44 $\pm$ 0.01 \\
J1147$-$2145 & 11:47:38.87 & $-$21:45:07.95 & 0.219 & 5.26 & 11.31 & 4.16 & 35 & 2151.0 $\pm$ 31.0 & 2044.0 $\pm$ 16.0 & 40.21 $\pm$ 0.01 & 40.19 $\pm$ 0.01 \\
J1225$-$0418 & 12:25:27.18 & $-$04:18:57.40 & 0.137 & 3.13 & 6.74 & 4.83 & 145 & 316.0 $\pm$ 11.0 & 272.2 $\pm$ 5.8 & 38.82 $\pm$ 0.03 & 38.75 $\pm$ 0.03 \\
J1337$-$0902 & 13:37:39.58 & $-$09:02:27.88 & 0.080 & 1.77 & 6.98 & 2.76 & 27 & 194.8 $\pm$ 8.6 & 165.7 $\pm$ 4.1 & 38.18 $\pm$ 0.03 & 38.11 $\pm$ 0.03 \\
J1345$-$0259 & 13:45:24.69 & $-$02:59:39.75 & 0.085 & 1.88 & < 4.10 & < 3.39 & - & < 383.0 $\pm$ 11.0 & 383.0 $\pm$ 11.0 & < 38.53 $\pm$ 0.02 & 38.53 $\pm$ 0.02 \\
J1423$-$0923 & 14:23:50.25 & $-$09:23:17.02 & 0.068 & 1.49 & 3.62 & 3.23 & 158 & 158.0 $\pm$ 17.0 & 80.5 $\pm$ 5.9 & 37.97 $\pm$ 0.06 & 37.68 $\pm$ 0.05 \\
J1511$-$2119 & 15:11:59.80 & $-$21:19:01.46 & 0.044 & 0.95 & 0.89 & 0.68 & 149 & 18187.0 $\pm$ 71.0 & 17598.0 $\pm$ 36.0 & 39.66 $\pm$ 0.01 & 39.65 $\pm$ 0.01 \\
J1522$-$0644 & 15:22:28.76 & $-$06:44:41.73 & 0.083 & 1.83 & 6.37 & 3.12 & 152 & 5476.0 $\pm$ 34.0 & 4400.0 $\pm$ 17.0 & 39.70 $\pm$ 0.01 & 39.60 $\pm$ 0.01 \\
J1615$-$0936 & 16:15:19.06 & $-$09:36:13.33 & 0.065 & 1.42 & 1.87 & 1.49 & 150 & 682.0 $\pm$ 26.0 & 527.0 $\pm$ 12.0 & 38.59 $\pm$ 0.02 & 38.47 $\pm$ 0.01 \\
J1616$-$1014 & 16:16:10.79 & $-$10:14:06.40 & 0.078 & 1.70 & 4.31 & 1.86 & 162 & 104.5 $\pm$ 9.1 & 72.1 $\pm$ 3.8 & 37.89 $\pm$ 0.04 & 37.73 $\pm$ 0.03 \\
J1628$-$0304 & 16:28:48.39 & $-$03:04:08.01 & 0.093 & 2.06 & 5.53 & 1.75 & 151 & 96.0 $\pm$ 10.0 & 71.4 $\pm$ 4.1 & 37.95 $\pm$ 0.07 & 37.83 $\pm$ 0.05 \\
J1638$-$2055 & 16:38:30.90 & $-$20:55:24.39 & 0.027 & 0.57 & 0.97 & 0.46 & 150 & 3698.0 $\pm$ 26.0 & 3514.0 $\pm$ 13.0 & 38.52 $\pm$ 0.01 & 38.50 $\pm$ 0.01 \\
J1646$-$1124 & 16:46:10.39 & $-$11:24:03.90 & 0.074 & 1.62 & 1.61 & 0.96 & 36 & 10837.0 $\pm$ 59.0 & 10526.0 $\pm$ 31.0 & 39.90 $\pm$ 0.01 & 39.89 $\pm$ 0.01 \\
J1937$-$0613 & 19:37:33.00 & $-$06:13:04.99 & 0.010 & 0.22 & 1.07 & 0.86 & 36 & 12120.0 $\pm$ 730.0 & 5160.0 $\pm$ 230.0 & 38.20 $\pm$ 0.03 & 37.83 $\pm$ 0.02 \\
J2021$-$2235 & 20:21:04.35 & $-$22:35:18.52 & 0.185 & 4.35 & 5.58 & 4.05 & 179 & 9469.0 $\pm$ 48.0 & 8924.0 $\pm$ 25.0 & 40.70 $\pm$ 0.01 & 40.67 $\pm$ 0.01 \\
J2143$-$2958 & 21:43:06.12 & $-$29:58:16.26 & 0.120 & 2.72 & 22.99 & 6.99 & 9 & 212.0 $\pm$ 17.0 & 133.3 $\pm$ 6.6 & 38.70 $\pm$ 0.05 & 38.50 $\pm$ 0.04 \\
\hline
\end{tabular}
\end{scriptsize}
\label{radio_measurement}
\flushleft{\textbf{Notes.} Columns: (1) name, (2) right ascension, (3) declination, (4) redshift, (5) scale, (6) core major axis, (7) core minor axis, (8) core position angle, (9) integrated flux density, (10) peak flux density, (11) integrated luminosity, (12) peak luminosity.}
\end{table}
\end{landscape}

\begin{table*}
\centering
\caption{The radio and optical properties of the southern NLS1 sample}
\begin{tabular}{ccccccccc}
\hline
\hline
Name & \multicolumn{2}{c}{Radio loudness} & \multicolumn{2}{c}{Flux concentration} & \multicolumn{2}{c}{Spectral slope} & $S_{4400\rm{\AA}}$ & $\log M_{\rm{BH}}$ \\
- & $R_{5.5}$ & RL/RQ & $f$ & C/D & $\alpha_{\rm{in-band}}$ & F/S & (10$^{-17}$ erg \, s$^{-1}$ cm$^{-2}$ $\rm{\AA}^{-1}$) & ($M_{\odot}$) \\
\hline
J0000$-$0541 & 5.1 $\pm$ 1.2 & RQ & 0.91 $\pm$ 0.01 & C & 1.20 $\pm$ 0.11 & S & 36.80 $\pm$ 8.74 & 6.36 $\pm$ 0.14 \\
J0022$-$1039 & 3.1 $\pm$ 0.7 & RQ & 0.83 $\pm$ 0.05 & C & 1.07 $\pm$ 0.40 & S & 26.64 $\pm$ 6.02 & 7.23 $\pm$ 0.12 \\
J0122$-$2646 & 38.2 $\pm$ 19.3 & RL & 0.88 $\pm$ 0.11 & C & 1.08 $\pm$ 0.50 & S & 4.00 $\pm$ 1.97 & 6.59 $\pm$ 0.17 \\
J0138$-$0109 & 0.7 $\pm$ 0.1 & RQ & 0.91 $\pm$ 0.06 & C & 0.09 $\pm$ 0.47 & F & 54.24 $\pm$ 6.55 & 7.29 $\pm$ 0.11 \\
J0159$-$1128 & 0.4 $\pm$ 0.1 & RQ & 0.79 $\pm$ 0.07 & C & $-$0.71 $\pm$ 0.50 & F & 44.76 $\pm$ 12.30 & 6.99 $\pm$ 0.12 \\
J0203$-$1247 & 0.7 $\pm$ 0.1 & RQ & 0.88 $\pm$ 0.03 & C & 0.92 $\pm$ 0.36 & S & 80.40 $\pm$ 7.89 & 6.46 $\pm$ 0.11 \\
J0212$-$0201 & 6.4 $\pm$ 1.7 & RQ & 0.97 $\pm$ 0.04 & C & 0.35 $\pm$ 0.29 & F & 5.82 $\pm$ 1.52 & 7.00 $\pm$ 0.17 \\
J0213$-$0551 & 1.1 $\pm$ 0.2 & RQ & 0.82 $\pm$ 0.05 & C & 0.67 $\pm$ 0.49 & S & 29.53 $\pm$ 5.37 & 6.76 $\pm$ 0.12 \\
J0230$-$0859 & 0.3 $\pm$ 0.1 & RQ & 0.83 $\pm$ 0.02 & C & $-$1.91 $\pm$ 0.73 & F & 628.89 $\pm$ 62.85 & 6.29 $\pm$ 0.11 \\
J0239$-$1118 & 4.0 $\pm$ 1.2 & RQ & 1.00 $\pm$ 0.02 & C & 0.66 $\pm$ 0.24 & S & 13.19 $\pm$ 3.78 & 6.82 $\pm$ 0.12 \\
J0347$-$1325 & 3.0 $\pm$ 0.8 & RQ & 1.00 $\pm$ 0.01 & C & 0.78 $\pm$ 0.10 & S & 34.22 $\pm$ 9.38 & 6.87 $\pm$ 0.11 \\
J0350$-$1025 & 0.3 $\pm$ 0.1 & RQ & 1.00 $\pm$ 0.04 & C & 3.11 $\pm$ 0.53 & S & 29.82 $\pm$ 6.59 & 6.91 $\pm$ 0.12 \\
J0351$-$0526 & 0.3 $\pm$ 0.1 & RQ & 0.76 $\pm$ 0.08 & C & $-$0.04 $\pm$ 0.72 & F & 65.58 $\pm$ 6.08 & 6.53 $\pm$ 0.11 \\
J0354$-$1340 & 5.7 $\pm$ 0.6 & RQ & 1.00 $\pm$ 0.01 & C & 0.12 $\pm$ 0.06 & F & 145.36 $\pm$ 15.70 & 6.99 $\pm$ 0.11 \\
J0400$-$2500 & 2.2 $\pm$ 0.1 & RQ & 1.00 $\pm$ 0.02 & C & $-$0.81 $\pm$ 0.22 & F & 91.20 $\pm$ 5.71 & 6.67 $\pm$ 0.11 \\
J0413$-$0050 & 0.6 $\pm$ 0.1 & RQ & 0.87 $\pm$ 0.07 & C & 0.28 $\pm$ 0.68 & F & 43.82 $\pm$ 10.74 & 6.46 $\pm$ 0.12 \\
J0422$-$1854 & 2.3 $\pm$ 0.3 & RQ & 0.95 $\pm$ 0.02 & C & 0.72 $\pm$ 0.22 & S & 78.90 $\pm$ 9.66 & 6.57 $\pm$ 0.11 \\
J0435$-$1643 & 0.4 $\pm$ 0.1 & RQ & 0.94 $\pm$ 0.04 & C & $-$1.39 $\pm$ 0.79 & F & 60.29 $\pm$ 10.60 & 6.77 $\pm$ 0.11 \\
J0436$-$1022 & 0.9 $\pm$ 0.1 & RQ & 0.87 $\pm$ 0.05 & C & 0.60 $\pm$ 0.33 & S & 869.95 $\pm$ 120.41 & 6.90 $\pm$ 0.11 \\
J0447$-$0403 & 0.1 $\pm$ 0.1 & RQ & 0.36 $\pm$ 0.05 & D & 0.61 $\pm$ 1.12 & S & 231.28 $\pm$ 21.70 & 6.63 $\pm$ 0.11 \\
J0447$-$0508 & 2.5 $\pm$ 0.3 & RQ & 0.89 $\pm$ 0.01 & C & 0.83 $\pm$ 0.13 & S & 260.87 $\pm$ 27.98 & 6.85 $\pm$ 0.11 \\
J0452$-$2953 & 33.6 $\pm$ 4.0 & RL & 0.76 $\pm$ 0.01 & C & 0.95 $\pm$ 0.07 & S & 16.29 $\pm$ 1.94 & 6.97 $\pm$ 0.11 \\
J0455$-$1456 & 0.2 $\pm$ 0.1 & RQ & 0.98 $\pm$ 0.04 & C & 0.22 $\pm$ 0.38 & F & 103.37 $\pm$ 13.08 & 7.02 $\pm$ 0.11 \\
J0549$-$2425 & 2.1 $\pm$ 0.2 & RQ & 1.00 $\pm$ 0.01 & C & 0.63 $\pm$ 0.08 & S & 118.90 $\pm$ 12.51 & 6.58 $\pm$ 0.11 \\
J0622$-$2317 & 0.6 $\pm$ 0.1 & RQ & 0.30 $\pm$ 0.02 & D & 0.31 $\pm$ 0.45 & F & 238.19 $\pm$ 26.21 & 6.47 $\pm$ 0.11 \\
J0842$-$0349 & 0.8 $\pm$ 0.1 & RQ & 0.83 $\pm$ 0.05 & C & 1.50 $\pm$ 0.56 & S & 70.58 $\pm$ 11.82 & 7.43 $\pm$ 0.12 \\
J0845$-$0732 & 0.7 $\pm$ 0.1 & RQ & 0.49 $\pm$ 0.06 & D & 1.68 $\pm$ 0.88 & S & 60.75 $\pm$ 9.49 & 6.71 $\pm$ 0.11 \\
J0846$-$1214 & 14.3 $\pm$ 1.6 & RL & 0.99 $\pm$ 0.01 & C & 0.90 $\pm$ 0.03 & S & 60.24 $\pm$ 6.78 & 6.75 $\pm$ 0.11 \\
J0849$-$2351 & 1.6 $\pm$ 0.2 & RQ & 0.81 $\pm$ 0.02 & C & 0.97 $\pm$ 0.19 & S & 91.99 $\pm$ 12.53 & 6.73 $\pm$ 0.11 \\
J0850$-$0318 & 0.6 $\pm$ 0.1 & RQ & 0.84 $\pm$ 0.04 & C & 2.15 $\pm$ 0.27 & S & 114.10 $\pm$ 19.29 & 7.17 $\pm$ 0.11 \\
J0952$-$0136 & 5.4 $\pm$ 0.9 & RQ & 0.96 $\pm$ 0.01 & C & 0.91 $\pm$ 0.03 & S & 634.94 $\pm$ 104.25 & 6.33 $\pm$ 0.11 \\
J1014$-$0418 & 0.1 $\pm$ 0.1 & RQ & 0.65 $\pm$ 0.05 & C & 0.79 $\pm$ 0.53 & S & 400.27 $\pm$ 31.77 & 6.86 $\pm$ 0.11 \\
J1015$-$1652 & 0.2 $\pm$ 0.1 & RQ & 0.95 $\pm$ 0.14 & C & 0.47 $\pm$ 1.00 & F & 58.22 $\pm$ 13.34 & 7.41 $\pm$ 0.12 \\
J1044$-$1826 & 5.1 $\pm$ 1.1 & RQ & 0.96 $\pm$ 0.03 & C & 0.84 $\pm$ 0.17 & S & 50.73 $\pm$ 10.59 & 6.49 $\pm$ 0.11 \\
J1147$-$2145 & 6.9 $\pm$ 0.7 & RQ & 0.95 $\pm$ 0.02 & C & 0.88 $\pm$ 0.10 & S & 50.53 $\pm$ 5.15 & 6.99 $\pm$ 0.11 \\
J1225$-$0418 & 0.5 $\pm$ 0.1 & RQ & 0.86 $\pm$ 0.04 & C & $-$1.19 $\pm$ 0.42 & F & 99.81 $\pm$ 19.62 & 6.90 $\pm$ 0.11 \\
J1337$-$0902 & 1.3 $\pm$ 0.2 & RQ & 0.85 $\pm$ 0.04 & C & $-$0.41 $\pm$ 0.71 & F & 24.62 $\pm$ 3.48 & 6.28 $\pm$ 0.12 \\
J1345$-$0259 & 0.8 $\pm$ 0.1 & RQ & 1.00 $\pm$ 0.04 & C & $-$0.31 $\pm$ 0.26 & F & 82.22 $\pm$ 10.76 & 6.68 $\pm$ 0.11 \\
J1423$-$0923 & 0.6 $\pm$ 0.1 & RQ & 0.51 $\pm$ 0.07 & C & 0.20 $\pm$ 1.33 & F & 42.12 $\pm$ 6.99 & 6.39 $\pm$ 0.11 \\
J1511$-$2119 & 9.8 $\pm$ 1.2 & RQ & 0.97 $\pm$ 0.01 & C & 0.81 $\pm$ 0.01 & S & 299.76 $\pm$ 37.46 & 6.63 $\pm$ 0.11 \\
J1522$-$0644 & 6.0 $\pm$ 1.1 & RQ & 0.80 $\pm$ 0.01 & C & 0.64 $\pm$ 0.06 & S & 148.74 $\pm$ 27.95 & 6.66 $\pm$ 0.11 \\
J1615$-$0936 & 0.2 $\pm$ 0.1 & RQ & 0.77 $\pm$ 0.03 & C & 1.15 $\pm$ 0.36 & S & 712.21 $\pm$ 147.41 & 6.94 $\pm$ 0.11 \\
J1616$-$1014 & 0.1 $\pm$ 0.1 & RQ & 0.69 $\pm$ 0.07 & C & $-$0.27 $\pm$ 0.75 & F & 125.07 $\pm$ 15.68 & 6.69 $\pm$ 0.11 \\
J1628$-$0304 & 0.1 $\pm$ 0.1 & RQ & 0.74 $\pm$ 0.09 & C & $-$1.72 $\pm$ 1.24 & F & 139.85 $\pm$ 22.90 & 6.72 $\pm$ 0.11 \\
J1638$-$2055 & 0.6 $\pm$ 0.1 & RQ & 0.95 $\pm$ 0.01 & C & 0.86 $\pm$ 0.06 & S & 992.50 $\pm$ 150.57 & 6.66 $\pm$ 0.11 \\
J1646$-$1124 & 3.9 $\pm$ 0.3 & RQ & 0.97 $\pm$ 0.01 & C & 1.03 $\pm$ 0.03 & S & 450.90 $\pm$ 31.46 & 7.07 $\pm$ 0.11 \\
J1937$-$0613 & 1.6 $\pm$ 0.2 & RQ & 0.43 $\pm$ 0.03 & D & 0.64 $\pm$ 0.39 & S & 1253.46 $\pm$ 127.85 & 6.03 $\pm$ 0.11 \\
J2021$-$2235 & 35.0 $\pm$ 3.9 & RL & 0.94 $\pm$ 0.01 & C & 0.95 $\pm$ 0.03 & S & 43.99 $\pm$ 4.88 & 6.65 $\pm$ 0.11 \\
J2143$-$2958 & 0.3 $\pm$ 0.1 & RQ & 0.63 $\pm$ 0.06 & C & 2.25 $\pm$ 0.68 & S & 123.17 $\pm$ 13.41 & 7.08 $\pm$ 0.11 \\
\hline
\end{tabular}
\label{classify+optical}
\flushleft{\textbf{Notes.} Columns: (1) name, (2) radio loudness based on 5.5 GHz flux, (3) radio classification based on 5.5 GHz flux, (4) flux concentration, (5) concentration classification, (6) in-band spectral index between 5 and 6 GHz, (7) spectral classification based on in-band spectral index, (8) flux density in optical B-band, (9) black hole mass.}
\end{table*}

\begin{table*}
\centering
\caption{The radio measurements of 21 NLS1s detected at 1.4 GHz}
\begin{tabular}{cccccccc}
\hline
\hline
Name & Instrument & $S_{1.4}$ & $\log L_{1.4}$ & \multicolumn{2}{c}{Radio loudness} & \multicolumn{2}{c}{Spectral slope} \\
- & - & (mJy) & (erg \, s$^{-1}$) & $R_{1.4}$ & RL/RQ & $\alpha_{\rm{1.4-5.5}}$ & F/S \\
\hline
J0000$-$0541 & FIRST & 3.5 $\pm$ 0.3 & 39.03 $\pm$ 0.04 & 7.79 $\pm$ 1.98 & RQ & 0.80 $\pm$ 0.07 & S \\
J0022$-$1039 & FIRST & 1.4 $\pm$ 0.1 & 40.04 $\pm$ 0.04 & 4.31 $\pm$ 1.06 & RQ & 0.73 $\pm$ 0.08 & S \\
J0122$-$2646 & NVSS & 3.8 $\pm$ 0.5 & 40.52 $\pm$ 0.06 & 77.86 $\pm$ 39.73 & RL & 1.02 $\pm$ 0.12 & S \\
J0230$-$0859 & FIRST & 1.2 $\pm$ 0.2 & 36.97 $\pm$ 0.06 & 0.15 $\pm$ 0.02 & RQ & $-$0.04 $\pm$ 0.10 & F \\
J0354$-$1340 & NVSS & 14.9 $\pm$ 1.1 & 39.46 $\pm$ 0.03 & 8.40 $\pm$ 1.10 & RQ & 0.56 $\pm$ 0.05 & S \\
J0400$-$2500 & NVSS & 4.1 $\pm$ 0.6 & 39.13 $\pm$ 0.06 & 3.68 $\pm$ 0.59 & RQ & 0.87 $\pm$ 0.11 & S \\
J0422$-$1854 & NVSS & 2.8 $\pm$ 0.5 & 38.58 $\pm$ 0.08 & 2.91 $\pm$ 0.63 & RQ & 0.67 $\pm$ 0.13 & S \\
J0436$-$1022 & NVSS & 17.0 $\pm$ 0.7 & 38.83 $\pm$ 0.02 & 1.60 $\pm$ 0.23 & RQ & 0.95 $\pm$ 0.05 & S \\
J0452$-$2953 & NVSS & 9.5 $\pm$ 0.5 & 40.51 $\pm$ 0.02 & 47.79 $\pm$ 6.24 & RL & 0.76 $\pm$ 0.04 & S \\
J0549$-$2425 & NVSS & 2.4 $\pm$ 0.5 & 38.19 $\pm$ 0.09 & 1.65 $\pm$ 0.39 & RQ & 0.31 $\pm$ 0.15 & F \\
J0622$-$2317 & NVSS & 4.3 $\pm$ 0.5 & 38.30 $\pm$ 0.05 & 1.48 $\pm$ 0.24 & RQ & 1.12 $\pm$ 0.09 & S \\
J0846$-$1214 & NVSS & 15.3 $\pm$ 0.7 & 39.79 $\pm$ 0.02 & 20.81 $\pm$ 2.53 & RL & 0.77 $\pm$ 0.03 & S \\
J0952$-$0136 & FIRST & 59.8 $\pm$ 0.1 & 38.86 $\pm$ 0.01 & 7.72 $\pm$ 1.27 & RQ & 0.76 $\pm$ 0.01 & S \\
J1044$-$1826 & NVSS & 4.4 $\pm$ 0.5 & 39.30 $\pm$ 0.05 & 7.11 $\pm$ 1.69 & RQ & 0.74 $\pm$ 0.08 & S \\
J1147$-$2145 & NVSS & 5.7 $\pm$ 0.5 & 40.03 $\pm$ 0.04 & 9.24 $\pm$ 1.24 & RQ & 0.71 $\pm$ 0.06 & S \\
J1511$-$2119 & NVSS & 46.9 $\pm$ 1.5 & 39.48 $\pm$ 0.01 & 12.82 $\pm$ 1.65 & RL & 0.69 $\pm$ 0.02 & S \\
J1522$-$0644 & FIRST & 14.4 $\pm$ 0.1 & 39.53 $\pm$ 0.01 & 7.92 $\pm$ 1.49 & RQ & 0.71 $\pm$ 0.01 & S \\
J1638$-$2055 & NVSS & 6.8 $\pm$ 0.5 & 38.19 $\pm$ 0.03 & 0.56 $\pm$ 0.09 & RQ & 0.45 $\pm$ 0.05 & F \\
J1646$-$1124 & NVSS & 38.3 $\pm$ 1.6 & 39.85 $\pm$ 0.02 & 6.96 $\pm$ 0.57 & RQ & 0.92 $\pm$ 0.03 & S \\
J1937$-$0613 & NVSS & 42.2 $\pm$ 1.7 & 38.15 $\pm$ 0.02 & 2.76 $\pm$ 0.30 & RQ & 0.91 $\pm$ 0.05 & S \\
J2021$-$2235 & NVSS & 24.6 $\pm$ 0.9 & 40.50 $\pm$ 0.02 & 45.82 $\pm$ 5.35 & RL & 0.70 $\pm$ 0.03 & S \\
\hline
\end{tabular}
\label{first+nvss}
\flushleft{\textbf{Notes.} Columns: (1) name, (2) instrument, (3) flux density, (4) luminosity, (5) radio loudness based on 1.4 GHz flux, (6) radio classification based on 1.4 GHz flux, (7) spectral index between 1.4 and 5.5 GHz, (8) spectral classification based on spectral index between 1.4 and 5.5 GHz.}
\end{table*}

\label{lastpage}
\end{document}